\documentclass[preprint2]{aastex6}
\usepackage{graphicx}
\usepackage{amsmath}
\usepackage{longtable}

\newbox\rotatingbox
\def\rotatetablecbw{\setbox\rotatingbox=\vbox\bgroup}
\def\endrotatetablecbw{\egroup\rotatebox{90}{\hbox to \textheight{\vbox{\unvbox\rotatingbox}}}}

\newcommand{\nhi}{$N_{\rm H\,I}$}

\newcommand{\mathnhi}{N_{\rm H\,I}}
\newcommand{\nmgii}{$N_{\rm Mg\,II}$}

\newcommand{\hi}{{\rm H}$\;${\small\rm I}\relax}

\newcommand{\neviii}{{\rm Ne}$\;${\small\rm VIII}\relax}

\newcommand{\cii}{{\rm C}$\;${\small\rm II}\relax}

\newcommand{\ciii}{{\rm C}$\;${\small\rm III}\relax}

\newcommand{\oii}{{\rm O}$\;${\small\rm II}\relax}
\newcommand{\oiii}{{\rm O}$\;${\small\rm III}\relax}

\newcommand{\ovi}{{\rm O}$\;${\small\rm VI}\relax}

\newcommand{\mgi}{{\rm Mg}$\;${\small\rm I}\relax}
\newcommand{\mgii}{{\rm Mg}$\;${\small\rm II}\relax}

\newcommand{\siii}{{\rm Si}$\;${\small\rm II}\relax}

\newcommand{\siiii}{{\rm Si}$\;${\small\rm III}\relax}

\newcommand{\feii}{{\rm Fe}$\;${\small\rm II}\relax}

\newcommand{\hit}{{\rm H}$\;${\scriptsize\rm I}\relax}

\newcommand{\ciit}{{\rm C}$\;${\scriptsize\rm II}\relax}

\newcommand{\ciiit}{{\rm C}$\;${\scriptsize\rm III}\relax}
\newcommand{\civt}{{\rm C}$\;${\scriptsize\rm IV}\relax}

\newcommand{\niit}{{\rm N}$\;${\scriptsize\rm II}\relax}

\newcommand{\oiit}{{\rm O}$\;${\scriptsize\rm II}\relax}
\newcommand{\oiiit}{{\rm O}$\;${\scriptsize\rm III}\relax}
\newcommand{\oivt}{{\rm O}$\;${\scriptsize\rm IV}\relax}

\newcommand{\ovit}{{\rm O}$\;${\scriptsize\rm VI}\relax}

\newcommand{\mgiit}{{\rm Mg}$\;${\scriptsize\rm II}\relax}

\newcommand{\siiit}{{\rm Si}$\;${\scriptsize\rm II}\relax}
\newcommand{\svt}{{\rm S}$\;${\scriptsize\rm V}\relax}
\newcommand{\siiiit}{{\rm Si}$\;${\scriptsize\rm III}\relax}

\newcommand{\aliit}{{\rm Al}$\;${\scriptsize\rm II}\relax}
\newcommand{\Siiit}{{\rm S}$\;${\scriptsize\rm III}\relax}
\newcommand{\siivt}{{\rm Si}$\;${\scriptsize\rm IV}\relax}
\newcommand{\sivt}{{\rm S}$\;${\scriptsize\rm IV}\relax}
\newcommand{\svit}{{\rm S}$\;${\scriptsize\rm VI}\relax}

\newcommand{\hst}{{\it HST}}

\newcommand{\numWPLLS}{20}

\newcommand{\WPLLSLowmet}{-1.96} 
\newcommand{\WPLLSLowmetError}{0.16} 
\newcommand{\WPLLSLowmetPercent}{1.1\%} 
\newcommand{\WPLLSHighmet}{-0.36} 
\newcommand{\WPLLSHighmetError}{0.05} 
\newcommand{\WPLLSHighmetPercent}{44\%} 

\newcommand{\pDipWPLLS}{97.4\%}
\newcommand{\pGMMWPLLS}{99.8\%}
\newcommand{\pGMMhomoscedasticLowmetWPLLS}{-1.74\pm0.09}
\newcommand{\pGMMhomoscedasticHighmetWPLLS}{-0.42\pm0.06}
\newcommand{\pGMMhomoscedasticDispersionWPLLS}{0.23\pm0.05}

\newcommand{\LPLLSLowmet}{-1.68} 
\newcommand{\LPLLSLowmetError}{0.08} 
\newcommand{\LPLLSHighmet}{-0.32} 
\newcommand{\LPLLSHighmetError}{0.11} 

\newcommand{\numLWPLLS}{44}
\newcommand{\numLWPLLSLowmet}{25} 
\newcommand{\numLWPLLSHighmet}{19} 
\newcommand{\fLowmetLWPLLS}{($57\pm8$)\%}

\newcommand{\LWPLLSLowmet}{-1.87} 
\newcommand{\LWPLLSLowmetError}{0.11} 
\newcommand{\LWPLLSLowmetPercent}{1.3\%} 
\newcommand{\LWPLLSHighmet}{-0.32} 
\newcommand{\LWPLLSHighmetError}{0.07} 
\newcommand{\LWPLLSHighmetPercent}{48\%} 

\newcommand{\pDipLWPLLS}{95.3\%} 
\newcommand{\pGMMLWPLLS}{$>$99.9\%}

\newcommand{\numWLLS}{10}

\newcommand{\numLWLLS}{11}

\newcommand{\LWLLSAllmet}{-1.00} 
\newcommand{\LWLLSAllmetError}{0.15} 

\newcommand{\numL}{25}

\newcommand{\numW}{30}
\newcommand{\numWUpperLower}{10} 
\newcommand{\numWExcluded}{17} 

\newcommand{\numLW}{55} 
\newcommand{\numLWLowmet}{31} 
\newcommand{\numLWHighmet}{24} 
\newcommand{\fLowmetLW}{($56\pm8$)\%}

\newcommand{\LWLowmet}{-1.76} 
\newcommand{\LWLowmetError}{0.09} 
\newcommand{\LWHighmet}{-0.33} 
\newcommand{\LWHighmetError}{0.07} 

\newcommand{\pDipLW}{82.4\%}
\newcommand{\pGMMLW}{99.6\%}

\newcommand{\numLymanBreaks}{47} 
\newcommand{\numMetalcomparison}{22}

\submitted{Received by ApJ 2016 May 25; accepted by ApJ 2016 August 5}
\shortauthors{Wotta et al.}
\shorttitle{The Metallicity Distribution of the CGM at $z\lesssim1$}

\begin{document}
\title{Low-metallicity Absorbers Account for Half of the Dense Circumgalactic Gas at $\lowercase{z}\lesssim1$\altaffilmark{1,2}}

\author{Christopher B.\ Wotta\altaffilmark{3},
    Nicolas Lehner\altaffilmark{3},
    J.\ Christopher Howk\altaffilmark{3},
    John M.\ O'Meara\altaffilmark{4},
    and J.\ Xavier Prochaska\altaffilmark{5}
    }
\altaffiltext{1}{Based on observations made with the NASA/ESA Hubble Space Telescope,
obtained at the Space Telescope Science Institute, which is operated by the Association of Universities for Research in Astronomy, Inc.\ under NASA contract No.\ NAS5-26555.}
\altaffiltext{2}{Based on observations made with the Large Binocular Telescope (LBT). The LBT is an international collaboration among institutions in the United States, Italy and Germany. LBT Corporation partners are: The University of Arizona on behalf of the Arizona university system; Istituto Nazionale di Astrofisica, Italy; LBT Beteiligungsgesellschaft, Germany, representing the Max-Planck Society, the Astrophysical Institute Potsdam, and Heidelberg University; The Ohio State University, and The Research Corporation, on behalf of The University of Notre Dame, University of Minnesota and University of Virginia.}
\altaffiltext{3}{Department of Physics, University of Notre Dame, Notre Dame, IN 46556}
\altaffiltext{4}{Department of Physics, Saint Michael's College, Colchester, VT 05439}
\altaffiltext{5}{UCO/Lick Observatory, University of California, Santa Cruz, CA 95064}

\begin{abstract}
We present an analysis of the metallicity distribution of the dense circumgalactic medium (CGM) of galaxies at $0.1\lesssim z\lesssim1.1$ as probed by partial Lyman limit systems (pLLSs, $16.1<\log\mathnhi<17.2$) and LLSs ($17.2\le\log\mathnhi<17.7$ in our sample). The new \hi-selected sample, drawn from our {\it HST} COS G140L snapshot survey of 61 QSOs, has \numWPLLS\ pLLSs and \numWLLS\ LLSs. Combined with our previous survey, we have a total of \numLWPLLS\ pLLSs and \numLWLLS\ LLSs. We find that the metallicity distribution of the pLLSs is bimodal at $z\lesssim1$, with a minimum at $[{\rm X/H}]=-1$. The low-metallicity peak comprises \fLowmetLWPLLS\ of the pLLSs and is centered at $[{\rm X/H}]\simeq\LWPLLSLowmet$ (\LWPLLSLowmetPercent\ solar metallicity), while the high-metallicity peak is centered at $[{\rm X/H}]\simeq\LWPLLSHighmet$ (\LWPLLSHighmetPercent\ solar metallicity). Although the sample of LLSs is still small, there is some evidence that the metallicity distributions of the LLSs and pLLSs are different, with a far lower fraction of very metal-poor ($[{\rm X/H}]<-1.4$) LLSs than pLLSs. The fraction of LLSs with $[{\rm X/H}]<-1$ is similar to that found in pLLSs ($\sim$56\%). However, higher \hi\ column density absorbers ($\log\mathnhi>19.0$) show a much lower fraction of metal-poor gas; therefore, the metallicity distribution of gas in and around galaxies depends sensitively on \nhi\ at $z\lesssim1$. We interpret the high-metallicity ($[{\rm X/H}]\ge-1$) pLLSs and LLSs as arising in outflows, recycling winds, and tidally-stripped gas around galaxies. The low-metallicity pLLSs and LLSs imply that the CGM of $z\lesssim1$ galaxies is also host to a substantial mass of cool, dense, low-metallicity gas that may ultimately accrete onto the galaxies.
\end{abstract}


\defcitealias{Lehner2013}{\rm L13}
\newcommand{\Lthirteen}{{\rm L13}}
\defcitealias{Haardt1996}{HM05}
\defcitealias{Haardt2012}{HM12}


\section{Introduction}\label{s-intro}

Large-scale flows of gas through the circumgalactic medium (CGM) into and out of galaxies are critical to shaping the evolution of the galaxies. Such flows are invoked to explain the galactic mass--metallicity relation, galaxy color bimodalities, and the maintenance of star formation in galaxies over billions of years \citep[e.g.][]{Keres2005,Dekel2006,Faucher-Giguere2011}. Given the importance of such flows in the CGM, their properties can be used as tests of our theories for galaxy evolution. The installation of the Cosmic Origins Spectrograph (COS) on the {\it Hubble Space Telescope} (\hst) has provided a huge advance in our ability to empirically characterize the CGM of low-redshift galaxies. COS has allowed us to assess the connection between the CGM and critical components of galaxy evolution, but also to demonstrate that the CGM represents a major reservoir of baryons and metals, with at least as many metals and baryons as found in stars within galaxies \citep[e.g.,][]{Tumlinson2011, Stocke2013, Werk2014, Peeples2014, Bordoloi2014, Liang2014}.

Characterizing gas flows through the CGM requires, however, methods to differentiate outflows from inflows. The most direct method is to study the gas directly toward galaxies using absorption against the galaxies' own stellar light (``down-the-barrel" experiment). Material seen blueshifted or redshifted relative to a galaxy's systemic velocity is a signature of outflow or inflow, respectively \citep[e.g.,][]{Heckman2000, Martin2005, Tremonti2007, Weiner2009, Steidel2010, Rubin2011, Chisholm2016}. Through absorption observed in the strong transitions of \mgii\ and \feii, this method has revealed that outflows are nearly ubiquitous in star-forming galaxies, although only a few galaxies having significant redshifted (infalling) absorption \citep{Rubin2012, Martin2012}. The difficulty with this method is that it almost never provides information on the metallicity of the observed gas. In fact the experiment requires metal-enriched gas and is therefore biased against the accretion of low-metallicity material. Metal-poor cold streams of gas as observed in simulations may be missed in this approach, and inflows observed in ``down-the-barrel" galaxies may trace metal-enriched gas being recycled onto the central galaxy.

\setcitestyle{notesep={; }}
Another method for distinguishing outflowing from inflowing matter relies on determining if the properties of CGM absorption seen in the spectra of background QSOs depend on the geometry of the central galaxy with respect to the sightline. Simulations show a preference for inflowing streams to align with the major axis of galaxies and for metal-enriched outflows to align with the minor axis \citep[e.g.,][]{Brook2011, Stewart2011a, vandeVoort2012a, Shen2013}. Thus strong metal absorption seen projected along the minor axis of the central galaxy is potentially consistent with an origin in feedback-driven outflows. \citet[see also \citealt{Bouche2012}; \citealt{Kacprzak2012a}]{Bordoloi2011} have demonstrated that the properties of CGM absorption depend on the geometry of the central galaxy with respect to the line of sight, with strong \mgii\ absorbers at small impact parameters ($\rho\lesssim40$ kpc) being primarily found along galaxies' minor axes (along the pole). However, as with ``down-the-barrel'' experiments, this technique has so far only been applied to significant samples of \mgii-selected absorbers. There is no information on the intrinsic metallicity of the gas, and the selection based on strong \mgii\ is potentially biased toward relatively high-metallicity gas (at least in the column density regime expected for many of the flows through the CGM; see \citealt{vandeVoort2012a}, but see also \citealt{Kacprzak2012c}).
\setcitestyle{notesep={, }}

Finally, to distinguish potentially infalling matter from outflows, recycling winds, or tidally-stripped gas, the metallicity of the CGM gas seen in absorption toward QSOs can be used as a diagnostic of its origin. Selection of CGM absorbers on the basis of an \hi\ column density criterion provides a census of the metallicity of CGM gas that should be free of the metallicity biases that may be inherent to the other approaches. This approach does not provide any information on the direction of the flows relative to the galaxy; it uses the metallicity as an indirect discriminator of potentially infalling gas from other forms.

Using the metallicity as a discriminant of potentially infalling material requires: (1) the selection of absorbers seen in QSO spectra that are known to probe the CGM of galaxies; and (2) the selection of absorbers where the metallicities of the metal-enriched outflows can be distinguished from those of infalling gas. The former condition requires selection of absorbers with column densities at least $\log\mathnhi\gtrsim 14.5$, those which show strong association with galaxy halos \citep{Prochaska2011b}. For the characteristic spectra collected at $z\lesssim1$, the latter condition requires sufficiently high \hi\ column densities for metal lines to be well detected in UV spectra, typically $\log\mathnhi\gtrsim16.0$ for 17 km s$^{-1}$ resolution COS spectra. At these column densities, absorbers characterized by $10^{16} \le \mathnhi < 10^{17.2}$ cm$^{-2}$ --- the partial-Lyman limit systems (hereafter pLLSs) --- and $10^{17.2} \le \mathnhi < 10^{19.0}$ cm$^{-2}$ --- the Lyman limit systems (hereafter LLSs) --- are expected to probe cool, dense streams through the CGM of galaxies based on empirical evidence \citep[e.g.,][]{Lanzetta1995, Penton2002, Bowen2002, Chen2005, Lehner2009, Lehner2013} and theoretical evidence \citep[e.g.,][]{Fumagalli2011a, Faucher-Giguere2011, vandeVoort2012b}. Furthermore, the overdensity required to produce a pLLS or LLS places these absorbers in the CGM. From \citet{Lehner2013}, the density of a pLLS/LLS is typically $n_{\rm H}\sim10^{-2.5}$ cm$^{-3}$. Thus, the baryon overdensity required to produce a pLLS or LLS at $z=0.7$ is $\delta\rho_{m}\equiv\rho_{m}/(\Omega_m\rho_{\rm c})\sim900$, where $\delta\rho_{m}$ is the mass overdensity relative to the mean matter density of the Universe, $\rho_{m}$ is the mass density of the pLLS/LLS, $\Omega_m$ is the matter density parameter, and $\rho_{\rm c}$ is the critical density of the Universe. This is in line with the overdensities of galaxy halos seen in simulations and analytically \citep{Schaye2001b}, with $\delta\rho_{m}\sim{\rm few}\times10^2$--$10^3$.

Using a sample of 28 \hi-selected absorbers (23 pLLSs and 5 LLSs) at $0.1\lesssim z\lesssim1$, \citet[hereafter \citetalias{Lehner2013}]{Lehner2013} have recently demonstrated that the dense gas in the CGM of $z\lesssim1$ galaxies has a bimodal metallicity distribution function (MDF), with an equal number of absorbers in the low-metallicity ($\left\langle Z\right\rangle\lesssim 0.03 Z_\sun$) and high-metallicity ($\left\langle Z\right\rangle\sim 0.4 Z_\sun$) branches. This is strong empirical evidence for low-metallicity gas well within the dark matter halos of galaxies at $z\lesssim 1$. This metal-poor gas is not rare, and it has at least as much mass density in the $z\lesssim1$ Universe as the metal-enriched gas, at least for CGM gas with $16 \lesssim \log\mathnhi \lesssim 17.5$. \citetalias{Lehner2013} argue that the high-metallicity branch of the population likely traces galactic winds, recycled outflows, and tidally-stripped gas, while all the observable properties of the low-metallicity pLLSs+LLSs are consistent with those expected of cold accretion flows \citep[e.g.,][]{Fumagalli2011a, Faucher-Giguere2011, vandeVoort2012a}.


In this paper we now reexamine the MDF of the dense CGM, doubling the sample of pLLSs and LLSs and improving the \nhi\ coverage at higher \hi\ column densities. With the selection criterion used in this work, there is no bias against strong pLLSs or LLSs in the UV spectra of QSOs. As we demonstrate in this work, the combined \Lthirteen\ and new (hereafter W16) sample contains \numLW\ pLLS+LLS absorbers at $z\lesssim 1$ with $16.1<\log\mathnhi<17.7$, covering a broader range of \hi\ column densities than \Lthirteen\ with a more representative sampling of the column density distribution function at these redshifts.

We also introduce a new approach to estimating the metallicities of pLLSs/LLSs.\footnote{We use the notation ``pLLSs+LLSs'' to refer to ``pLLSs and LLSs, together'' (e.g., when discussing a sample). We use the notation ``pLLSs/LLSs'' when referring to ``pLLSs and/or LLSs'' (e.g., when discussing how we estimate the metallicities of the absorbers).}$^,$\footnote{Our new method estimates the gas-phase abundance of Mg (an $\alpha$ element) relative to H. We hereafter use ``abundance'' and ``metallicity'' interchangeably.} Since pLLSs and LLSs are almost entirely ionized and only \nhi\ is known, large ionization corrections are required to estimate their metallicities \citepalias{Lehner2013}. However, we demonstrate in this paper that with a previously-derived distribution of the ionization parameters for pLLSs and LLSs at $z\lesssim 1$, we can accurately estimate the metallicity of an absorber using only observations of \nhi\ (from our COS G140L observations of the break at the Lyman limit) and \nmgii\ (obtained from ground-based observations).


Our paper is organized as follows. In \S\ref{s-uvobservations} and \S\ref{s-mgiiobservations}, we describe the new sample of pLLSs and LLSs and the space- and ground-based observations of the QSOs. In \S\ref{s-analysis} we present the analysis procedure to determine the column densities of \hi\ and \mgii. We present the adopted sample of pLLSs and LLSs in \S\ref{s-adoptedsample}. The photoionization models and our new low-resolution method to determine the metallicities of the pLLSs and LLSs are described \S\ref{s-llsmet} (in the Appendix, we test the effect of changing the ionizing background on the determination of the metallicity). Our main results are presented in \S\ref{s-results} and discussed in \S\ref{s-discussion}. In \S\ref{s-summary} we summarize our main conclusions.


\section{Observations and Analysis}\label{s-observationsandanalysis}

The goal of this work is to increase the sample of $z\lesssim1$ pLLSs and LLSs and to improve the \nhi\ coverage compared to previous studies. The \Lthirteen\ sample is biased toward lower \nhi\ pLLSs ($16.1<\log\mathnhi\lesssim16.8$). This is due to the selection of UV-bright QSOs for the high-resolution COS observations used in \citetalias{Lehner2013} that were known {\it a priori} to have no strong LLSs in their spectra (at least over most redshifts, corresponding to the wavelengths used to assess the UV flux of the QSOs in previous observations). As we will show, the sample of CGM absorbers studied here are {\it optically}-selected QSOs at $0.75\le z_{em} \le 1.25$. With this selection there is no bias against weak pLLSs or strong LLSs in the UV spectra of QSOs. Although due to the low resolution and low signal-to-noise ratio (S/N) of the COS G140L spectra, we cannot detect weak pLLSs below $\log\mathnhi\lesssim16.4$. As we demonstrate in this work, the combined \Lthirteen\ and new (hereafter W16) sample contains \numLW\ pLLS+LLS absorbers at $z\lesssim 1$ with $16.1<\log\mathnhi<17.7$, covering a broader range of \hi\ column densities than \Lthirteen\ with a more representative sampling of the column density distribution function at these redshifts.


\subsection{UV Observations and Sample Definition}\label{s-uvobservations}

We make use of our Cycle 18 {\it HST} snapshot program with an input target list of the 140 $g$-band brightest ($15.5\lesssim g \lesssim 17.15$) Sloan Digital Sky Survey (SDSS) QSOs at $0.75\le z_{em} \le 1.25$ (PID: 12289; PI: Howk) with the Cosmic Origins Spectrograph (COS). The main science goal was to determine the redshift density of the pLLSs+LLSs at $z\lesssim1$. The QSOs were selected to be bright enough that 15-minute exposures with the low-resolution G140L grating provided high enough S/N to detect Lyman limit absorption with an optical depth at the Lyman limit of $\tau_{\rm LL} \gtrsim 0.2$ (i.e., $\log\mathnhi\gtrsim16.4$). Of the 140 proposed QSOs, 61 were observed; the redshift, RA, and DEC for each observed QSO are summarized in Table~\ref{t-qso}.

Information on COS can be found in \citet{Green2012} and \citet{Holland2014}. In short, we used the G140L central wavelength setting $\lambda_{\rm c}=1280$ \AA, giving wavelength coverage of $\lambda =900$--$2400$ \AA\ (although the usable range is typically $\lambda =1100$--$2100$ \AA). A detector gap from $\lambda= 1165$--$1280$ \AA\ is present, blocking access to the Lyman limit for absorbers at $z\sim0.22$--$0.29$. The data were processed with CALCOS (v 2.13.6). The resolution of the G140L grating is $R\approx2300$ at $\lambda=1280$ \AA, and it has a dispersion of 80.3 m\AA\ pixel$^{-1}$. We rebin the data during analysis to $R\approx1250$, yielding a typical signal-to-noise ratio of ${\rm S/N} \sim 2$--$10$ per binned pixel (320 m\AA) in unabsorbed regions. This is sufficient to reveal \hi\ Lyman continuum absorption at $\tau_{\rm LL}\gtrsim0.2$ and strong Lyman series lines.

We searched for one or more pLLSs/LLSs in the continuum of each QSO using the break at the Lyman limit as the indicator. We use the Lyman series lines to confirm the presence of a break, especially when a second break is present at a similar redshift, since the higher-redshift break can partially obscure the depth of the flux decrement of the lower-redshift break. We found 32 sightlines with at least one pLLS or LLS, and have a total of \numLymanBreaks\ pLLSs and LLSs. Our search is sensitive to absorbers with $\log\mathnhi\gtrsim16.4$, although it is incomplete below $\log\mathnhi\approx16.8$. Despite this, the study is still sensitive to all metallicities across the entire \nhi\ range (i.e., we are not biased against low- or high-metallicity systems at low \nhi). Thus, our results are not biased due to this incompleteness.

\begin{figure}[tbp]
\epsscale{1.2}
\plotone{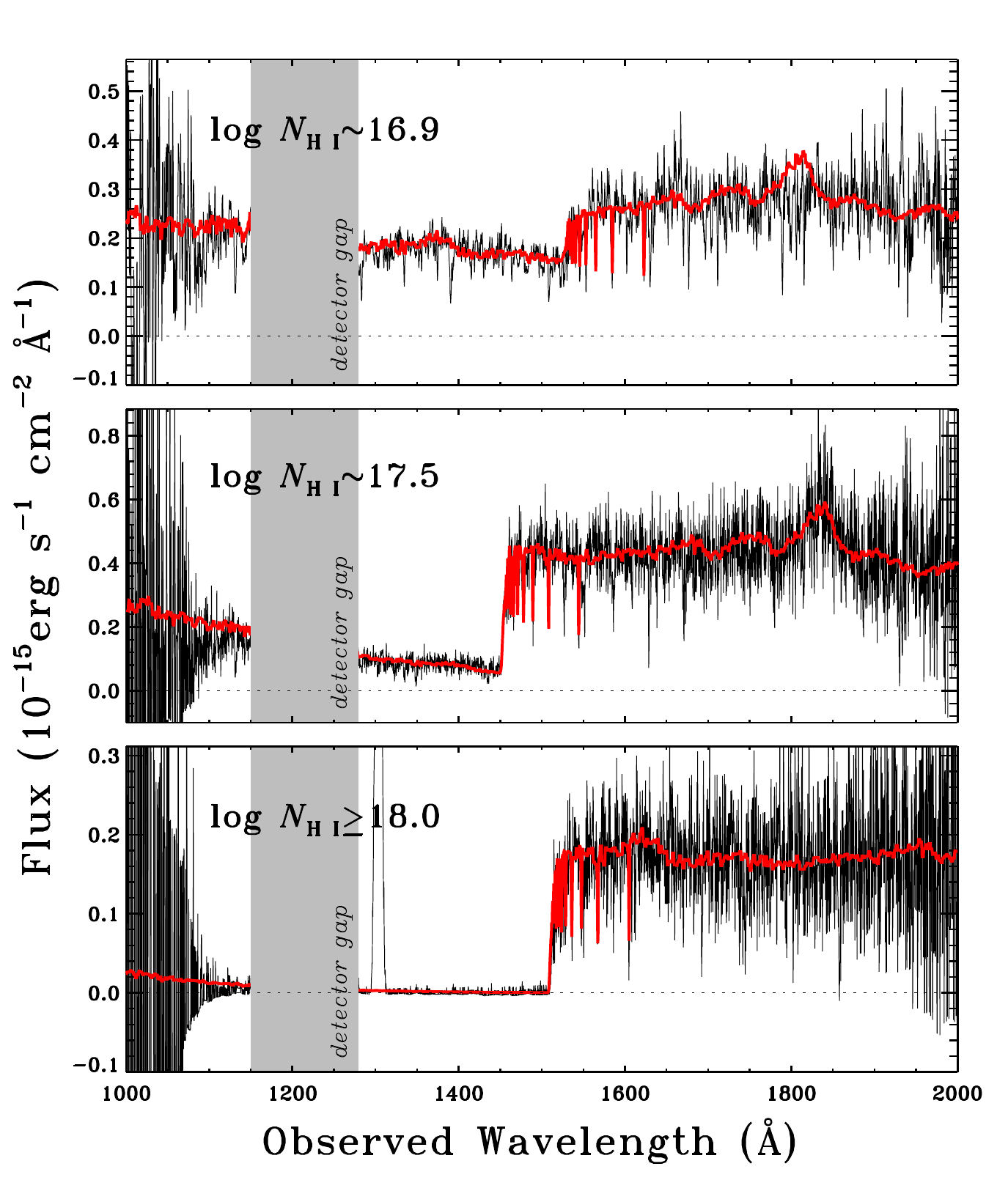}
    \caption{Examples of low-resolution COS G140L spectra from our snapshot survey, showing a weak (top), moderate (middle), and strong (bottom) pLLS/LLS. The recovery of flux near the blue end is useful in constraining the strength of the flux decrement and hence the \nhi\ of the pLLS/LLS. The lack of a recovery in the strongest LLS only allows us to determine a lower limit for the \hit\ column density. The flux at $1150\lesssim\lambda\lesssim1300$ \AA\ is missing due to the gap between the COS detectors.
    \label{f-lybreaks}}
\end{figure}

In Figure~\ref{f-lybreaks}, we show the G140L spectra of 3 QSOs with examples of weak, moderate, and strong pLLSs/LLSs along their sightlines. The overplotted red curve is a composite QSO spectrum \citep{Telfer2002} with the best adopted absorption model applied (see \S\ref{s-hianalysis}). The S/N is not sufficient enough to model the Lyman series lines to determine \nhi, but is sufficient to accurately estimate \nhi\ from the strength of the break at the Lyman limit (see \S\ref{s-hianalysis}). The recovery of flux at the blue end of the spectrum can provide useful constraints on the \hi\ column density even when the absorption near the Lyman limit is strongly saturated. At the highest column densities ($\log\mathnhi\gtrsim18.0$, depending on redshift), however, the recovery is not present, and we can only derive a lower limit on the \hi\ column density. We finally note that the data at $\lambda<1165$ \AA, blueward of the detector gap, generally do not have enough continuous wavelength coverage at high enough S/N to allow us to reliably find pLLSs and LLSs and measure their properties. Therefore, our survey only covers $0.4 \le z_{\rm abs} \le 1.0$.

\subsection{Ground-Based Observations}\label{s-mgiiobservations}

We targeted all sightlines containing at least one pLLS or LLS for ground-based follow-up observations to measure the strength of the \mgii\ $\lambda\lambda$2796, 2803 doublet. We had a completion rate of 97\% (31/32 pLLS- or LLS-bearing sightlines). We used three spectrographs to collect the \mgii\ observations: the Multi-Object Double Spectrograph (MODS) on the Large Binocular Telescope, the Magellan Echellette Spectrograph (MagE) on Magellan, and the High Resolution Echelle Spectrometer (HIRES) on Keck I.

In Table~\ref{t-mgiihiraw}, we summarize the new sample of pLLSs+LLSs with basic measurements ($z$; \nhi; \mgii\ equivalent widths; \nmgii; velocity integration ranges, $[v_1,v_2]$; \S\S\ref{s-analysis} and \ref{s-adoptedsample} present these measurements and adopted results), the source of the \mgii\ data, and the absorbers that will be excluded from our W16 pLLSs+LLS sample (see \S\ref{s-adoptedsample}).

\begin{figure}[tbp]
\epsscale{1.2}
\plotone{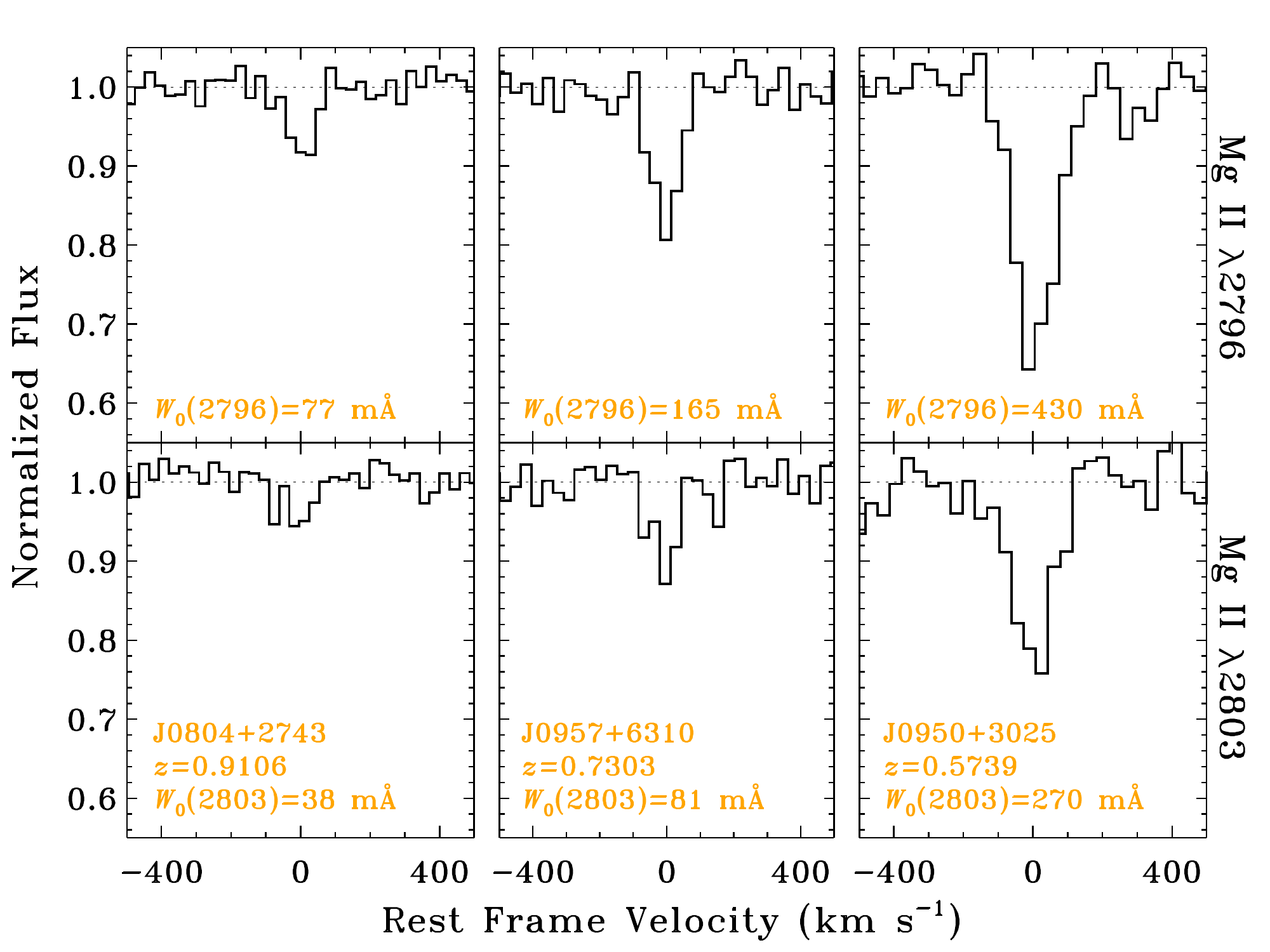}
    \caption{Examples of low-, moderate-, and high-column density \mgiit\ absorption observed with the MODS spectrograph. With these low-resolution ($R\approx4000$), good S/N MODS spectra, we can detect \mgiit\ absorption even for weak absorbers, and can perform moderate saturation correction using the ratio of the doublet lines.
    \label{f-mg2spectra}}
\end{figure}


\subsubsection{S/N and Detection Limits}\label{s-mgiiobslimits}

We acquired sufficient S/N in our spectra so any non-detections of \mgii\ were sensitive to metallicities $[{\rm X/H}]\le-1.3$. Since this sensitivity limit depends on \nhi\ and the brightness of the QSO, there is a broad range of S/N for each instrument (see the sections that follow). Furthermore, for some MODS sightlines, we acquired additional follow-up exposures for non-detections with insufficient S/N; however, we did not need to follow up any sightlines if there was already a clear detection of \mgii. This further increases the range in S/N seen in the MODS data.

With this strategy, the metallicities in our sample are unbiased despite using spectrographs with different resolutions. {\it A posteriori}, we checked that neither the upper limits nor the lower limits on the metallicity were affected by the resolution and we saw no instrumental bias. Thus, the metallicity spread of the pLLSs and LLSs is an intrinsic property of these absorbers, not an artifact from using different instruments. We note, however, that for most pLLSs and a reasonable amount of telescope time, we are typically sensitive to the metallicity range $-2\lesssim[{\rm X/H}]\lesssim-1.3$; the only exception is toward J1500+4836 where, thanks to the relatively high \nhi\ of this pLLS, we managed to push the limit to $<$$-$2.5 with additional HIRES observations.


\subsubsection{MagE and HIRES Observations}\label{s-mgiiobsmagehires}

The MagE spectrograph was used to observe 5 sightlines, which represents 6 pLLSs+LLSs. Of these, 4 pLLSs+LLSs are included in the W16 sample; 1/4 ultimately yielded a limit on \nmgii. The 2 excluded pLLSs/LLSs are above our \nhi\ cutoff (see \S\ref{s-adoptedsample}). For the MagE data, exposure times varied from 600 to 900 s. The final spectra were single exposures. The resolution of MagE is $R\approx4100$. The S/N of the MagE spectra range from 17 to 51 (median 35) per pixel. To reduce the MagE data, we used the MASE reduction pipeline written for the MagE spectrograph \citep{Bochanski2009}. The HIRES spectrograph was used to observe 14 sightlines, which represents 20 pLLSs+LLSs. Of these, 11 pLLSs+LLSs are included in the W16 sample; 6/11 ultimately yielded a limit on \nmgii. Of the excluded 9 pLLSs/LLSs, 2 are above our \nhi\ cutoff and 7 have lower limits on both \nhi\ and \nmgii\ simultaneously (see \S\ref{s-adoptedsample}). For the HIRES data, exposure times varied from 1000 to 3000 s and typically was 1800 s. The final spectra were typically S/N-weighted coadditions of multiple exposures (occasionally they were single exposures). The resolution of HIRES using the C1 decker is $R\approx45000$ and using the C5 decker is $R\approx34000$ (about half of the data were collected with each). The S/N of the HIRES spectra range from 6 to 27 (median 14) per pixel. We used the XIDL HIRES Redux pipeline\footnote{http://www.ucolick.org/$\sim$xavier/HIRedux/} to reduce the HIRES data (see, e.g., \citealt{Lehner2014} and \citealt{O'Meara2015} for more details). Details of the reduction pipeline can also be found in \citet{Bernstein2015}.


\subsubsection{MODS Observations}\label{s-mgiiobsmods}

The MODS spectrograph was used to observe 14 sightlines, which represents 21 pLLSs+LLSs.\footnote{In two cases we employ both MODS and HIRES observations for a given sightline. In one case we do not have wavelength coverage of the LLS with HIRES. For the other absorber, the S/N in the region of interest is poor in the HIRES spectrum.} Of these, 15 pLLSs+LLSs are included in the W16 sample (see \S\ref{s-adoptedsample}); 6/15 ultimately yielded a limit on \nmgii. Of the excluded 6 pLLSs/LLSs, 3 are above our \nhi\ cutoff and 3 have lower limits on both \nhi\ and \nmgii\ simultaneously (see \S\ref{s-adoptedsample}). Because the data reduction pipeline for the MODS spectrograph has been developed more recently, we describe the MODS setup and data reduction process in more detail; information can also be found in the MODS Instrument Manual.\footnote{\label{fnote-mods}http://www.astronomy.ohio-state.edu/MODS/index.html} The MODS double spectrograph has two modes of operation: single-channel (blue or red) and dual, in which a dichroic splits the beam to simultaneously collect blue and red spectra. When in blue-only (red-only) mode, the G400L (G670L) grating is used with a central wavelength of 4000 \AA\ (7600 \AA) and has wavelength coverage $\lambda=3200$--$6000$ \AA\ ($5000$--$10000$ \AA). When in place, the dichroic greatly reduces the sensitivity in the region where the blue and red bandpasses overlap near 5500 \AA, so MODS is used in single-channel mode when the \mgii\ absorption lines would lie near these wavelengths. With the dichroic in place, the blue G400L and red G670L gratings have effective wavelength coverages of $\lambda=3200$--$5650$ \AA\ and $5650$--$10000$ \AA, respectively.

Our data were acquired with the LS5$\times$60$\times$0.3 slitmask, a 0.3-arcsecond-wide longslit, with exposure times ranging from 2$\times$600 s to 3$\times$600 s. The resolution of MODS using the 0.3-arcsecond longslit is $R\approx4000$. The S/N in the reduced MODS spectra are roughly ${\rm S/N} \sim 20$--$110$ (median 47) per pixel. There is a large range in S/N because the clear detection of \mgii\ in a spectrum (i.e., not an upper limit) does not require high S/N to estimate its column density. Figure~\ref{f-mg2spectra} shows examples of MODS spectra with low, moderate, and high \mgii\ column densities. The ratio of the lines of the doublet is useful for moderate saturation correction, as discussed in \S\ref{s-mgiianalysis}. For example, the leftmost plot in Figure~\ref{f-mg2spectra} shows extremely weak (but detectable) unsaturated absorption, the middle plot shows unsaturated absorption, and the rightmost plot shows saturated \mgii\ absorption at a level we are able to correct.

To reduce the MODS data, we followed the MODS data preprocessing procedure (modsCCDRed), then processed the data with the reduction pipeline, modsIDL, developed at The Ohio State University.$^{\ref{fnote-mods}}$ The modsIDL pipeline is based on the LowREDUX reduction pipeline\footnote{http://www.ucolick.org/$\sim$xavier/LowRedux} written for longslit spectra, with additional features specific to the MODS spectrograph, including better treatment of sky subtraction for MODS data and mask slit-finding. When multiple spectra of an object were taken in consecutive exposures (on a single night), the raw images were coadded (S/N-weighted average) prior to being sent through the reduction pipeline. When multiple spectra of an object were collected on different nights, the final spectra were coadded post-reduction to allow for variations in each night's calibrations.


\begin{deluxetable}{lccc}
\tabcolsep=3pt
\tablecolumns{4}
\tablewidth{0pc}
\tablecaption{QSOs Observed with COS G140L\label{t-qso}}
\tabletypesize{\footnotesize}
\tablehead{\colhead{Sightline} & \colhead{$z_{\rm QSO}$} & \colhead{RA} & \colhead{DEC} \\ \colhead{} & \colhead{} & \colhead{(J2000)} & \colhead{(J2000)} }
\startdata
  J023521.87$-$083229.3 &         1.23430 &         +02:35:21.87 &       $-$08:32:29.3 \\
  J033001.64$-$071828.5 &         1.24824 &         +03:30:01.64 &       $-$07:18:28.5 \\
    J074358.27+323512.5 &         0.90498 &         +07:43:58.27 &         +32:35:12.5 \\
    J074451.37+292005.9 &         1.18342 &         +07:44:51.37 &         +29:20:05.9 \\
    J074815.44+220059.5 &         1.05938 &         +07:48:15.44 &         +22:00:59.5 \\
    J075222.91+273823.1 &         1.05691 &         +07:52:22.91 &         +27:38:23.1 \\
    J075514.58+230607.2 &         0.85502 &         +07:55:14.58 &         +23:06:07.2 \\
    J080003.90+421253.2 &         0.99407 &         +08:00:03.90 &         +42:12:53.2 \\
    J080424.97+274323.2 &         1.21924 &         +08:04:24.97 &         +27:43:23.2 \\
    J080630.30+144242.4 &         1.21268 &         +08:06:30.30 &         +14:42:42.4 \\
    J081002.69+502538.7 &         1.20573 &         +08:10:02.69 &         +50:25:38.7 \\
    J081007.64+542443.7 &         1.12307 &         +08:10:07.64 &         +54:24:43.7 \\
    J081050.90+350828.7 &         1.23661 &         +08:10:50.90 &         +35:08:28.7 \\
    J081520.65+273617.0 &         0.90834 &         +08:15:20.65 &         +27:36:17.0 \\
    J081740.13+232731.5 &         0.89135 &         +08:17:40.13 &         +23:27:31.5 \\
    J082045.38+130618.9 &         1.12479 &         +08:20:45.38 &         +13:06:18.9 \\
    J085723.99+090349.0 &         1.04768 &         +08:57:23.99 &         +09:03:49.0 \\
    J090533.30+513507.7 &         0.89764 &         +09:05:33.30 &         +51:35:07.7 \\
    J090603.64+194142.2 &         1.20555 &         +09:06:03.64 &         +19:41:42.2 \\
    J093602.10+200542.9 &         1.18096 &         +09:36:02.10 &         +20:05:42.9 \\
    J094133.68+594811.2 &         0.96713 &         +09:41:33.68 &         +59:48:11.2 \\
    J095045.71+302518.4 &         1.19464 &         +09:50:45.71 &         +30:25:18.4 \\
    J095711.78+631010.1 &         0.91755 &         +09:57:11.78 &         +63:10:10.1 \\
    J100502.34+465927.3 &         0.95429 &         +10:05:02.34 &         +46:59:27.3 \\
    J100906.35+023555.3 &         1.09990 &         +10:09:06.35 &         +02:35:55.3 \\
    J101548.27+404008.7 &         0.92295 &         +10:15:48.27 &         +40:40:08.7 \\
    J101557.05+010913.6 &         0.77987 &         +10:15:57.05 &         +01:09:13.6 \\
    J102005.99+033308.4 &         0.93886 &         +10:20:05.99 &         +03:33:08.4 \\
    J104244.24+164656.1 &         0.97572 &         +10:42:44.24 &         +16:46:56.1 \\
    J104411.44+015850.5 &         1.01007 &         +10:44:11.44 &         +01:58:50.5 \\
    J114343.09+674455.5 &         0.79485 &         +11:43:43.09 &         +67:44:55.5 \\
    J115027.26+665848.0 &         1.03615 &         +11:50:27.26 &         +66:58:48.0 \\
    J122222.55+041315.7 &         0.96555 &         +12:22:22.55 &         +04:13:15.7 \\
    J130631.63+435100.4 &         0.75480 &         +13:06:31.63 &         +43:51:00.4 \\
    J132652.44+292534.8 &         1.20704 &         +13:26:52.44 &         +29:25:34.8 \\
    J132909.25+480109.6 &         0.92816 &         +13:29:09.25 &         +48:01:09.6 \\
    J132957.14+540505.9 &         0.94844 &         +13:29:57.14 &         +54:05:05.9 \\
    J134719.40+590232.8 &         0.76700 &         +13:47:19.40 &         +59:02:32.8 \\
    J135559.88+260039.0 &         0.85110 &         +13:55:59.88 &         +26:00:39.0 \\
    J140819.59+603617.9 &         0.80071 &         +14:08:19.59 &         +60:36:17.9 \\
    J141359.55+485120.7 &         0.91837 &         +14:13:59.55 &         +48:51:20.7 \\
  J141528.76$-$002633.2 &         1.15549 &         +14:15:28.76 &       $-$00:26:33.2 \\
    J143621.29+072720.8 &         0.88894 &         +14:36:21.29 &         +07:27:20.8 \\
    J150031.80+483646.8 &         1.02848 &         +15:00:31.80 &         +48:36:46.8 \\
    J150420.99+543610.3 &         1.16584 &         +15:04:20.99 &         +54:36:10.3 \\
    J151006.75+034908.7 &         0.90414 &         +15:10:06.75 &         +03:49:08.7 \\
    J151907.61+440424.5 &         0.78174 &         +15:19:07.61 &         +44:04:24.5 \\
    J152843.91+520517.7 &         1.22788 &         +15:28:43.91 &         +52:05:17.7 \\
    J153602.47+393207.0 &         0.78354 &         +15:36:02.47 &         +39:32:07.0 \\
    J155846.98+043802.6 &         1.20156 &         +15:58:46.98 &         +04:38:02.6 \\
    J160500.55+353949.6 &         1.04006 &         +16:05:00.55 &         +35:39:49.6 \\
    J160744.82+184648.2 &         1.08214 &         +16:07:44.82 &         +18:46:48.2 \\
    J163156.13+435943.6 &         0.79337 &         +16:31:56.13 &         +43:59:43.6 \\
    J170648.06+321422.8 &         1.06979 &         +17:06:48.06 &         +32:14:22.8 \\
    J171654.20+302701.4 &         0.75369 &         +17:16:54.20 &         +30:27:01.4 \\
    J171704.68+281400.4 &         1.07684 &         +17:17:04.68 &         +28:14:00.4 \\
    J214937.85+120546.1 &         0.79619 &         +21:49:37.85 &         +12:05:46.1 \\
  J223817.26$-$093859.2 &         0.84452 &         +22:38:17.26 &       $-$09:38:59.2 \\
    J225350.09+140210.3 &         0.86390 &         +22:53:50.09 &         +14:02:10.3 \\
    J225541.64+145715.9 &         0.80742 &         +22:55:41.64 &         +14:57:15.9 \\
    J234403.10+003804.2 &         1.23256 &         +23:44:03.10 &         +00:38:04.2
\enddata
\end{deluxetable}

\begin{deluxetable*}{lcccccccccc}
\tabcolsep=3pt
\tablecolumns{11}
\tablewidth{0pc}
\tablecaption{pLLS and LLS Properties\label{t-mgiihiraw}}
\tabletypesize{\footnotesize}
\tablehead{\colhead{Target} & \colhead{$z_{\rm abs}$\tablenotemark{a}} & \colhead{$\log N_{\rm H\,I}$} & \colhead{$W_0(2796)$} & \colhead{$\log N_a(2796)$} & \colhead{$W_0(2803)$} & \colhead{$\log N_a(2803)$} & \colhead{$[v_1,v_2]$} & \colhead{$\log N_{\rm Mg\,II}$} & \colhead{Instrument} & \colhead{Included in} \\ \colhead{} & \colhead{} & \colhead{[cm$^{-2}$]} & \colhead{(m\AA)} & \colhead{[cm$^{-2}$]} & \colhead{(m\AA)} & \colhead{[cm$^{-2}$]} & \colhead{(km s$^{-1}$)} & \colhead{[cm$^{-2}$]} & \colhead{} & \colhead{W16?} }
\startdata
  J0235$-$0832                  &     0.7552 &                   $>$18.10 &           $600\pm 8$ &                   $>$13.62 &               $540\pm 15$ &                   $>$13.91 &   [$-85,+85$] &                   $>$14.06 &  HIRES &  N    \\
    J0743+3235\tablenotemark{b} &     0.7182 &                   $>$18.05 &          $1702\pm 9$ &                   $>$14.81 &              $1600\pm 40$ &                   $>$15.03 &  [$-115,+95$] &                   $>$15.03 &  HIRES &  N    \\
    J0744+2920\tablenotemark{b} &     1.0622 &                   $>$18.20 &          $1165\pm 9$ &                   $>$13.93 &              $1070\pm 10$ &                   $>$14.21 &  [$-58,+118$] &                   $>$14.21 &  HIRES &  N    \\
    J0800+4212                  &     0.598  &            $16.95\pm 0.20$ &           $24\pm 12$ &    $11.75^{+0.17}_{-0.30}$ &                   $<$27.1 &                   $<$12.10 & [$-100,+100$] &    $11.75^{+0.17}_{-0.30}$ &   MODS &  Y    \\
    J0800+4212                  &     0.5772 &            $18.02\pm 0.07$ &          $472\pm 22$ &            $13.13\pm 0.02$ &               $347\pm 18$ &            $13.28\pm 0.02$ & [$-308,+247$] &                   $>$13.43 &   MODS &  N    \\
    J0804+2743                  &     0.9106 &            $17.45\pm 0.10$ &           $77\pm 13$ &    $12.27^{+0.07}_{-0.08}$ &                $38\pm 10$ &    $12.26^{+0.11}_{-0.14}$ &  [$-103,+79$] &    $12.27^{+0.06}_{-0.08}$ &   MODS &  Y    \\
    J0804+2743                  &     0.7543 &                   $>$18.05 &          $750\pm 10$ &                   $>$14.49 &                $693\pm 9$ &                   $>$14.43 &   [$-51,+44$] &                   $>$14.49 &  HIRES &  N    \\
    J0806+1442                  &     1.0943 &            $17.40\pm 0.20$ &          $220\pm 13$ &                   $>$13.44 &               $148\pm 12$ &                   $>$13.51 &   [$-26,+16$] &                   $>$13.51 &  HIRES &  Y    \\
    J0806+1442                  &     0.9238 &            $17.20\pm 0.20$ &            $90\pm 8$ &    $12.53^{+0.07}_{-0.08}$ &                 $64\pm 9$ &    $12.61^{+0.13}_{-0.16}$ &    [$-14,+9$] &    $12.57^{+0.08}_{-0.09}$ &  HIRES &  Y    \\
    J0810+5025                  &     0.650  &            $16.82\pm 0.15$ &              $<$24.2 &                   $<$11.75 &                   $<$23.6 &                   $<$12.05 & [$-100,+100$] &                   $<$11.75 &   MODS &  Y    \\
    J0810+5025\tablenotemark{b} &     0.5606 &                   $>$17.70 &          $500\pm 40$ &            $13.13\pm 0.03$ &               $307\pm 29$ &            $13.20\pm 0.04$ & [$-280,+125$] &    $13.28^{+0.07}_{-0.08}$ &   MODS &  N    \\
    J0810+5424                  &     0.8570 &            $17.50\pm 0.15$ &          $584\pm 15$ &            $13.26\pm 0.01$ &               $428\pm 11$ &            $13.40\pm 0.01$ & [$-115,+115$] &                   $>$13.55 &   MODS &  Y    \\
    J0950+3025                  &     0.5876 &            $16.95\pm 0.15$ &          $403\pm 24$ &    $13.05^{+0.02}_{-0.03}$ &               $296\pm 24$ &    $13.20^{+0.03}_{-0.04}$ & [$-150,+150$] &                   $>$13.35 &   MODS &  Y    \\
    J0950+3025                  &     0.5739 &            $17.50\pm 0.15$ &          $430\pm 30$ &            $13.07\pm 0.03$ &               $270\pm 30$ &    $13.15^{+0.05}_{-0.06}$ & [$-142,+191$] &    $13.24^{+0.09}_{-0.10}$ &   MODS &  Y    \\
    J0957+6310                  &     0.9100 &            $16.75\pm 0.20$ &              $<$33.2 &                   $<$11.89 &                   $<$31.2 &                   $<$12.17 & [$-100,+100$] &                   $<$11.89 &   MODS &  Y    \\
    J0957+6310                  &     0.7303 &            $17.90\pm 0.10$ &          $165\pm 16$ &    $12.62^{+0.04}_{-0.05}$ &                $81\pm 18$ &    $12.60^{+0.09}_{-0.11}$ &  [$-100,+92$] &    $12.61^{+0.05}_{-0.06}$ &   MODS &  N    \\
    J1005+4659                  &     0.8413 &            $16.92\pm 0.15$ &          $725\pm 26$ &            $13.31\pm 0.02$ &               $470\pm 29$ &            $13.39\pm 0.03$ & [$-255,+160$] &            $13.49\pm 0.05$ &   MODS &  Y    \\
    J1005+4659                  &     0.8187 &                   $>$18.20 &         $1763\pm 23$ &            $13.85\pm 0.01$ &              $1437\pm 25$ &            $14.01\pm 0.01$ & [$-330,+150$] &                   $>$14.16 &   MODS &  N    \\
    J1009+0235                  &     1.0875 &            $17.50\pm 0.10$ &          $181\pm 21$ &            $12.68\pm 0.05$ &               $126\pm 24$ &    $12.80^{+0.07}_{-0.09}$ & [$-135,+133$] &    $12.94^{+0.16}_{-0.19}$ &   MAGE &  Y    \\
    J1009+0235                  &     0.488  &            $16.90\pm 0.15$ &           $66\pm 26$ &    $12.20^{+0.14}_{-0.22}$ &                $63\pm 27$ &    $12.49^{+0.15}_{-0.24}$ & [$-100,+100$] &    $12.37^{+0.11}_{-0.17}$ &   MAGE &  Y    \\
    J1015+0109                  &     0.5880 &            $17.50\pm 0.05$ &          $324\pm 22$ &            $12.94\pm 0.03$ &               $226\pm 20$ &            $13.06\pm 0.04$ & [$-150,+130$] &            $13.20\pm 0.09$ &   MAGE &  Y    \\
    J1020+0333\tablenotemark{b} &     0.8726 &            $18.02\pm 0.05$ &         $1235\pm 24$ &            $13.75\pm 0.01$ &              $1120\pm 40$ &            $14.01\pm 0.01$ & [$-320,+140$] &                   $>$14.16 &   MAGE &  N    \\
    J1222+0413\tablenotemark{c} &     0.6547 &            $17.55\pm 0.10$ &          $240\pm 30$ &            $12.81\pm 0.07$ &               $170\pm 29$ &    $12.95^{+0.15}_{-0.18}$ &   [$-57,+77$] &                   $>$13.50 &   MAGE &  Y    \\
       \nodata                  &     0.6598 &                    \nodata &          $426\pm 26$ &            $13.14\pm 0.04$ &               $370\pm 30$ &            $13.35\pm 0.08$ & [$+233,+388$] &                    \nodata &   MAGE &  Y    \\
    J1306+4351                  &     0.6686 &            $16.85\pm 0.10$ &          $413\pm 28$ &            $13.05\pm 0.03$ &               $227\pm 19$ &            $13.07\pm 0.04$ & [$-150,+115$] &            $13.09\pm 0.06$ &   MODS &  Y    \\
    J1326+2925                  &     0.7275 &            $17.90\pm 0.10$ &           $227\pm 7$ &                   $>$13.07 &                $189\pm 6$ &            $13.21\pm 0.02$ &   [$-25,+40$] &                   $>$13.36 &  HIRES &  N    \\
    J1326+2925\tablenotemark{b} &     0.7324 &            $17.20\pm 0.30$ &           $263\pm 5$ &            $13.07\pm 0.03$ &                $179\pm 6$ &            $13.10\pm 0.04$ &   [$-27,+26$] &            $13.08\pm 0.03$ &  HIRES &  Y    \\
    J1355+2600                  &     0.5360 &            $16.75\pm 0.10$ &          $260\pm 10$ &            $13.01\pm 0.02$ &               $176\pm 10$ &            $13.06\pm 0.02$ &   [$-77,+52$] &            $13.12\pm 0.04$ &  HIRES &  Y    \\
    J1355+2600                  &     0.4852 &            $16.70\pm 0.15$ &            $29\pm 7$ &    $11.95^{+0.08}_{-0.10}$ &                 $22\pm 6$ &    $12.06^{+0.11}_{-0.14}$ &   [$-18,+18$] &    $12.19^{+0.22}_{-0.29}$ &  HIRES &  Y    \\
    J1355+2600                  &     0.2582 &                   $>$17.80 &          $221\pm 23$ &            $12.75\pm 0.05$ &               $167\pm 23$ &    $12.92^{+0.06}_{-0.07}$ &   [$-96,+95$] &                   $>$13.07 &   MODS &  N    \\
    J1500+4836                  &     0.898  &            $17.15\pm 0.15$ &              $<$11.7 &                   $<$11.44 &                   $<$11.8 &                   $<$11.74 &   [$-30,+30$] &                   $<$11.44 &  HIRES &  Y    \\
    J1500+4836                  &     0.8896 &                   $>$18.10 &         $1237\pm 13$ &                   $>$13.89 &               $953\pm 14$ &                   $>$14.02 & [$-115,+237$] &                   $>$14.17 &  HIRES &  N    \\
    J1519+4404\tablenotemark{c} &     0.6042 &            $17.05\pm 0.10$ &          $500\pm 60$ &    $13.12^{+0.05}_{-0.06}$ &               $310\pm 60$ &    $13.19^{+0.08}_{-0.09}$ & [$-220,+150$] &    $13.36^{+0.12}_{-0.18}$ &   MODS &  Y    \\
       \nodata                  &     0.6111 &                    \nodata &          $260\pm 60$ &    $12.82^{+0.09}_{-0.11}$ &               $120\pm 60$ &    $12.77^{+0.17}_{-0.29}$ & [$+339,+679$] &                    \nodata &   MODS &  Y    \\
    J1528+5205                  &     0.5809 &            $17.27\pm 0.10$ &           $427\pm 9$ &                   $>$13.35 &               $300\pm 10$ &                   $>$13.50 &  [$-103,+90$] &                   $>$13.65 &  HIRES &  Y    \\
    J1536+3932                  &     0.454  &            $16.65\pm 0.20$ &              $<$13.1 &                   $<$11.49 &                   $<$12.7 &                   $<$11.78 &   [$-30,+30$] &                   $<$11.49 &  HIRES &  Y    \\
    J1558+0438\tablenotemark{b} &     0.8468 &                   $>$18.10 &          $541\pm 17$ &            $13.23\pm 0.01$ &               $378\pm 18$ &            $13.34\pm 0.04$ &  [$-152,+90$] &            $13.48\pm 0.08$ &   MAGE &  N    \\
    J1605+3539                  &     0.7501 &                   $>$18.10 &          $124\pm 17$ &            $12.49\pm 0.06$ &                $90\pm 18$ &    $12.64^{+0.08}_{-0.10}$ &  [$-100,+75$] &                   $>$12.79 &   MODS &  N    \\
    J1631+4359                  &     0.5196 &            $16.75\pm 0.10$ &          $120\pm 14$ &    $12.47^{+0.05}_{-0.06}$ &                $93\pm 11$ &    $12.66^{+0.05}_{-0.06}$ & [$-100,+125$] &                   $>$12.81 &   MODS &  Y    \\
    J1706+3214                  &     0.6505 &            $18.05\pm 0.10$ &          $633\pm 10$ &                   $>$13.50 &               $435\pm 10$ &            $13.50\pm 0.01$ &  [$-100,+90$] &            $13.51\pm 0.01$ &  HIRES &  N    \\
    J1716+3027                  &     0.756  &            $16.50\pm 0.20$ &              $<$10.3 &                   $<$11.38 &                   $<$10.2 &                   $<$11.68 &   [$-30,+30$] &                   $<$11.38 &  HIRES &  Y    \\
    J1716+3027                  &     0.7103 &            $16.40\pm 0.25$ &              $<$11.6 &                   $<$11.43 &                   $<$11.6 &                   $<$11.74 &   [$-30,+30$] &                   $<$11.43 &  HIRES &  Y    \\
    J1716+3027                  &     0.3995 &            $16.90\pm 0.20$ &            $37\pm 6$ &    $12.01^{+0.07}_{-0.08}$ &                 $13\pm 5$ &    $11.85^{+0.13}_{-0.20}$ &   [$-11,+20$] &    $11.94^{+0.07}_{-0.09}$ &  HIRES &  Y    \\
    J1717+2814                  &     0.2835 &                   $>$17.80 &          $2161\pm 4$ &                   $>$14.97 &              $2104\pm 27$ &                   $>$15.21 & [$-122,+126$] &                   $>$15.36 &  HIRES &  N    \\
  J2238$-$0938                  &     0.3613 &                   $>$17.85 &          $848\pm 11$ &                   $>$13.77 &               $705\pm 12$ &                   $>$13.93 & [$-100,+100$] &                   $>$14.08 &  HIRES &  N    \\
    J2253+1402                  &     0.5737 &            $16.75\pm 0.15$ &          $424\pm 17$ &            $13.06\pm 0.02$ &               $280\pm 23$ &    $13.16^{+0.03}_{-0.04}$ & [$-220,+190$] &    $13.28^{+0.06}_{-0.07}$ &   MODS &  Y    \\
    J2253+1402                  &     0.327  &            $16.75\pm 0.20$ &           $38\pm 18$ &    $11.96^{+0.17}_{-0.28}$ &                   $<$30.5 &                   $<$12.16 & [$-100,+100$] &    $11.96^{+0.17}_{-0.28}$ &   MODS &  Y    \\
    J2255+1457                  &     0.8051 &            $16.85\pm 0.10$ &          $315\pm 26$ &    $12.93^{+0.03}_{-0.04}$ &               $200\pm 10$ &            $13.01\pm 0.02$ & [$-100,+135$] &            $13.11\pm 0.06$ &   MODS &  Y    \\
    J2255+1457                  &     0.446  &            $16.60\pm 0.30$ &              $<$21.2 &                   $<$11.70 &                   $<$23.2 &                   $<$12.04 & [$-100,+100$] &                   $<$11.70 &   MODS &  Y    
\enddata
\tablecomments{
In cases where \mgiit\ is detected, $z_{\rm abs}$ is set at the center of the \mgiit\ absorption line, with an error of $\pm0.0001$. When there is a\\non-detection of \mgiit, $z_{\rm abs}$ is assumed from the redshift of the fitted break at the Lyman limit, $z_{\rm abs}$, with an error of $\pm0.001$.\\Non-detections are denoted by ``$<$'' and are 2$\sigma$ upper limits. Lower limits are denoted by ``$>$''.
}
\tablenotetext{a}{For each pLLS/LLS, we first estimate the redshift from the break at the Lyman limit and then check for \mgii\ $\lambda$$\lambda$ 2796, 2803 absorption\\near this redshift. If there is absorption of \mgii, we adopt the more accurate value of $z_{\rm MgII}$ ($\pm 0.0001$); otherwise, we adopt $z_{\rm LLS}$ with a\\larger error ($\pm 0.001$).}
\tablenotetext{b}{There is evidence for several \mgiit\ components at the redshift of the absorbers.}
\tablenotetext{c}{We were unable to separate two breaks at the Lyman limit, despite evidence for two \mgiit\ absorbers separated by $\sim$1000 km s$^{-1}$. Since we\\were unable to independently measure the \hit\ column densities, we combined the \mgiit\ column densities of the two systems to estimate the\\adopted \nmgii, and therefore the metallicity. If we were to only include the absorption from the \mgiit\ system nearer the redshift of the\\predominant break at the Lyman limit in each case, the metallicity would only change by $\sim$0.1 dex. The velocity integration ranges for both\\systems are based on the redshift given for the first system.}
\end{deluxetable*}


\subsection{Analysis}\label{s-analysis}


\subsubsection{\hi\ Column Density}\label{s-hianalysis}

We measure the optical depth at the Lyman limit, $\tau_{\rm LL}$, to estimate the \hi\ column density from the break at the Lyman limit ($\mathnhi=\tau/\sigma$, where $\sigma = 6.3\times 10^{-18}$ cm$^2$ is the absorption cross-section of the hydrogen atom at the Lyman limit \citep{Spitzer1978}). We first identify the presence of a pLLS or LLS using the break at the Lyman limit, often with confirmation from Lyman series lines; no other lines (e.g., metal lines) are used in the identification or measurement of \hi\ absorption. Next we determine the redshift using the Lyman series lines. To estimate \nhi, we define the flux continuum on the red side of the break ($\lambda\sim950$--$1250$ \AA) using a composite, low-redshift, QSO spectrum \citep{Telfer2002}, scaled to match the continuum level and spectral slope of the data.\footnote{The spectrum from \citet{Telfer2002} is a composite of 184 $z>0.33$ QSO spectra. This composite spectrum therefore shows features common to most, but not necessarily all, QSOs (e.g., the emission line seen at $\lambda\approx1800$ \AA\ in Figure~\ref{f-lybreaks}). Thus, the composite spectrum is not a perfect match to all spectral features in our UV data, as seen in the top panel of Figure~\ref{f-lybreaks}.} A model break is applied at the Lyman limit to the composite QSO spectrum and the flux decrement on the blue side of the break is adjusted to match the observations, yielding an estimate for \nhi. The errors on \nhi\ were estimated by adjusting \nhi\ higher and lower than the best fit until the model was clearly inconsistent with the observed flux; we took these to represent 2$\sigma$ extrema (the errors reported in Table~\ref{t-mgiihiraw} are 1$\sigma$). Thus, since the redshift is set using the Lyman series lines and \nhi\ is measured from the depth of the flux decrement of the break at the Lyman limit, $z$ and \nhi\ are independent measurements (and the uncertainties associated with each are independent, as well).

Two of us (C.~B.~W.\ and J.~C.~H.) independently identified and fit pLLSs/LLSs in the spectra. They then compared and reconciled their results, using the independent fits as a guideline to come to a consensus, for the number of pLLSs/LLSs in each QSO spectrum and their redshifts and column densities. There was generally a very good agreement between the independent identifications and measurements, with all redshifts and \nhi\ agreeing to within 1$\sigma$. In one case there was significant disagreement regarding the presence of a secondary pLLS, so we consulted a third collaborator.

One challenge inherent in the process is fitting multiple breaks; for example, the break in flux of a pLLS or LLS (especially if it is a weak pLLS with $\log\mathnhi\sim16.5$) may be obscured if it has a similar redshift to the convergence of the Lyman series lines (in which the \hi\ lines from 910--925 \AA\ blend together) from a lower-redshift pLLS or LLS. The Lyman series lines can be used to confirm the presence of a second absorber in a given sightline, although blending can sometimes make this difficult or impossible. Thus we have a bias against finding low \nhi\ systems ($\log\mathnhi<17.2$) within $\Delta z \lesssim0.01$ of a lower-redshift LLS --- the so-called ``pLLS bias'' \citep{Prochaska2010}.

\begin{figure}[tbp]
\epsscale{1.2}
\plotone{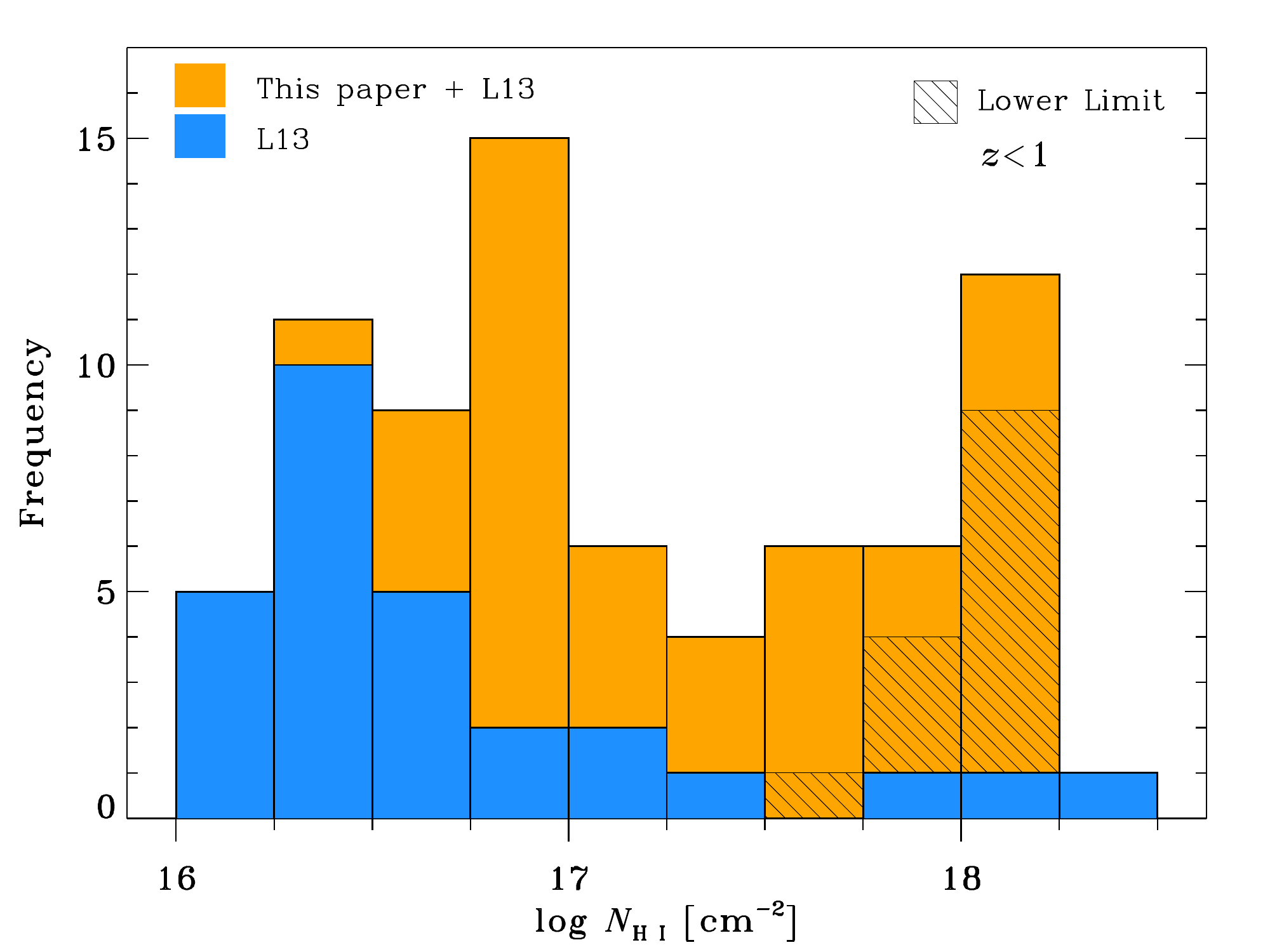}
    \caption{\hit\ column density distribution for the \Lthirteen\ sample (blue) and this work (orange). The hatched regions signify lower limits. The \hit\ column densities now more-uniformly cover a broad range ($16.1<\log \mathnhi\lesssim18.5$).
    \label{f-hihisto}}
\end{figure}

\begin{figure}[tbp]
\epsscale{1.2}
\plotone{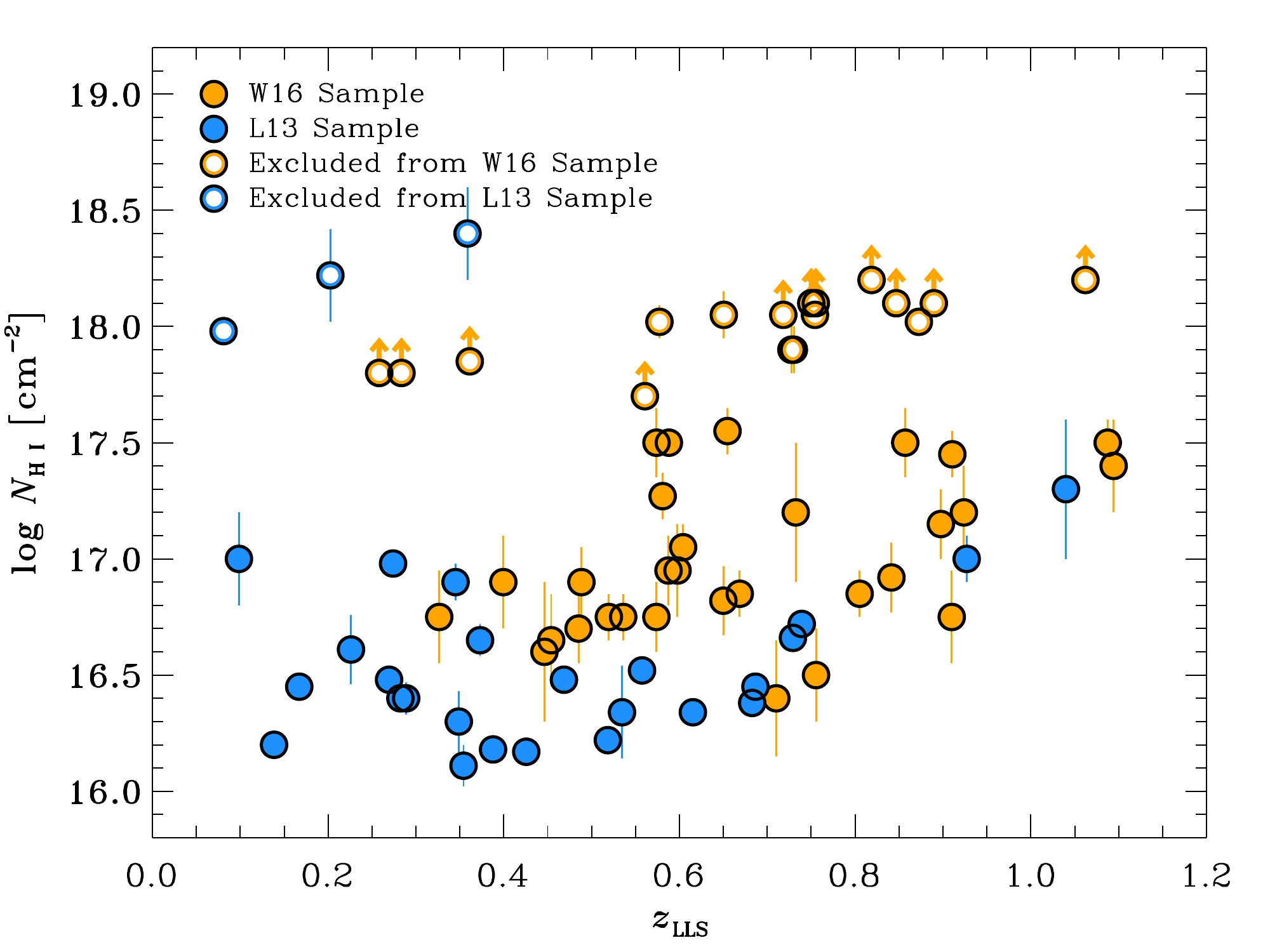}
    \caption{\hit\ column density vs.\ redshift, $z$, for the \Lthirteen\ sample (blue) and this work (orange). The filled circles represent our adopted statistical sample (see \S\ref{s-adoptedsample}), while excluded pLLSs/LLSs are shown as open circles. We are less sensitive to low-\nhi\ absorbers above $z\simeq0.8$--1.1 due to lower S/N above $\lambda\sim1700$ \AA. Note that 3 systems from the \Lthirteen\ data set are not shown here, since they are above the $\log\mathnhi=17.7$ cutoff.
    \label{f-nhivsz}}
\end{figure}

Our adopted \hi\ column densities are summarized in Table~\ref{t-mgiihiraw}, and the distribution of column densities is shown in Figure~\ref{f-hihisto}. Overall, since the W16 sample was \hi-selected, the \hi\ column densities are driven by the column density distribution function \citep[see, e.g.,][]{Ribaudo2011b,Prochaska2010}; i.e., there was no {\it a priori} bias against weak or strong absorbers, so there is a much larger number of pLLSs than LLSs. Combining our sample with that of \citetalias{Lehner2013}, we have a stronger weighting to $\log\mathnhi > 16.7$ absorbers than the \Lthirteen\ sample alone. In Figure~\ref{f-nhivsz} we show the \nhi\ distribution as a function of redshift. There is a small bias against weak \nhi\ absorbers at high redshift ($z\simeq0.8$--$1.1$), but this should not affect our results. Data included in our adopted sample (see \S\ref{s-adoptedsample}) are plotted as filled circles, while excluded pLLSs/LLSs are shown as open circles.

\subsubsection{\mgii\ Column Density}\label{s-mgiianalysis}

Measurement of the \mgii\ $\lambda$$\lambda$2796, 2803 column densities was done using the apparent optical depth (AOD) method described in \citet{Savage1991},
\begin{equation}
N_a(v) = \frac{3.768\times 10^{14}}{f\lambda[\text{\AA}]} \ln\left[\frac{F_c(v)}{F_{\rm obs}(v)}\right] \quad\left(\frac{{\rm cm}^{-2}}{{\rm km\,s}^{-1}}\right),
\end{equation}
where $N_a(v)$ is the apparent column density per unit velocity, $f$ is the oscillator strength and $\lambda$ is the wavelength of the transition in \AA\ from \citet{Morton2003}, and $F_{\rm obs}(v)$ and $F_c(v)$ are the observed and modeled continuum flux, respectively. To estimate $F_c(v)$, a continuum is applied to the spectrum by fitting absorption-free regions on either side of the \mgii\ absorption with a Legendre polynomial (typical orders of 2--4). Integrating over the $N_a(v)$ profile gives the total apparent column density. The integration is performed over the total width of the lines, except where no \mgii\ is found. In those cases, we assume the redshift from the Lyman limit (with an uncertainty of $\pm$0.001, or $\sim$300 km s$^{-1}$; see \S\ref{s-mgiiobslimits}) and integrate the \mgii\ spectrum between characteristic velocities for each instrument (MODS: $\pm100$ km s$^{-1}$, MagE: $\pm100$ km s$^{-1}$, HIRES: $\pm30$ km s$^{-1}$) to get a 2$\sigma$ upper limit, as described by \citet{Lehner2008}. These characteristic velocity full-widths were set based on the full-widths of the weakest \mgii\ absorbers observed with each spectrograph. For the MODS and MagE absorbers, our exact choice of redshift (within the uncertainty) does not affect the upper limit estimate. For the HIRES absorbers, we estimate the upper limits by statistically sampling over a redshift range defined by a Gaussian (with a full width at half maximum of 300 km s$^{-1}$) and integrating over the characteristic $\pm$30 km s$^{-1}$ to derive an upper limit; we perform this 100 times for each HIRES sightline with an upper limit (4), and use the median values as our adopted upper limits. The velocity integration limits, equivalent widths, and column densities for \mgii\ $\lambda$$\lambda$2796, 2803 are listed in Table~\ref{t-mgiihiraw}.

We take the following steps to determine the final adopted \mgii\ column density in Table~\ref{t-mgiihiraw}. If the \mgii\ $\lambda$$\lambda$2796, 2803 measurements are both lower (upper) limits, the adopted column density is the higher (lower) apparent column density of the two transitions. If one line gives an upper limit and the other gives a detection consistent with that limit, the column density from the detected transition is adopted. For detections with non-zero flux in the absorption, we compare the \mgii\ column densities derived using the $\lambda$2803 and $\lambda$2796 lines. If the two column densities agree to within 1$\sigma$ (the measurement error), we take the average of the two as our adopted column density. If the difference between the $\lambda$2803 and $\lambda$2796 lines is greater than 1$\sigma$ and less than 0.127 dex, this is evidence for saturation at a level we can correct using the method described by \citet{Savage1991}. In that case, the adopted \nmgii\ reflects this correction. If the separation is 0.127 dex or greater between the strong and weak transitions, the saturation level is too high to reliably correct. In that case, we apply the maximum correction of 0.15 dex to the apparent column density of the $\lambda$2803 line, and treat it as a lower limit.


\subsection{Adopted Sample}\label{s-adoptedsample}

\begin{figure}[tbp]
\epsscale{1.24}
\plotone{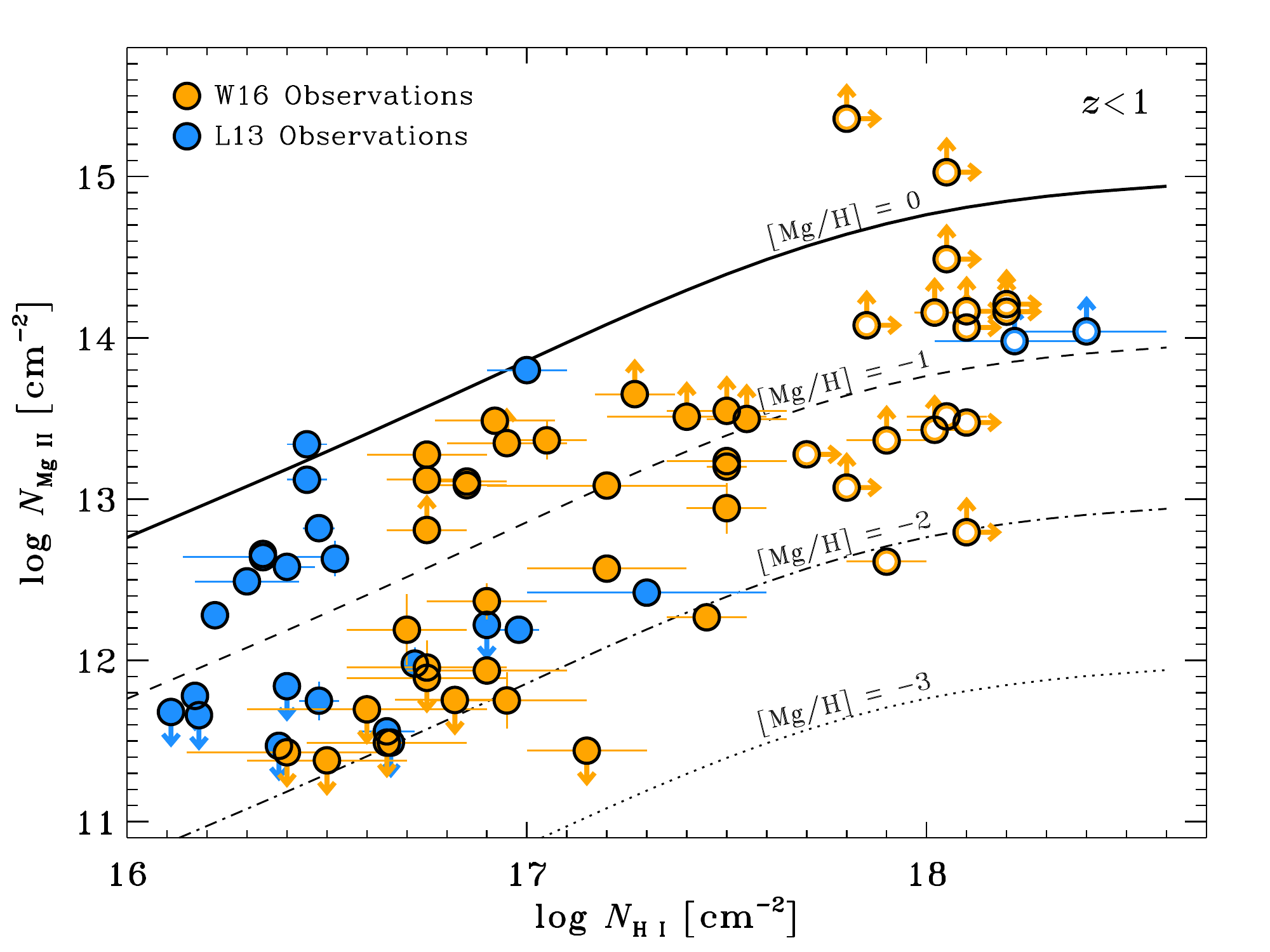}
    \caption{\mgiit\ vs.\ \hit\ column densities for each system. The \hit\ column density coverage of this sample (orange dots) plus the \Lthirteen\ sample (blue dots) is roughly double that of the \Lthirteen\ sample. The solid and dashed curves are metallicity estimates assuming $z=0.7$. Above $\log\mathnhi\gtrsim17.7$, we cannot reliably estimate the metallicity of the absorbers in the low-resolution sample because (1) the \mgiit\ lines easily saturate, giving us a lower limit on \nmgii, and (2) we are unable to reliably estimate the \hit\ column densities, giving us also a lower limit on \nhi. These systems (open circles) are excluded from our statistical sample (\S\ref{s-adoptedsample}).
    \label{f-mg2h1}}
\end{figure}

In Figure~\ref{f-mg2h1}, we plot the adopted column densities of \mgii\ against those of \hi\ for all of our absorbers (orange circles) along with those of \Lthirteen\ (blue circles). Our goal is to estimate the metallicity, so absorbers with lower limits on \nhi\ cannot be used (very high \hi\ column density systems have much different ionization corrections than those in the $\log\mathnhi<18$ regime). Additionally, for the remainder of this paper, absorbers with $\log \mathnhi \ge 17.7$ are excluded from the analysis. Beyond $\log \mathnhi\gtrsim17.7$, our ability to accurately derive \hi\ and \mgii\ column densities simultaneously is very limited (and is redshift- and metallicity-dependent). First, determining \nhi\ for these column densities requires $z>0.5$ LLSs to place the recovery on the blue end of the COS spectrum (see \S\ref{s-uvobservations}). In most cases, we are therefore only able to assign lower limits to \nhi\ beyond 17.7 dex, and hence we cannot exclude that they are damped Ly$\alpha$ absorbers (DLAs; $\log\mathnhi\ge20.3$) or super-Lyman limit systems (SLLSs, a.k.a.\ sub-DLAs; $19\lesssim\log\mathnhi<20.3$). Second, at these \hi\ column densities, \mgii\ becomes saturated even for relatively low-metallicity absorbers, providing little or no information on the shape of the MDF. Since unsaturated \mgii\ could only be observed in the lowest-metallicity absorbers (at these \nhi), this effect is metallicity-dependent. Thus, we exclude these data from our statistical sample; these LLSs are shown with open circles in Figure~\ref{f-mg2h1}. For the highest S/N COS data, we can accurately estimate \nhi\ down to $\log\mathnhi\ge16.4$, but we are incomplete below $\log\mathnhi=16.8$ owing to the inhomogeneous S/N of the COS G140L data.

In Table~\ref{t-mgiihiraw}, we indicate the absorbers that are excluded from the measurable pLLS+LLS metallicity sample of this work. In total, \numWExcluded\ absorbers were excluded, so our ``W16'' statistical sample has \numW\ pLLSs+LLSs with measurable metallicities at $0.4\lesssim z\lesssim 1.1$. In Figure~\ref{f-mg2h1}, we also display the \Lthirteen\ sample of \numL\ pLLSs+LLSs at $0.1\lesssim z\lesssim1.1$ (3 of the 28 pLLSs+LLSs in \citetalias{Lehner2013} were above $\log \mathnhi=17.7$, so they are removed from the combined sample for consistency with the W16 sample definition). Thus, the combined ``\Lthirteen+W16'' sample consists of \numLWPLLS\ pLLSs and \numLWLLS\ LLSs at $0.1\lesssim z\lesssim1.1$.

\section{Determining the Metallicities of the \lowercase{p}LLS\lowercase{s} and LLS\lowercase{s}}\label{s-llsmet}

Since pLLSs and LLSs are almost entirely ionized and the column densities of only select metal ions and of \hi\ are known, large ionization corrections are required to estimate their metallicities \citepalias{Lehner2013}. Owing to the low-resolution, low-S/N COS G140L spectra of our new sample, detailed ionization models requiring several ions to constrain the ionization parameter ($U$; defined as the ionizing photon density/total hydrogen number density) and the metallicity cannot be applied. We introduce a new approach to estimating the metallicities of pLLSs/LLSs, and demonstrate that with a previously-derived distribution of the ionization parameters for pLLSs and LLSs at $z\lesssim 1$, we can accurately estimate the metallicity of an absorber using only observations of \nhi\ (from our COS G140L observations of the break at the Lyman limit) and \nmgii\ (obtained from ground-based observations). Given that this method can make use of only low-resolution spectroscopy and provides a less-precise metallicity estimate than detailed modeling, we refer to these metallicity estimates hereafter as the ``low-resolution metallicities.''


\subsection{Ionization Corrections}\label{s-typicalioncorrect}

To determine the metallicity of an absorber, one would ideally use observations of all ionization states of a given species and of hydrogen. For example, to obtain the ratio of Mg/H, one would ideally use
\begin{equation}
\frac{N_{{\rm Mg}}}{N_{\rm H}} =
   \left(\frac{N_{{\rm Mg}^{0}}+N_{{\rm Mg}^{+}}+
   N_{{\rm Mg}^{+2}}+\ldots}{N_{{\rm H}^{0}}+N_{{\rm H}^{+}}}\right).
\end{equation}
However, we neither have access to all of the ionization states of the metals (in this case Mg$^{0}$, Mg$^{+2}$, Mg$^{+3}$, Mg$^{+4}$, \ldots) nor to H$^{+}$ in absorption. In the predominantly-photoionized pLLSs/LLSs \citep{Fumagalli2016}, the vast majority of the total H column density is in the form of H$^{+}$. Thus, the available species, e.g., \mgii\ and \hi, probe only a small fraction of the total metals and hydrogen in these systems.

Given that we do not have access to a majority of the ions that dominate the total column densities, we would be required make ionization corrections to the observed metal and H tracers to determine the metallicities of pLLSs/LLSs. We would do this by constructing photoionization models \citep[e.g., using the Cloudy photoionization code;][]{Ferland2013}. We would use ratios of observed metal ions \citep[e.g., \oii, \oiii, \cii, \ciii, \siii, \siiii; see][]{Lehner2013, Werk2014, Crighton2015, Fumagalli2016} to constrain the properties of the models. We would then use the output of these models to make corrections for unobserved ionization states, for example, writing the ratio of Mg/H:
\begin{align}
\frac{N_{{\rm Mg}}}{N_{\rm H}} &= \left(\frac{N_{{\rm Mg}^{+}}}{N_{{\rm H}^{0}}}\right)  \cdot  \left(\frac{N_{{\rm Mg}}}{N_{{\rm Mg}^{+}}}\right)  \cdot   \left(\frac{N_{{\rm H}^{0}}}{N_{{\rm H}}}\right) \\
&= \left(\frac{N_{{\rm Mg}^{+}}}{N_{{\rm H}^{0}}}\right)  \cdot  x({\rm Mg}^{+})^{-1}  \cdot   x({\rm H}^{0}),
\end{align}
where $x({\rm X}^{+i})$ is the ionization fraction for the atom or ion ${\rm X}^{+i}$, defined
\begin{equation}
x({\rm X}^{+i})\equiv\frac{N_{{\rm X}^{+i}}}{N_{{\rm X}}},
\end{equation}
a quantity that is derived from photoionization modeling. We would write the ionization correction factor (ICF) for converting the observed column densities to the ratio of the ionization fractions of H$^{0}$ and the ion of interest, ${\rm X}^{+i}$, as:
\begin{equation}
{\rm ICF}({\rm X}^{+i})\equiv\frac{x({\rm H}^{0})}{x({\rm X}^{+i})}.
\end{equation}
In the case of Mg/H, then, the gas-phase abundance of Mg relative to its solar system abundance is:
\begin{equation}
\label{e-metcalc}
\begin{split}
[{\rm Mg/H}] =& \log\left(\frac{N_{{\rm Mg}^{+}}}{N_{{\rm H}^{0}}}\right) + \log\bigg({\rm ICF}({\rm Mg}^{+})\bigg) \\
&- \log\left(\frac{\rm Mg}{\rm H}\right)_\sun.
\end{split}
\end{equation}
In what follows we adopt solar system abundances from \citet{Asplund2009}, i.e., for Mg, $\log({\rm Mg/H})_\sun = -4.47$.

This approach to estimating gas-phase abundances would be done on an absorber-by-absorber basis. Determining accurate column densities usually requires high resolution and high S/N for the detection of often-weak metal ion absorption and to assess potential line saturation, which can affect the validity of the column densities used for both modeling the ionization and ultimately determining the abundance of the gas. To sum up, high resolution, high S/N observations of multiple ionization states are required to constrain simultaneously the ionization parameter ($U$) and the metallicity ($[{\rm X/H}]$).


\subsection{``Low-Resolution'' Metallicities}\label{s-lowres_method}

Determining the ionization correction as described in \S\ref{s-typicalioncorrect} requires high-resolution observations (collected from space for absorbers with redshifts $z\lesssim1$) of several metal ions and \hi\ to constrain $U$. While we can accurately estimate \nhi\ in the COS G140L spectra (\S\ref{s-hianalysis}), their resolution and S/N are too low to sensitively detect absorption from metal ions. \mgii\ is, however, accessible with ground-based observations and, as discussed in \citetalias{Lehner2013}, it provides strong constraints on the metallicity of a pLLSs or LLSs once $U$ is known.

\begin{figure}[tbp]
\epsscale{1.2}
\plotone{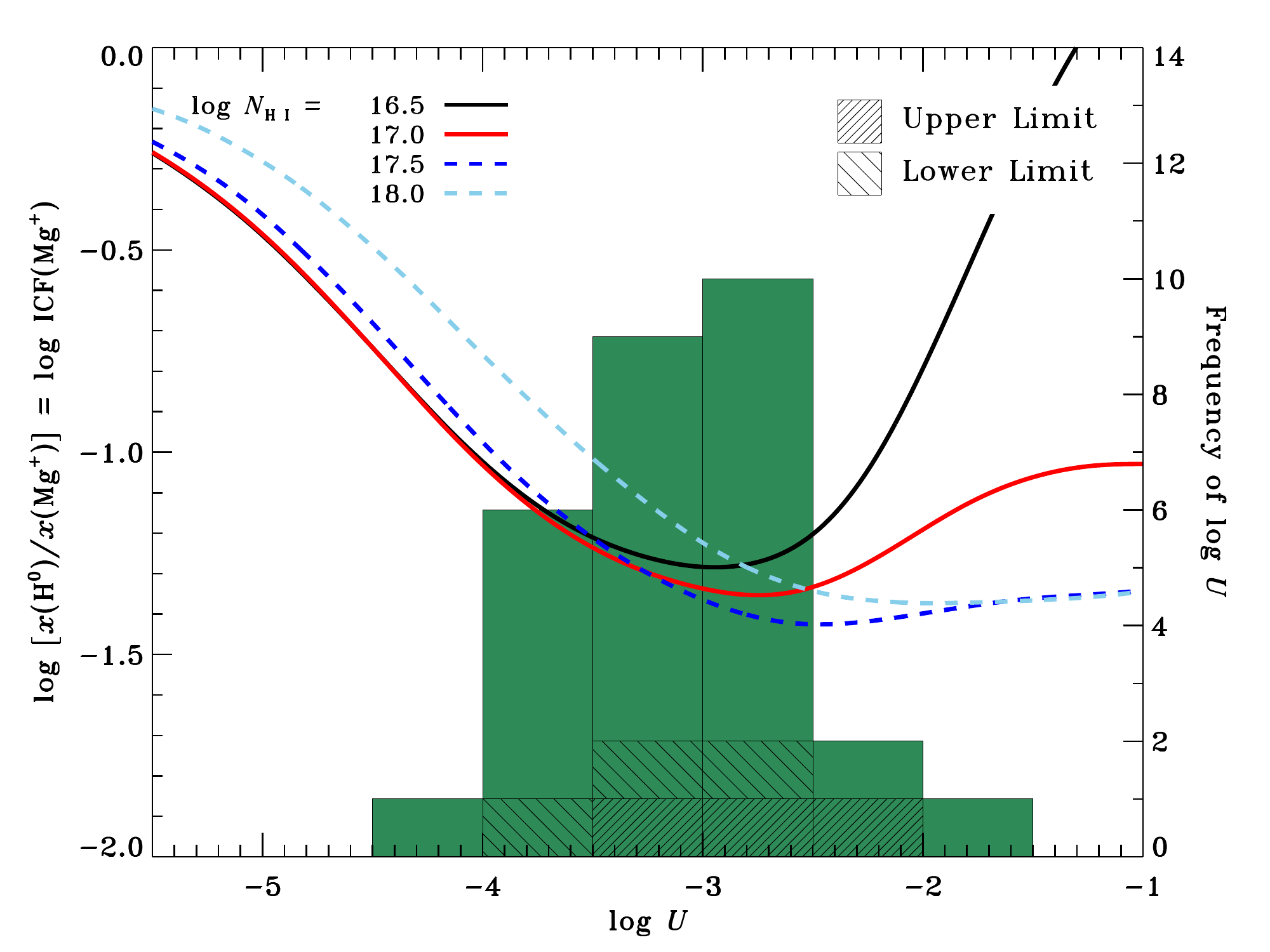}
    \caption{ICF(Mg$^+$) vs.\ the ionization parameter, $\log U$. The solid and dashed lines show the ICF of \mgiit\ for various \nhi. Across $-4\lesssim\log U\lesssim-2$, the ICF is relatively flat, only varying by $<$0.5 dex. The histogram shows the $\log U$ distribution of the \Lthirteen\ sample, derived using detailed ionization corrections. It is relatively narrow, with the 68\% confidence interval at $\pm$0.6 dex. That is, the ionization parameter for a pLLS or LLS at $z\lesssim1$ is likely $\log U=-3.1\pm0.6$ (standard deviation). Thus, rather than using detailed corrections to constrain the ICF of an absorber, we can use the expectation value of the ICF probability distribution. We can apply this to our entire sample because the ICF remains relatively flat at each \nhi\ in our sample for this range of $U$.
    \label{f-logu}}
\end{figure}

With only a single metal ion in most cases (\mgi\ is rarely measurable, except in the strongest absorbers), we cannot constrain the ionization parameter with Cloudy simulations. However, from the work of \citetalias{Lehner2013} we know the properties of the distribution function of $\log U$ for pLLSs/LLSs at $z\lesssim1$, which is shown in Figure~\ref{f-logu} as the green histogram.\footnote{There was a typo in \citetalias{Lehner2013} for the $\log U$ value of PG1634+706. It should read $\log U=-2.7\pm0.4$, based on \citet{Zonak2004}, which we have now corrected.} This distribution is relatively well-confined, with $\left\langle\log U\right\rangle=-3.1\pm0.6$ (standard deviation). In addition, the curves in Figure~\ref{f-logu} show the ICF(Mg$^+$) for several \hi\ column densities at a redshift $z=0.58$ (the median redshift of our sample). The ICFs do not vary strongly over the range of $\log U$ probed by the $z\lesssim1$ pLLSs/LLSs; dropping the highest and lowest values for $\log U$, the maximum change for a given \nhi\ curve over the range of ionization parameters probed by the remaining \Lthirteen\ absorbers in Figure~\ref{f-logu} is $\sim$0.5 dex. The relatively small variations in the ICF are due to the similarity of the ionization potentials of \hi\ and \mgii\ and a lack of any strong spectral features in the typical UV radiation fields responsible for the ionization. Together these imply that instead of undertaking a detailed ionization model to match many constraints, we can derive a characteristic ICF for each pLLS/LLS based on the mean pLLS+LLS properties at $z\lesssim1$ that is not too far from the ICF that would be derived from detailed modeling. We will use the distribution of ionization parameters found by \citetalias{Lehner2013} as an input probability distribution of $\log U$ values for each pLLS/LLS. While we do not know the ionization parameters for any of these individual systems, the errors we make in assigning a metallicity by using this relatively narrow $\log U$ distribution are not large given the limited variation in the ICFs with $\log U$ (see Figure~\ref{f-logu}). Measurements of \nhi\ and \nmgii\ then suffice to estimate the metallicities of the pLLSs and LLSs at $z\lesssim1$ in our sample.

\begin{figure*}[tbp]
\epsscale{1.2}
\plotone{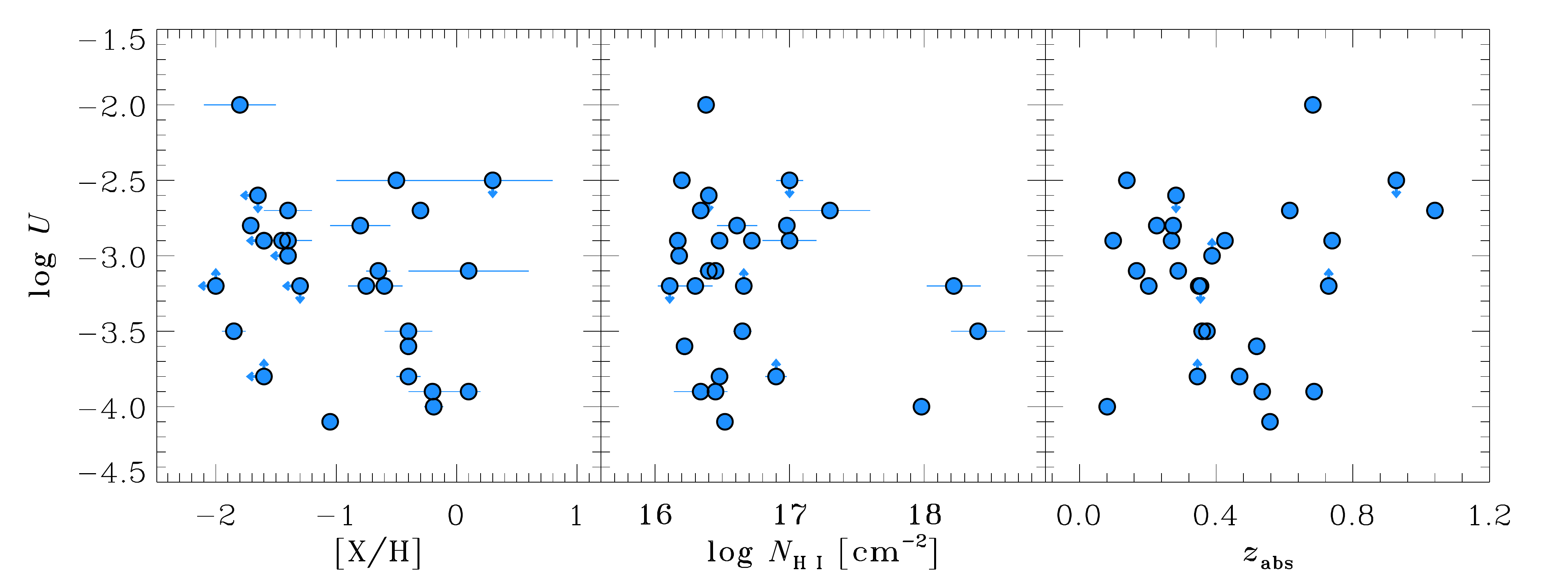}
    \caption{Ionization parameter, $\log U$, as a function of: metallicity (left); \hit\ column density (middle); and redshift (right) for the \Lthirteen\ pLLSs and LLSs at $z\lesssim1$. There is no trend between $\log U$ and any of these parameters.
    \label{f-logUtrends}}
\end{figure*}

For each absorber we calculate the ICF probability distribution, $P({\rm ICF}[z,\mathnhi])$, and from this the expectation value of the ICF, $\left\langle {\rm ICF}[z,\mathnhi] \right\rangle$. We assume the ICF predicted by a specific Cloudy model \citep[v13.03;][]{Ferland2013} has an intrinsic uncertainty of 0.1 dex, distributed normally. This assumption allows for uncertainties in the atomic data, input conditions, and simplifications of the models. The choice of the dispersion is motivated by the ability of \citet{Fumagalli2016} to recover input model results in their tests and by past experience modifying the assumptions in such models. We note that the shape and width of the output ICF distributions are not dominated by this choice. Thus, at each $\log U$ we assume the ICF is characterized by a Gaussian probability distribution, $\phi ({\rm ICF})$, with an intrinsic dispersion $\sigma=0.1$ dex. We further assume that each absorber's ionization parameter is given by the distribution seen in the \citetalias{Lehner2013}, $\phi (\log U)$, which we model as a Gaussian fitted to the observed distribution.\footnote{We performed a Kolmogorov--Smirnov test and found that the \citetalias{Lehner2013} $\log U$ distribution was not statistically distinguishable from a Gaussian distribution ($p=0.56$). While these results do not imply that the $\log U$ distribution is Gaussian, its small variation from normal makes very little difference to the final results. We thus adopt a Gaussian with a mean $\left\langle\log U\right\rangle=-3.094$ and a standard deviation $\sigma=0.552$ for simplicity.} Finally, we note that the adopted $\log U$ distribution is independent of metallicity, \nhi, and $z$, as demonstrated in Figure~\ref{f-logUtrends}, where we show $\log U$ as a function of these parameters for the \Lthirteen\ sample (using the \citetalias{Lehner2013} results). We note that 3 LLSs with $\log\mathnhi>17.7$ have $U$ values smaller than the mean, but the sample of these strong LLSs is small and those absorbers are not included in our statistical sample.

The resulting ICF probability distribution for each absorber is derived by integrating over the ionization parameter:
\begin{equation}
\label{e-PICF}
\begin{split}
P&({\rm ICF}[z, \mathnhi]) = \\
&\int \phi(\log U) \;\phi({\rm ICF}[z, \mathnhi, \log U]) \;d(\log U) \Big/ \\
&\int \phi(\log U) \; d(\log U),
\end{split}
\end{equation}
thus collapsing the equivalent of Figure~\ref{f-logu} over the $\log U$ dimension. The expectation value of ICF for each absorber is then:
\begin{equation}
\label{e-meanICF}
\begin{split}
\left\langle {\rm ICF}[z,\mathnhi] \right\rangle =&
\int {\rm ICF}[z,\mathnhi] \; P({\rm ICF}) \; d({\rm ICF}) \Big/ \\
&\int P({\rm ICF}) \; d({\rm ICF}).
\end{split}
\end{equation}
We use these expectation values to estimate the metallicity of each absorber according to Equation~(\ref{e-metcalc}).

\begin{figure}[tbp]
\epsscale{1.24}
\plotone{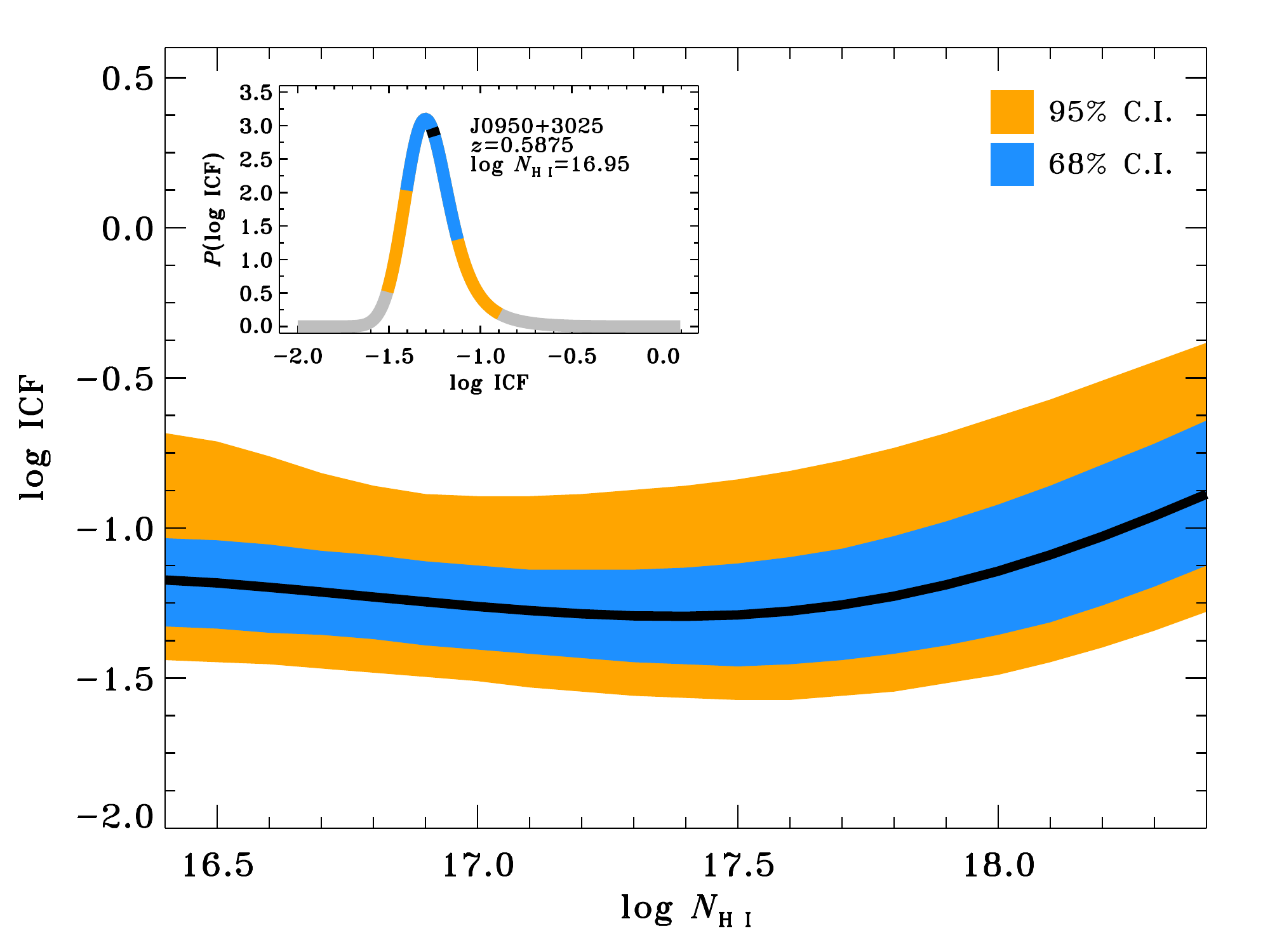}
    \caption{Mean ICF as a function of \nhi\ at $z=0.58$ (the median redshift of our pLLSs+LLSs). The black line shows the ICF expectation value that would be adopted for an absorber at $z=0.58$. The 68\% and 95\% confidence intervals are marked in blue and orange, respectively. The one-sidedness of the errors are due to the shape of the ICF as a function of $\log U$. There is little variation in the ICF across \nhi, rising by just $\sim$0.3 dex at higher \hit\ column densities ($\log\mathnhi > 18.0$, beyond the \nhi\ range of the \Lthirteen+W16 sample). The adopted ICF is determined from the expectation value of the ICF probability distribution function, $P({\rm ICF})$. The inset shows an example of $P({\rm ICF})$ as a function of ICF for a specific absorber, with the ICF expectation value for that absorber marked as a black point.
    \label{f-ICFvNHI}}
\end{figure}

To demonstrate the properties of the predicted corrections, we show in Figure~\ref{f-ICFvNHI} the expectation value of ICF as a function of \nhi\ as the thick black line (for the median redshift of our survey). The 68\% and 95\% confidence intervals are marked in blue and orange, respectively. Neither the expectation values nor the confidence intervals of the ICF vary strongly with \nhi, with a full range of just $\sim$0.3 dex in the expectation values at the 68\% confidence interval. At a fixed \nhi, we expect only a mild evolution in the ICF with redshift, as the radiation field changes. The inset shows a probability distribution in log ICF for the absorber toward J0950+3025, which has both a redshift and \hi\ column density very close to the median values from our sample. This $P({\rm ICF})$ distribution is typical of those derived for our absorbers. The expectation value is shown in black, while the 68\%\ and 95\%\ confidence intervals are again shown in blue and orange, respectively. The distribution is notably asymmetric, an effect largely due to concave shapes of the ICFs as a function of $\log U$ (see Figure~\ref{f-logu}). The confidence intervals are also asymmetric, although the sense of the asymmetry is often different for the 68\% and 95\% confidence intervals. Due to the extended tail seen at high ICF, the expectation value is at a slightly higher ICF than the maximum value (peak) of the $P({\rm ICF})$ distribution, as seen in the inset of Figure~\ref{f-ICFvNHI}.

\begin{figure}[tbp]
\epsscale{1.2}
\plotone{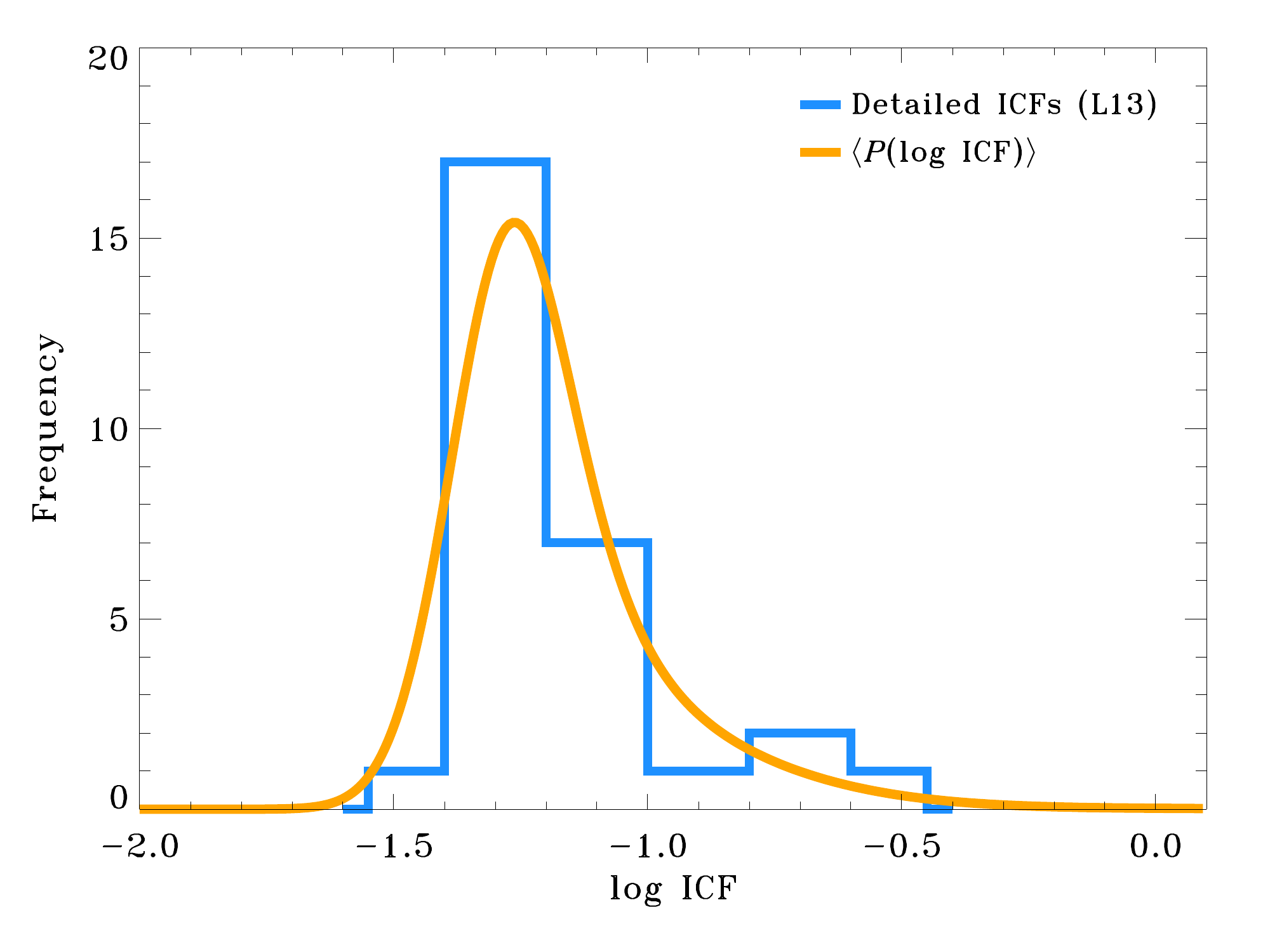}
    \caption{ICFs derived using detailed modeling (blue histogram) and probabilistically-determined ICFs (orange curve) for the \Lthirteen\ data. There is a very good agreement between these methods, suggesting that the characteristic $\left\langle {\rm ICF} \right\rangle$ for each pLLS/LLS based on the mean pLLS+LLS properties at $z\lesssim1$ is an appropriate estimate of the ionization properties of the absorbers. The $\left\langle P(\log {\rm ICF}) \right\rangle$ curve has been scaled to contain the same total area as the \Lthirteen\ ICF histogram.
    \label{f-PICFvICFL13}}
\end{figure}

To test this probabilistic approach to determining ICF against the typical detailed method, we show in Figure~\ref{f-PICFvICFL13} the ICFs of the \Lthirteen\ sample derived using the typical method (blue histogram) and the sum of the predicted ICF distributions using the low-resolution method (orange curve) for the same sample. There is a very good agreement between these methods, suggesting that the derived ICF is primarily driven by redshift due to a changing ultraviolet background (UVB) radiation field. This suggests that the characteristic $\left\langle {\rm ICF} \right\rangle$ for each pLLS/LLS based on the mean pLLS+LLS properties at $z\lesssim1$ is an appropriate estimate of the ionization properties of the absorbers.


\subsection{Validating the ``Low-Resolution'' Method}\label{s-lowres_nlvscbw}

To test the low-resolution method outlined above, we estimate the metallicity of a sample of \numMetalcomparison\ pLLSs/LLSs from \citetalias{Lehner2013} using our low-resolution method and compare those with the metallicities derived in \citetalias{Lehner2013} using detailed ionization corrections. We use 18 of the 19 \mgii\ measurements available in \citetalias{Lehner2013} (one absorber is above $\log\mathnhi=17.7$ so is not included in the comparison), notably supplemented with new \mgii\ column densities for 5 additional absorbers for which \citetalias{Lehner2013} did not have \mgii\ measurements (one of these absorbers is also above $\log\mathnhi=17.7$ so is not included in the comparison). The metallicities of these 5 systems were derived by \citetalias{Lehner2013} using other ions (i.e., without knowledge of \mgii); \mgii\ was significant in constraining the majority of the other 19 systems. The properties of the 5 new \mgii\ measurements are listed in Table~\ref{t-mgiihinl}. The remaining systems in \citetalias{Lehner2013} do not have any \mgii\ observations. To sum up, of the 19+5 absorbers with \mgii\ column density measurements, 18+4 are not beyond our cutoff of $\log\mathnhi=17.7$, contributing 22 systems to this comparison. The remaining 4 absorbers in \citetalias{Lehner2013} do not have \mgii\ so cannot be used in this comparison.

\begin{deluxetable*}{lccccccccc}
\tabcolsep=3pt
\tablecolumns{10}
\tablewidth{0pc}
\tablecaption{New \mgii\ Measurements of \Lthirteen\ Absorbers\label{t-mgiihinl}}
\tabletypesize{\footnotesize}
\tablehead{\colhead{Target} & \colhead{$z_{\rm abs}$} & \colhead{$\log N_{\rm H\,I}$} & \colhead{$W_0(2796)$} & \colhead{$\log N_a(2796)$} & \colhead{$W_0(2803)$} & \colhead{$\log N_a(2803)$} & \colhead{$[v_1,v_2]$} & \colhead{$\log N_{\rm Mg\,II}$} & \colhead{Instrument} \\ \colhead{} & \colhead{} & \colhead{[cm$^{-2}$]} & \colhead{(m\AA)} & \colhead{[cm$^{-2}$]} & \colhead{(m\AA)} & \colhead{[cm$^{-2}$]} & \colhead{(km s$^{-1}$)} & \colhead{[cm$^{-2}$]} & \colhead{} }
\startdata
   HE0439$-$5254                  &     0.6150 &            $16.34\pm 0.03$ &          $120\pm 30$ &    $12.50^{+0.10}_{-0.12}$ &                $88\pm 27$ &    $12.66^{+0.11}_{-0.16}$ &    [$-75,+70$] &                   $>$12.66 &   MAGE    \\
      PG1216+069                  &     0.2823 &            $16.40\pm 0.05$ &                $<$29 &                   $<$11.84 &                     $<$29 &                   $<$12.15 &  [$-100,+100$] &                   $<$11.84 &   MODS    \\
    PKS0312$-$77\tablenotemark{a} &     0.2026 &            $18.22\pm 0.20$ &        $1300\pm 150$ &                   $>$13.72 &              $870\pm 120$ &    $13.84^{+0.09}_{-0.11}$ &  [$-105,+200$] &                   $>$13.98 &   MAGE    \\
   PKS0405$-$123                  &     0.1672 &            $16.45\pm 0.05$ &            $293\pm6$ &             $13.22\pm0.02$ &                 $203\pm7$ &             $13.28\pm0.02$ &    [$-44,+26$] &             $13.34\pm0.04$ &  HIRES    \\
     SBS1122+594\tablenotemark{b} &     0.5580 &            $16.52\pm 0.03$ &          $180\pm 40$ &    $12.63^{+0.08}_{-0.10}$ &                    $<$103 &    $12.63^{+0.08}_{-0.10}$ &  [$-209,+192$] &            $12.46\pm 0.11$ &   MODS   
\enddata
\tablecomments{
Redshifts and \nhi\ are from \citetalias{Lehner2013}. Non-detections are denoted by ``$<$'' and are listed at 2$\sigma$ upper limits. Lower limits are denoted\\by ``$>$''.
}
\tablenotetext{a}{There is evidence for several \mgiit\ components included in the adopted \nmgii. This absorber is excluded from the \Lthirteen+W16 sample\\because $\log\mathnhi>17.7$.}
\tablenotetext{b}{\citetalias{Lehner2013} included only the \hit\ component associated with the pLLS, estimated to be $\log\mathnhi=16.24\pm0.03$. Near that pLLS, there are\\two more components spreading from $-200$ to $+120$ km s$^{-1}$, with in particular a strong absorption in the metal lines (e.g., \ciiit\ and\\\oiiit) at $-125$ km s$^{-1}$. With the MODS observations, \mgiit\ is not resolved and it is apparent there is extra absorption at negative\\velocities. We therefore estimate the total \nhi\ in that absorber to compare with \mgiit\ and estimate the metallicity using the\\low-resolution method.}

\end{deluxetable*}

\begin{figure}[tbp]
\epsscale{1.2}
\plotone{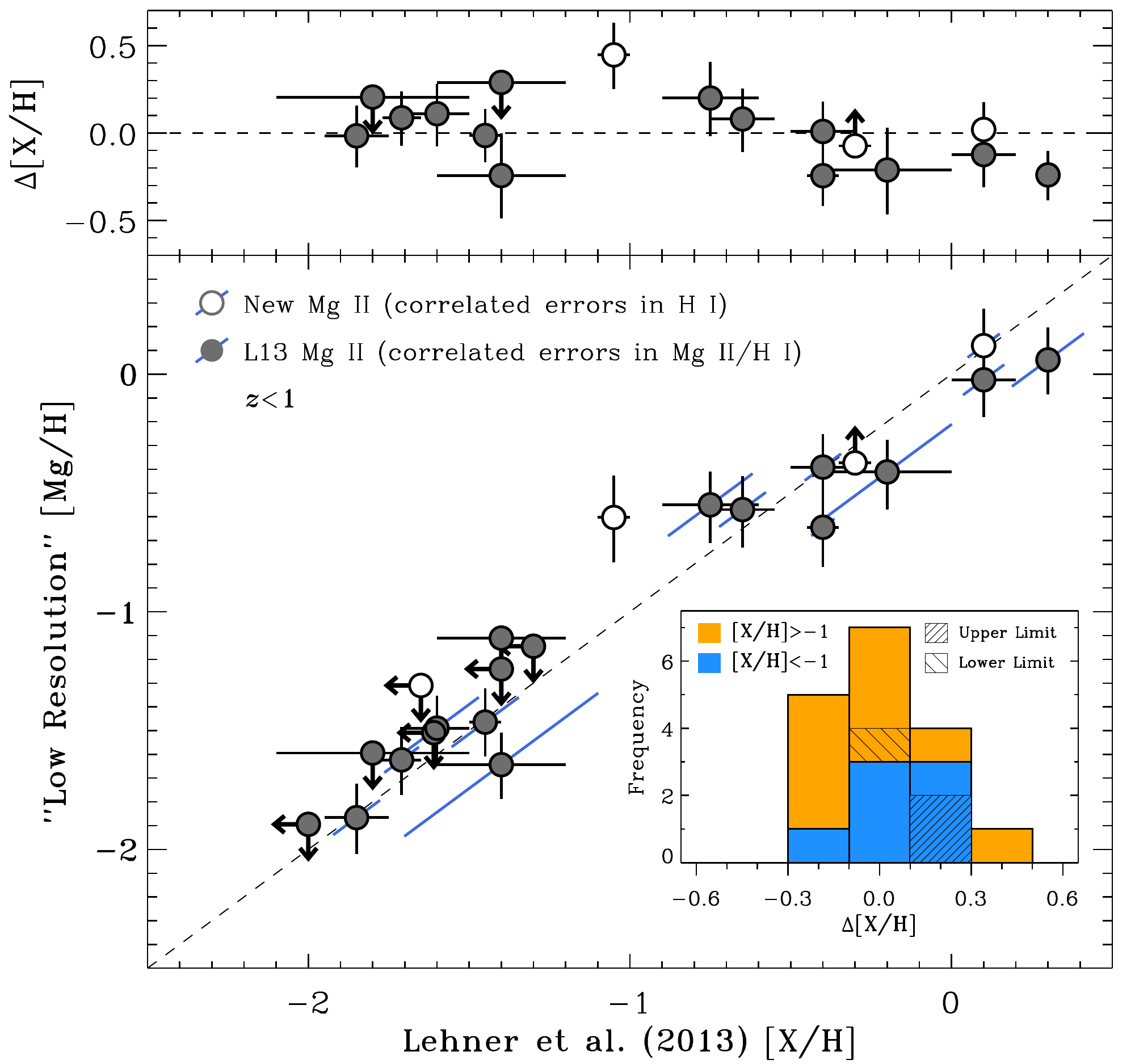}
    \caption{Comparison of low-resolution method with the detailed ionization correction of \citetalias{Lehner2013}. The error bars on the low-resolution [Mg/H] abundances represent the 68\% confidence interval in the applied ICFs added in quadrature with any uncertainty in the column density measurements not already included in the [X/H]$_{\Lthirteen}$ error bars (i.e., for new \mgiit\ measurements, the uncertainty in \nmgii\ is included). The diagonal error bars represent correlated uncertainty from column density measurements, affecting both the \Lthirteen\ and W16 metallicity estimates equally. Lower and upper limits are marked by arrows. Data points for which \mgiit\ was collected after the publication of \citetalias{Lehner2013} are indicated by open circles. Most of the points lie within 1$\sigma$ of the 1:1 line (dashed line). The inset shows the difference, $\Delta[{\rm X/H}]\equiv[{\rm Mg/H}]-[{\rm X/H}]_{\Lthirteen}$, for both low- and high-metallicity systems. The top panel shows the residuals as a function of metallicity. In the 5 cases where there are simultaneous limits on both $[{\rm Mg/H}]$ and $[{\rm X/H}]_{\Lthirteen}$, we do not display these data in the top panel.
    \label{f-nlvscbw}}
\end{figure}

Table~\ref{t-nl} lists the absorber redshifts, \mgii\ column densities, and metallicities derived in both ways for these \numMetalcomparison\ pLLSs/LLSs. Figure~\ref{f-nlvscbw} plots the metallicities derived using the low-resolution method against those derived with the detailed ionization models of \citetalias{Lehner2013}. In this figure, the error bars on the low-resolution [Mg/H] abundances represent the 68\% confidence interval in the applied ICFs added in quadrature with any uncertainty in the column density measurements not already included in the [X/H]$_{\Lthirteen}$ error bars (i.e., for new \mgii\ measurements, the uncertainty in \nmgii\ is included). The diagonal error bars represent correlated uncertainty from column density measurements, affecting both the \Lthirteen\ and W16 metallicity estimates equally.

The two methods produce very good agreement for both low- and high-metallicity systems. The 4 out of 5 pLLSs+LLSs with no \mgii\ in \citetalias{Lehner2013} exhibit similar metallicities using the two methods. Whereas \mgii\ played a role in the \citetalias{Lehner2013} metallicity determinations for most absorbers, the metallicities derived here for these 5 systems share only the \hi\ column densities with the \citetalias{Lehner2013} metallicities. The most discrepant value is toward SBS1122+594 where the metallicities differ by 0.35 dex between \citetalias{Lehner2013} and the low-resolution approach. This difference can be understood considering the high-resolution COS observations from \citetalias{Lehner2013}. Indeed, within 150 km s$^{-1}$ of this pLLS, there are 2 other absorbers with $\log\mathnhi = 15.90$ and 16.01. The metallicity in the latter one is similar to the pLLS metallicity, but the one with $\log\mathnhi = 15.90$ has a metallicity $[{\rm X/H}] \simeq -0.35$ \citep[in prep.]{Ribaudo2016}. For the low-resolution method, we have to consider these 3 absorbers together, yielding a metallicity intermediate between the actual pLLS metallicity and the higher metallicity absorber. This demonstrates a limitation of the low-resolution method, but on the other hand this only affected 1 absorber in 22.

Table~\ref{t-nl} and Figure~\ref{f-nlvscbw} indicate that our low-resolution method provides an estimate of pLLS/LLS metallicities that is consistent with those derived from detailed calculations. The differences $\Delta[{\rm X/H}]\equiv[{\rm Mg/H}]-[{\rm X/H}]_{\Lthirteen}$, shown in the final column of Table~\ref{t-nl} and in the inset of Figure~\ref{f-nlvscbw}, are consistent with our expectations given the ICF probability distributions; the 68\% confidence interval for the ICFs spans 0.29 dex on average, while the 68\% confidence interval for $\Delta[{\rm X/H}]$ spans 0.33 dex. The top panel of Figure~\ref{f-nlvscbw} shows the residuals $\Delta[{\rm X/H}]$ as a function of metallicity, where there is no evidence for any notable trend with metallicity. Note that where there are simultaneous limits on both $[{\rm Mg/H}]$ and $[{\rm X/H}]_{\Lthirteen}$ we did not plot these 5 in the top panel. Our exploration of parameter space demonstrates that this method is valid for pLLSs and LLSs with $16.1\lesssim\log\mathnhi\lesssim18.4$ at $z\lesssim1$.

\begin{deluxetable*}{lcccccccccc}
\tabcolsep=3pt
\tablecolumns{11}
\tablewidth{0pc}
\tablecaption{Comparison of the \citetalias{Lehner2013} and low-resolution LLS metallicities\label{t-nl}}
\tabletypesize{\footnotesize}
\tablehead{\colhead{Target} & \colhead{$z_{\rm abs}$} & \colhead{$\log N_{\rm H\,I}$} & \colhead{$\log N_{\rm Mg\,II}$} & \colhead{\citet{Lehner2013}} & \colhead{Low-Resolution} & \colhead{$\Delta[{\rm X/H}]\equiv$} \\ \colhead{} & \colhead{} & \colhead{[cm$^{-2}$]} & \colhead{[cm$^{-2}$]} & \colhead{$[{\rm X/H}]_{\Lthirteen}$} & \colhead{$[{\rm Mg/H}]$} & \colhead{$[{\rm Mg/H}]-[{\rm X/H}]_{\Lthirteen}$}}
\startdata
    J0943+0531 &   0.3542 &   $16.11\pm 0.09$ &          $<$11.68 &        $<$$-1.30$                  &        $<$$-1.14$   &   \nodata         \\
    J1419+4207 &   0.4256 &   $16.17\pm 0.06$ &          $<$11.78 &   $-1.40\pm 0.20$                  &        $<$$-1.11$   &   $<$$+0.29$      \\
    J1435+3604 &   0.3878 &   $16.18\pm 0.05$ &          $<$11.66 &        $<$$-1.40$                  &        $<$$-1.24$   &   \nodata         \\
    PG1522+101 &   0.5185 &   $16.22\pm 0.02$ &   $12.28\pm 0.03$ &   $-0.40\pm 0.05$                  &   $-0.64\pm 0.04$   &   $-0.24\pm 0.15$ \\
    PG1338+416 &   0.3488 &   $16.30\pm 0.13$ &   $12.49\pm 0.02$ &   $-0.75\pm 0.15$                  &   $-0.55\pm 0.13$   &   $+0.20\pm 0.21$ \\
   HE0439-5254 &   0.6153 &   $16.34\pm 0.03$ &          $>$12.66 &   $-0.30\pm 0.05$\tablenotemark{a} &        $>$$-0.37$   &   $>$$-0.07$      \\
    J1419+4207 &   0.5346 &   $16.34\pm 0.20$ &   $12.64\pm 0.02$ &   $-0.20\pm 0.20$                  &   $-0.41\pm 0.20$   &   $-0.21\pm 0.24$ \\
    PG1407+265 &   0.6828 &   $16.38\pm 0.02$ &          $<$11.47 &   $-1.80\pm 0.30$                  &        $<$$-1.59$   &   $<$$+0.21$      \\
    J1419+4207 &   0.2889 &   $16.40\pm 0.07$ &   $12.58\pm 0.02$ &   $-0.65\pm 0.10$                  &   $-0.57\pm 0.07$   &   $+0.08\pm 0.18$ \\
    PG1216+069 &   0.2823 &   $16.40\pm 0.05$ &          $<$11.84 &        $<$$-1.65$\tablenotemark{a} &        $<$$-1.31$   &   \nodata         \\
   PKS0405-123 &   0.1672 &   $16.45\pm 0.05$ &   $13.34\pm 0.06$ &            +0.10:\tablenotemark{a} &   $+0.12\pm 0.08$   &   $+0.02\pm 0.16$ \\
    PG1338+416 &   0.6865 &   $16.45\pm 0.05$ &   $13.12\pm 0.04$ &   $+0.10\pm 0.10$                  &   $-0.02\pm 0.06$   &   $-0.12\pm 0.17$ \\
    J1619+3342 &   0.2694 &   $16.48\pm 0.05$ &   $11.75\pm 0.12$ &   $-1.60\pm 0.10$                  &   $-1.49\pm 0.13$   &   $+0.11\pm 0.17$ \\
   PKS0637-752 &   0.4685 &   $16.48\pm 0.04$ &   $12.82\pm 0.04$ &   $-0.40\pm 0.10$                  &   $-0.39\pm 0.06$   &   $+0.01\pm 0.17$ \\
   SBS1122+594 &   0.5574 &   $16.52\pm 0.03$ &   $12.63\pm 0.11$ &   $-1.05\pm 0.05$\tablenotemark{a} &   $-0.60\pm 0.11$   &   $+0.45\pm 0.19$ \\
    J1435+3604 &   0.3730 &   $16.65\pm 0.07$ &   $11.56\pm 0.02$ &   $-1.85\pm 0.10$                  &   $-1.87\pm 0.07$   &   $-0.02\pm 0.17$ \\
    PG1522+101 &   0.7292 &   $16.66\pm 0.05$ &          $<$11.49 &        $<$$-2.00$                  &        $<$$-1.89$   &   \nodata         \\
       PHL1377 &   0.7392 &   $16.72\pm 0.03$ &   $11.98\pm 0.10$ &   $-1.45\pm 0.05$                  &   $-1.46\pm 0.10$   &   $-0.01\pm 0.15$ \\
   PKS0552-640 &   0.3451 &   $16.90\pm 0.08$ &          $<$12.22 &        $<$$-1.60$                  &        $<$$-1.50$   &   \nodata         \\
    PG1630+377 &   0.2740 &   $16.98\pm 0.05$ &   $12.19\pm 0.02$ &   $-1.71\pm 0.06$                  &   $-1.62\pm 0.05$   &   $+0.09\pm 0.15$ \\
    PG1206+459 &   0.9270 &   $17.00\pm 0.10$ &   $13.80\pm 0.05$ &            +0.30:                  &   $+0.06\pm 0.11$   &   $-0.24\pm 0.13$ \\
    PG1634+706 &   1.0400 &   $17.30\pm 0.30$ &   $12.42\pm 0.03$ &   $-1.40\pm 0.20$                  &   $-1.64\pm 0.30$   &   $-0.24\pm 0.25$
\enddata
\tablecomments{
The \citetalias{Lehner2013} metallicities marked with ``:'' indicate that the references drawn upon by \citetalias{Lehner2013} did not report an error in their column density measurements, and therefore \citetalias{Lehner2013} were unable to report an error in the derived metallicities. The errors on $[{\rm X/H}]_{\Lthirteen}$ reported here include uncertainties in the ICF and \nhi\ (and \nmgii, where appropriate). The errors on $[{\rm Mg/H}]$ reported here include uncertainties in the ICF, \nhi, and \nmgii\ added in quadrature. The errors on $\Delta[{\rm X/H}]$ reported here are based on standard propagation of errors in $[{\rm X/H}]_{\Lthirteen}$ and $[{\rm Mg/H}]$ (without double-counting measurement errors).  \\
}
\tablenotetext{a}{The \mgiit\ measurements were not present in \citetalias{Lehner2013}, but were collected later.}
\end{deluxetable*}

The low-resolution metallicity method described here provides a new approach to determining the metallicity of pLLSs and LLSs at $z\lesssim1$ without the need to measure a large suite of metal ions. The ion \mgii\ is especially advantageous for this method both because of its observational characteristics --- its doublet nature makes its identification reliable and straightforward, in addition to providing the ability to correct for moderate saturation --- and its ionization characteristics, which show that the derived ionization corrections do not depend strongly on the ionization parameter (Figure~\ref{f-logu}). Additionally, Mg is an $\alpha$-element, so its abundance should follow the same nucleosynthesis $\alpha$-element pattern as the metals used in \citetalias{Lehner2013} at $z\lesssim1$. Finally, dust depletion of Mg at these \nhi\ values is unlikely to be important (see \citetalias{Lehner2013} and \citealt{Fumagalli2016}).

There are some limitations to the method, however. While the strength of the \mgii\ $\lambda\lambda$2796, 2803 doublet is an asset for detecting low-metallicity gas at $z\lesssim1$, it saturates quickly at all metallicities for equivalent widths of $W_0 (2796) \gtrsim 200$ m\AA. Without the suite of ions available in the UV, the low-resolution approach only enables us to derive a lower limit for the metallicity in that situation. Finally, with low-resolution data, we are unable to study the detailed velocity structure of the absorbers. We compare the entire \mgii\ column density with the entire \hi\ column density, assuming a constant metallicity across all velocity components. This could lead to inaccurate metallicities if there are multiple systems with different metallicities very close in redshift. We in fact clearly observe that effect in 1 case (see above), and some absorbers with discrepancies of 0.2--0.3 dex may also be the result of a variation of the metallicity across the velocity profiles near the pLLSs/LLSs. However, this effect remains small (typically $<$0.2--0.3 dex on the metallicity) and hence the MDF of the pLLSs and LLSs can still be reliably determined using the low-resolution method at $z\lesssim1$.

We note that the low-resolution method would be more challenging at $z\gtrsim1$ because the MDF at higher redshift exhibits lower metallicities on average \citep[see, e.g.,][]{Lehner2016, Fumagalli2016}; it is much more difficult (i.e., it requires much higher S/N data) to place stringent limits on the \mgii\ column density for very low-metallicity absorbers with $[{\rm X/H}]<-2$ that are common at higher redshifts.


\section{The Metallicity of the CGM at $\lowercase{z}\lesssim1$}\label{s-results}


\subsection{Metallicities of the \lowercase{p}LLS\lowercase{s} and LLS\lowercase{s} in the W16 Sample}\label{s-results-W16}

\begin{figure}[tbp]
\epsscale{1.2}
\plotone{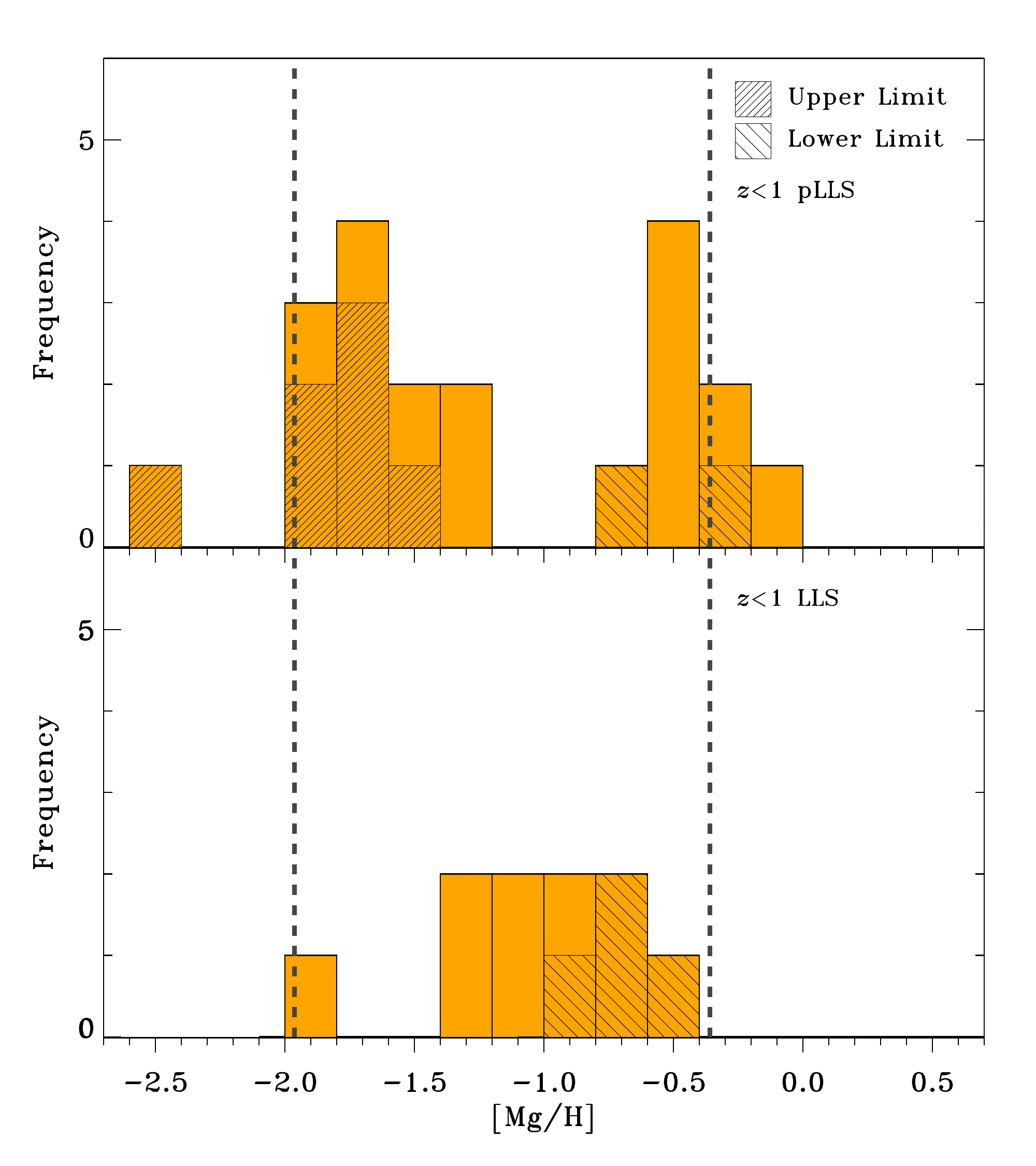}
    \caption{Metallicity distribution functions (MDFs) of the W16 pLLSs (20; top) and the LLSs (10; bottom) at $0.4\lesssim z\lesssim1.1$. The MDF of the pLLSs is double-peaked, with a \pDipWPLLS\ probability that it is bimodal. The shape of the W16 LLS MDF is unclear owing to the sparsity of data and the high fraction (40\%) of lower limits. The vertical lines are the means of the low- and high-metallicity branches of the pLLSs derived from a survival analysis.
    \label{f-zdistmcbwnhistack}}
\end{figure}

As described in \S\ref{s-lowres_nlvscbw}, the low-resolution method works well for the pLLSs and LLSs explored in \citetalias{Lehner2013}. Although the \nhi\ distributions in this work and \citetalias{Lehner2013} are not the same, there is significant overlap over the range of \nhi\ studied here (see Figure~\ref{f-hihisto}; all but one of the \Lthirteen\ absorbers in our adopted sample were pLLSs). There is no evidence from \citetalias{Lehner2013} that the $\log U$ distribution is different for the higher values of \nhi\ (see Figure~\ref{f-logUtrends}). We therefore apply the low-resolution method described in \S\ref{s-lowres_method} to our \numW\ pLLSs+LLSs to derive their metallicities (including 13 upper limits, confined to $[{\rm X/H}]<-1$, and 6 lower limits, confined to $[{\rm X/H}]\ge-1$). We summarize our results in Table~\ref{t-cbw}, and in Figure~\ref{f-zdistmcbwnhistack} we show the MDFs of the pLLSs and LLSs in our sample.\footnote{In two cases, we were unable to separate two Lyman limit breaks, despite evidence for two \mgiit\ absorbers separated by $\sim$1000 km s$^{-1}$. These are marked in Table~\ref{t-mgiihiraw} with ``\nhi\ = \ldots''. Since we were unable to independently measure the \hit\ column densities, we combined the \mgiit\ column densities of the two systems to estimate the adopted \nmgii, and therefore the metallicity. If we were to only include the absorption from the \mgiit\ system nearer the redshift of the predominant Lyman limit break in each case, the metallicity would only change by $\sim$0.1 dex.} While the \Lthirteen\ sample included both pLLSs and LLSs, the approach taken here is different. We will show that the MDF of the optically-thick LLSs does not necessarily follow that of the optically-thin pLLSs. For this reason, we discuss these two populations separately.

As seen in Figure~\ref{f-zdistmcbwnhistack}, the metallicities of the pLLSs range from $[{\rm X/H}]<-2.4$ to $-0.1$ (where ``X'' for the W16 sample is Mg), although absorbers with only lower or upper limits could reside beyond these bounds. The range of metallicities is similar to the \Lthirteen\ sample. The MDF of the W16 pLLSs is clearly bimodal, with a paucity of systems at $[{\rm X/H}]=-1$, consistent with \citetalias{Lehner2013}. We test the bimodality of the W16 pLLSs using the Gaussian Mixture Model \citep[GMM; see][]{Muratov2010} and the dip statistic \citep{Hartigan1985}, as performed in \citetalias{Lehner2013}. There is a \pDipWPLLS\ probability that the MDF of the W16 pLLSs is bimodal according to the dip test. The GMM test rejects a unimodal distribution at the \pGMMWPLLS\ confidence level. In both tests, limits on the metallicities are treated as actual values; the means of the two peaks would separate further if the true metallicities of the limits were known.

To determine the mean metallicities of the low-metallicity ($[{\rm X/H}]<-1$) and high-metallicity ($[{\rm X/H}]\ge-1$) branches of the bimodal pLLS MDF, we use a survival analysis, which takes into account the large number of limits in the W16 sample. We are able to make use of a survival analysis because we cover a broad range in \nhi\ in both the low- and high-metallicity branches. Except for the special case of J1500+4836 (where we obtained additional HIRES exposures to greatly improve the S/N in this $[{\rm X/H}]<-2.5$ pLLS; see \S\ref{s-discussion}), the upper and lower limits are evenly-distributed across their metallicity branches, $[{\rm X/H}]<-1$ and $[{\rm X/H}]\ge-1$, respectively. The limits are a mixture of MODS, MagE, and HIRES observations at similar metallicities. There is no evidence for a difference in the sensitivity to upper limits (and, to an extent, lower limits) based on the instrument used (i.e., based on the resolution of the spectra), which would otherwise cause a bias. We therefore apply the Kaplan--Meier product limit estimator as described by \citet{Feigelson1985} and \citet{Isobe1986} and implemented in ASURV Rev 1.2 program \citep{LaValley1992}.

The mean metallicity of the low-metallicity branch of the pLLS MDF is $\left\langle[{\rm X/H}]\right\rangle= \WPLLSLowmet \pm \WPLLSLowmetError$ (\WPLLSLowmetPercent\ solar) and of the high-metallicity branch is $\left\langle[{\rm X/H}]\right\rangle= \WPLLSHighmet \pm \WPLLSHighmetError$ (\WPLLSHighmetPercent\ solar) using the Kaplan--Meier estimator. The errors quoted are the Kaplan--Meier errors on the mean value. The mean of the low-metallicity branch is biased slightly high because the lowest-metallicity point, an upper limit, is considered a detection in the Kaplan--Meier estimator. To check for consistency with the survival analysis, we use the GMM algorithm to estimate the means of the two metallicity branches using a homoscedastic split. This yields $\left\langle\Delta[{\rm X/H}]\right\rangle \le \pGMMhomoscedasticLowmetWPLLS$ for the low-metallicity branch and $\left\langle\Delta[{\rm X/H}]\right\rangle \ge \pGMMhomoscedasticHighmetWPLLS$ for the high-metallicity branch of the W16 pLLSs. The dispersion in each branch is $\sigma_1 \equiv \sigma_2 = \pGMMhomoscedasticDispersionWPLLS$. Since the GMM does not properly treat the upper and lower limits, the means of the low- and high-metallicity peaks derived with the test are taken as upper and lower limits, respectively, consistent with the values derived using the Kaplan--Meier estimator.

In contrast to the pLLSs, the MDF of the optically-thick W16 LLSs is not clearly bimodal. While the number of LLSs is still too small to robustly describe the shape of the MDF, we infer some tentative results: (1) there are 3--4/10 LLSs between $-1.2\lesssim[{\rm X/H}]\lesssim-0.8$, in the gap seen in the MDF of the pLLSs; (2) only 1/10 of the LLSs has $[{\rm X/H}]<-1.4$, while 10/20 of the W16 pLLSs are at such low metallicities. These two conclusions suggest a difference in the MDFs of the pLLSs and LLSs, even if the statistics are too small to describe the LLS MDF robustly.

\begin{deluxetable*}{lccccccccccc}
\tabcolsep=3pt
\tablecolumns{12}
\tablewidth{0pc}
\tablecaption{Low-Resolution Metallicities of the pLLSs and LLSs\label{t-cbw}}
\tabletypesize{\footnotesize}
\tablehead{\colhead{Target} & \colhead{$z_{\rm abs}$} & \colhead{$\log N_{\rm H\,I}$} & \colhead{$\log N_{\rm Mg\,II}$} & \colhead{\rm [Mg/H]} & \colhead{\rm [Mg/H]} & \colhead{\rm [Mg/H]} \\ \colhead{} & \colhead{} & \colhead{[cm$^{-2}$]} & \colhead{[cm$^{-2}$]} & \colhead{Measurement Errors} & \colhead{68\% C.I.} & \colhead{95\% C.I.}}
\startdata
    J0800+4212 &     0.598  &      $16.95\pm 0.20$ &        $11.75^{+0.17}_{-0.30}$ &                 $-1.99\pm 0.3$ &             [$-2.13$, $-1.85$] &             [$-2.24$, $-1.62$] \\
    J0804+2743 &     0.9106 &      $17.45\pm 0.10$ &        $12.27^{+0.06}_{-0.08}$ &                $-1.96\pm 0.13$ &             [$-2.12$, $-1.79$] &             [$-2.24$, $-1.52$] \\
    J0806+1442 &     1.0943 &      $17.40\pm 0.20$ &                       $>$13.51 &                     $>$$-0.64$ &          [$>$$-0.80$, \nodata] &          [$>$$-0.92$, \nodata] \\
    J0806+1442 &     0.9238 &      $17.20\pm 0.20$ &                $12.57\pm 0.09$ &                $-1.40\pm 0.22$ &             [$-1.55$, $-1.26$] &             [$-1.66$, $-1.01$] \\
    J0810+5025 &     0.650  &      $16.82\pm 0.15$ &                       $<$11.75 &                     $<$$-1.81$ &          [\nodata, $<$$-1.67$] &          [\nodata, $<$$-1.43$] \\
    J0810+5424 &     0.8570 &      $17.50\pm 0.15$ &                       $>$13.55 &                     $>$$-0.73$ &          [$>$$-0.89$, \nodata] &          [$>$$-1.01$, \nodata] \\
    J0950+3025 &     0.5876 &      $16.95\pm 0.15$ &                       $>$13.35 &                     $>$$-0.39$ &          [$>$$-0.54$, \nodata] &          [$>$$-0.64$, \nodata] \\
    J0950+3025 &     0.5739 &      $17.50\pm 0.15$ &                $13.24\pm 0.10$ &                $-1.08\pm 0.18$ &             [$-1.26$, $-0.91$] &             [$-1.37$, $-0.63$] \\
    J0957+6310 &     0.9100 &      $16.75\pm 0.20$ &                       $<$11.89 &                     $<$$-1.56$ &          [\nodata, $<$$-1.42$] &          [\nodata, $<$$-1.18$] \\
    J1005+4659 &     0.8413 &      $16.92\pm 0.15$ &                $13.49\pm 0.05$ &                $-0.17\pm 0.16$ &             [$-0.31$, $-0.04$] &             [$-0.42$, $+0.18$] \\
    J1009+0235 &     1.0875 &      $17.50\pm 0.10$ &        $12.94^{+0.16}_{-0.19}$ &        $-1.31^{+0.19}_{-0.21}$ &             [$-1.48$, $-1.14$] &             [$-1.59$, $-0.88$] \\
    J1009+0235 &     0.488  &      $16.90\pm 0.15$ &        $12.37^{+0.11}_{-0.17}$ &        $-1.33^{+0.19}_{-0.23}$ &             [$-1.47$, $-1.19$] &             [$-1.57$, $-0.96$] \\
    J1015+0109 &     0.5880 &      $17.50\pm 0.05$ &                $13.20\pm 0.09$ &                $-1.12\pm 0.10$ &             [$-1.29$, $-0.95$] &             [$-1.40$, $-0.67$] \\
    J1222+0413 &     0.6547 &      $17.55\pm 0.10$ &                       $>$13.50 &                     $>$$-0.85$ &          [$>$$-1.02$, \nodata] &          [$>$$-1.14$, \nodata] \\
    J1306+4351 &     0.6686 &      $16.85\pm 0.10$ &                $13.09\pm 0.06$ &                $-0.52\pm 0.12$ &             [$-0.66$, $-0.38$] &             [$-0.77$, $-0.17$] \\
    J1326+2925 &     0.7324 &      $17.20\pm 0.30$ &                $13.08\pm 0.03$ &                $-0.92\pm 0.30$ &             [$-1.07$, $-0.77$] &             [$-1.18$, $-0.52$] \\
    J1355+2600 &     0.5360 &      $16.75\pm 0.10$ &                $13.12\pm 0.04$ &                $-0.41\pm 0.11$ &             [$-0.55$, $-0.27$] &             [$-0.66$, $-0.03$] \\
    J1355+2600 &     0.4852 &      $16.70\pm 0.15$ &        $12.19^{+0.22}_{-0.29}$ &        $-1.27^{+0.27}_{-0.33}$ &             [$-1.42$, $-1.13$] &             [$-1.52$, $-0.87$] \\
    J1500+4836 &     0.898  &      $17.15\pm 0.15$ &                       $<$11.44 &                     $<$$-2.48$ &          [\nodata, $<$$-2.34$] &          [\nodata, $<$$-2.09$] \\
    J1519+4404 &     0.6042 &      $17.05\pm 0.10$ &        $13.36^{+0.12}_{-0.18}$ &        $-0.49^{+0.16}_{-0.20}$ &             [$-0.63$, $-0.35$] &             [$-0.75$, $-0.11$] \\
    J1528+5205 &     0.5809 &      $17.27\pm 0.10$ &                       $>$13.65 &                     $>$$-0.44$ &          [$>$$-0.60$, \nodata] &          [$>$$-0.71$, \nodata] \\
    J1536+3932 &     0.454  &      $16.65\pm 0.20$ &                       $<$11.49 &                     $<$$-1.92$ &          [\nodata, $<$$-1.78$] &          [\nodata, $<$$-1.52$] \\
    J1631+4359 &     0.5196 &      $16.75\pm 0.10$ &                       $>$12.81 &                     $>$$-0.72$ &          [$>$$-0.86$, \nodata] &          [$>$$-0.97$, \nodata] \\
    J1716+3027 &     0.756  &      $16.50\pm 0.20$ &                       $<$11.38 &                     $<$$-1.79$ &          [\nodata, $<$$-1.66$] &          [\nodata, $<$$-1.31$] \\
    J1716+3027 &     0.7103 &      $16.40\pm 0.25$ &                       $<$11.43 &                     $<$$-1.65$ &          [\nodata, $<$$-1.51$] &          [\nodata, $<$$-1.16$] \\
    J1716+3027 &     0.3995 &      $16.90\pm 0.20$ &        $11.94^{+0.07}_{-0.09}$ &                $-1.77\pm 0.22$ &             [$-1.91$, $-1.63$] &             [$-2.02$, $-1.40$] \\
    J2253+1402 &     0.5737 &      $16.75\pm 0.15$ &                $13.28\pm 0.07$ &                $-0.23\pm 0.17$ &             [$-0.37$, $-0.09$] &             [$-0.49$, $+0.14$] \\
    J2253+1402 &     0.327  &      $16.75\pm 0.20$ &        $11.96^{+0.17}_{-0.28}$ &        $-1.60^{+0.26}_{-0.34}$ &             [$-1.74$, $-1.46$] &             [$-1.85$, $-1.22$] \\
    J2255+1457 &     0.8051 &      $16.85\pm 0.10$ &                $13.11\pm 0.06$ &                $-0.48\pm 0.12$ &             [$-0.62$, $-0.35$] &             [$-0.73$, $-0.13$] \\
    J2255+1457 &     0.446  &      $16.60\pm 0.30$ &                       $<$11.70 &                     $<$$-1.66$ &          [\nodata, $<$$-1.52$] &          [\nodata, $<$$-1.24$]
\enddata
\tablecomments{Non-detections are 2$\sigma$ upper limits. Lower limits are adopted when the saturation correction for \mgiit\ is beyond the adopted standard correction of 0.15 dex.}
\end{deluxetable*}

\subsection{The Metallicity Distribution Function of the \lowercase{p}LLS\lowercase{s} and LLS\lowercase{s} at $\lowercase{z}\lesssim1$}\label{s-results-L13_W16}

The goals of our study are to increase the original sample of \citetalias{Lehner2013} in order to more robustly determine the metallicity distribution of the pLLSs and LLSs at $z\lesssim1$ and to build a more representative sample in \hi\ column density (see \S\ref{s-results-W16} and Figure~\ref{f-hihisto}). The combined \Lthirteen+W16 sample consists of \numLW\ pLLSs and LLSs, doubling the \Lthirteen\ sample and providing a much better sampling in the range $16.7<\log\mathnhi<17.7$.

\begin{figure}[tbp]
\epsscale{1.2}
\plotone{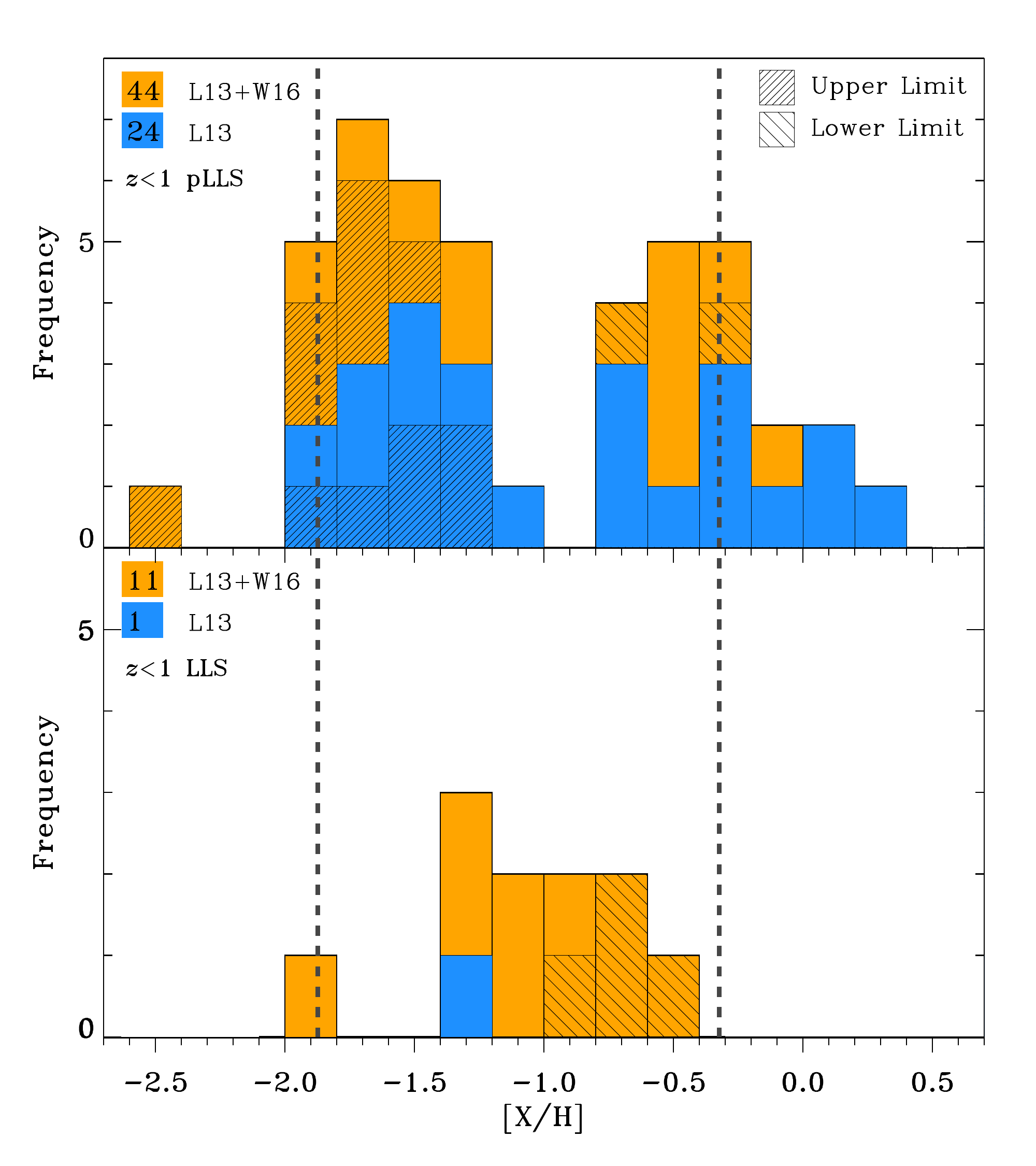}
    \caption{Metallicity distribution functions (MDFs) of the pLLSs (top) and the LLSs (bottom) at $z\lesssim1$ from the \Lthirteen+W16 sample. The MDF of the pLLSs is double-peaked, with a \pDipLWPLLS\ probability that it is bimodal. The shape of the LLS MDF is unclear owing to the sparsity of data and the high fraction (45\%) of lower limits. The vertical lines are the means of the low- and high-metallicity branches of the \Lthirteen+W16 pLLSs only, $\left\langle[{\rm X/H}]\right\rangle = \LWPLLSLowmet \pm \LWPLLSLowmetError$ and $\left\langle[{\rm X/H}]\right\rangle = \LWPLLSHighmet \pm \LWPLLSHighmetError$, respectively.
    \label{f-zdistmnlnhistack}}
\end{figure}

In Figure~\ref{f-zdistmnlnhistack} we show separately the MDFs of the \Lthirteen+W16 pLLSs and the LLSs. With \numLWPLLS\ absorbers, the MDF of the pLLSs is consistent with a bimodal distribution. The probability that the MDF of the joint \Lthirteen+W16 pLLS sample is bimodal is \pDipLWPLLS\ according to the dip test, and the GMM test rejects a unimodal distribution at the \pGMMLWPLLS\ confidence level. Limits on the metallicities are treated as values in both tests. There are \numLWPLLSLowmet\ pLLSs with $[{\rm X/H}]<-1$ and \numLWPLLSHighmet\ with $[{\rm X/H}]\ge-1$. Hence the fraction of pLLSs in the low-metallicity branch of the MDF is \fLowmetLWPLLS\ (68\% confidence interval using the Wilson score interval). Using the Kaplan--Meier statistic (see \S\ref{s-results-W16}), we derive $\left\langle[{\rm X/H}]\right\rangle = \LWPLLSLowmet \pm \LWPLLSLowmetError$ and $\left\langle[{\rm X/H}]\right\rangle = \LWPLLSHighmet \pm \LWPLLSHighmetError$ as the means of the low- and high-metallicity branches, respectively, of the \Lthirteen+W16 pLLS MDF. To compare this with the results of \citetalias{Lehner2013}, we recalculate the means of the \Lthirteen\ pLLSs using the Kaplan--Meier statistic and find $\left\langle[{\rm X/H}]\right\rangle = \LPLLSLowmet \pm \LPLLSLowmetError$ and $\left\langle[{\rm X/H}]\right\rangle = \LPLLSHighmet \pm \LPLLSHighmetError$ for the low- and high-metallicity branches, respectively. The means of the \Lthirteen+W16 pLLSs are consistent with those of the \Lthirteen\ pLLSs.\footnote{The mean of the low-metallicity branch of the W16 pLLSs is influenced by the strong outlier at $[{\rm X/H}]<-2.5$, which partly explains the difference between the meants of the \Lthirteen+W16 pLLSs and the \Lthirteen\ pLLSs. When this outlier is removed, we derive a mean for the low-metallicity branch of the \Lthirteen+W16 pLLSs of $\left\langle[{\rm X/H}]\right\rangle = -1.73 \pm 0.06$, in good agreement with that of the \Lthirteen\ low-metallicity branch.}

Since the adopted \Lthirteen\ sample only adds one LLS to the combined \Lthirteen+W16 sample, the shape of the LLS metallicity distribution is still difficult to assess. The conclusions drawn from the W16 sample in \S\ref{s-results-W16} still apply. We also note that, if we were to include the LLSs with $17.7<\log\mathnhi<18.5$ (see \S\ref{s-results-met_vs_h1}), we would reach a similar conclusion, strengthening the inference that the metallicity distribution of the LLSs is different from that of the pLLSs.

\begin{figure}[tbp]
\epsscale{1.2}
\plotone{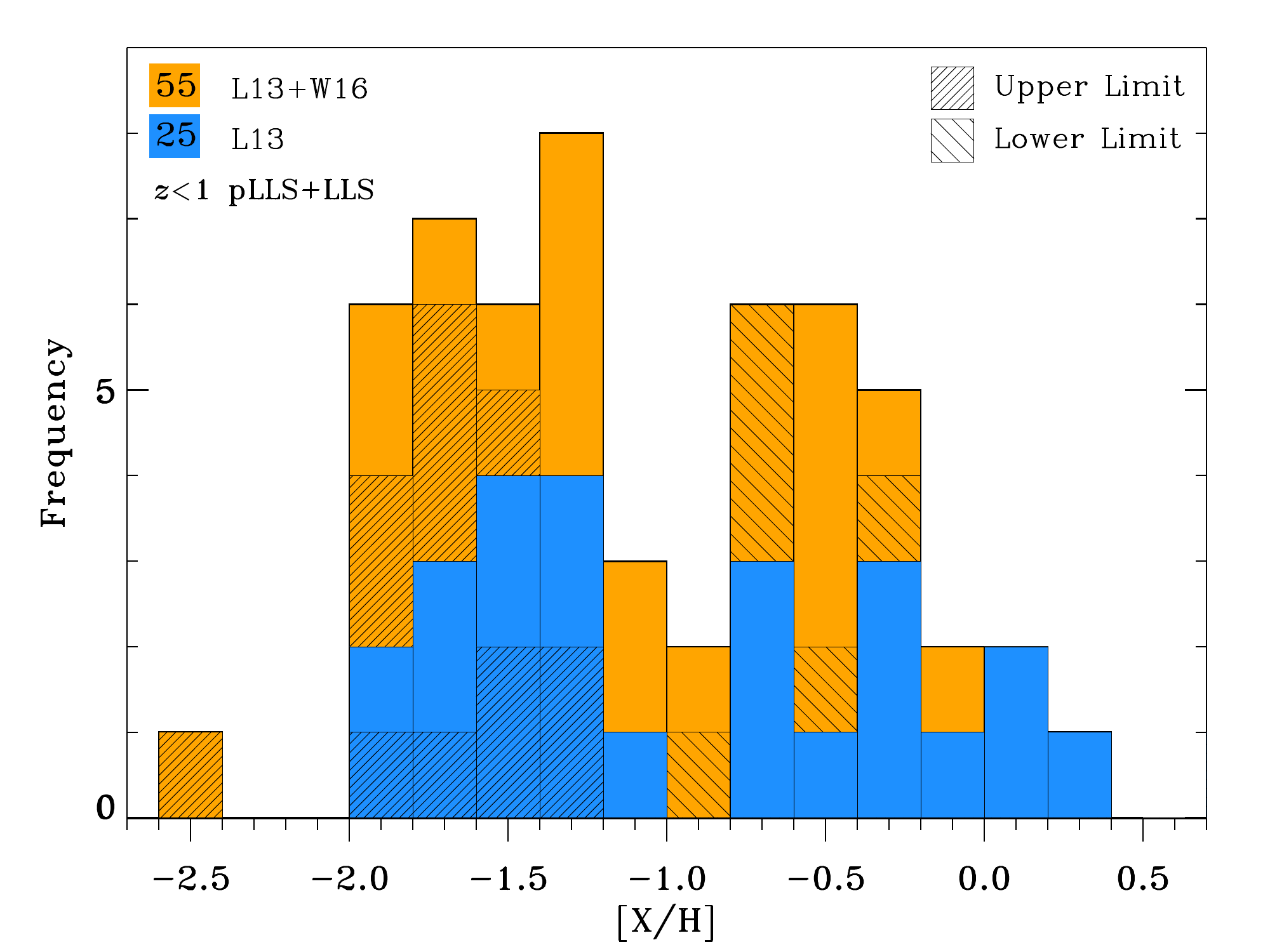}
    \caption{Metallicity distribution function (MDF) of the pLLSs+LLSs ($16.1<\log\mathnhi<17.7$) at $z\lesssim1$ from the \Lthirteen+W16 sample. The MDF is double-peaked with a dip around $[{\rm X/H}]\approx-1$. At $[{\rm X/H}]<-1$, there are \numLWLowmet\ pLLSs+LLSs and \numLWHighmet\ at $[{\rm X/H}]\ge-1$. Using the Kaplan--Meier statistic (see \S\ref{s-results-W16}), we derive $\left\langle[{\rm X/H}]\right\rangle = \LWLowmet \pm \LWLowmetError$ and $\left\langle[{\rm X/H}]\right\rangle = \LWHighmet \pm \LWHighmetError$ as the means of the low-metallicity and high-metallicity branches, respectively, of the combined pLLS+LLS MDF. The pLLS with $[{\rm X/H}]<-2.4$ is the lowest-metallicity gas known at $z\lesssim1$.
    \label{f-zdistmnl}}
\end{figure}

Since the MDF in \citetalias{Lehner2013} contained both pLLSs and LLSs, we show in Figure~\ref{f-zdistmnl} the MDF of the pLLSs+LLSs in the \Lthirteen+W16 sample. The MDF is double-peaked with a dip around $[{\rm X/H}]\approx-1$, consistent with the results of \citetalias{Lehner2013}. The fraction of pLLSs+LLSs in the low-metallicity branch of the MDF is \fLowmetLW\ (68\% confidence interval using the Wilson score interval). The probability that the \Lthirteen+W16 pLLS+LLS MDF is bimodal is \pDipLW\ according to the dip test (this number was 88\% in \citetalias{Lehner2013}). The GMM test rejects a unimodal distribution at the \pGMMLW\ confidence level. In both tests, limits on the metallicities are treated as actual values; the means of the two peaks would separate further if the true metallicities of the limits were known. While the dip test suggests that the pLLS+LLS distribution is bimodal, it has a lower level of confidence than the MDF of the pLLSs alone (or than that of \citetalias{Lehner2013}, which had only a few LLSs). This is primarily due to the differences between the pLLS and LLS populations, with 3--4 LLSs at $-1.2\lesssim[{\rm X/H}]\lesssim-0.8$, filling the metallicity gap of the pLLSs.

\subsection{Evolution of the CGM Metallicity as a Function of \nhi}\label{s-results-met_vs_h1}

Both simulations and observations suggest that the conditions of the gas being probed change as one looks at CGM absorbers of increasing \nhi, from pLLSs through DLAs. For example, in the studies of the low-redshift CGM compiled by \citetalias{Lehner2013}, the DLAs, SLLSs, and LLSs probed different ranges of impact parameters relative to their identified hosts, though with significant overlap (see their Figure 9). The pLLSs and LLSs in that sample were associated with galaxies projected 30--130 kpc from the QSO sightline (well within the virial radius, $R_{\rm vir}$); the SLLS were found 10--85 kpc from their host galaxies; and the DLAs were found at impact parameters of only 5--25 kpc. Simulations at $z\sim2$--3.5 \citep[e.g.,][]{Fumagalli2011a,Faucher-Giguere2011,vandeVoort2012a} show similar results: (1) most of the cross-section for DLAs is within the inner regions of halos; (2) gas with $\log\mathnhi\lesssim19$ (including pLLSs and LLSs) comprises most of the cross-section of cold accretion streams within $R_{\rm vir}$; and (3) less than half of the cross-section for SLLSs is in accretion streams, suggesting they are intermediate to LLSs and DLAs. Furthermore, the ionization levels observed in absorbers from pLLSs through DLAs imply the level of ionization generally decreases with increasing \hi\ column density. For photoionized gas, this corresponds to an increasing density with increased \hi\ column density. LLSs typically have lower densities than those found in higher-\nhi\ absorbers \citep[see, e.g.,][though at $z\sim3$]{Schaye2001b}. While the connection is somewhat crude, examining the MDF for absorbers covering a broad range of \hi\ column densities provides a measure of how the origins of the gas are changing with environment (impact parameter and density) within galaxy halos.

\begin{figure}[tbp]
\epsscale{1.2}
\plotone{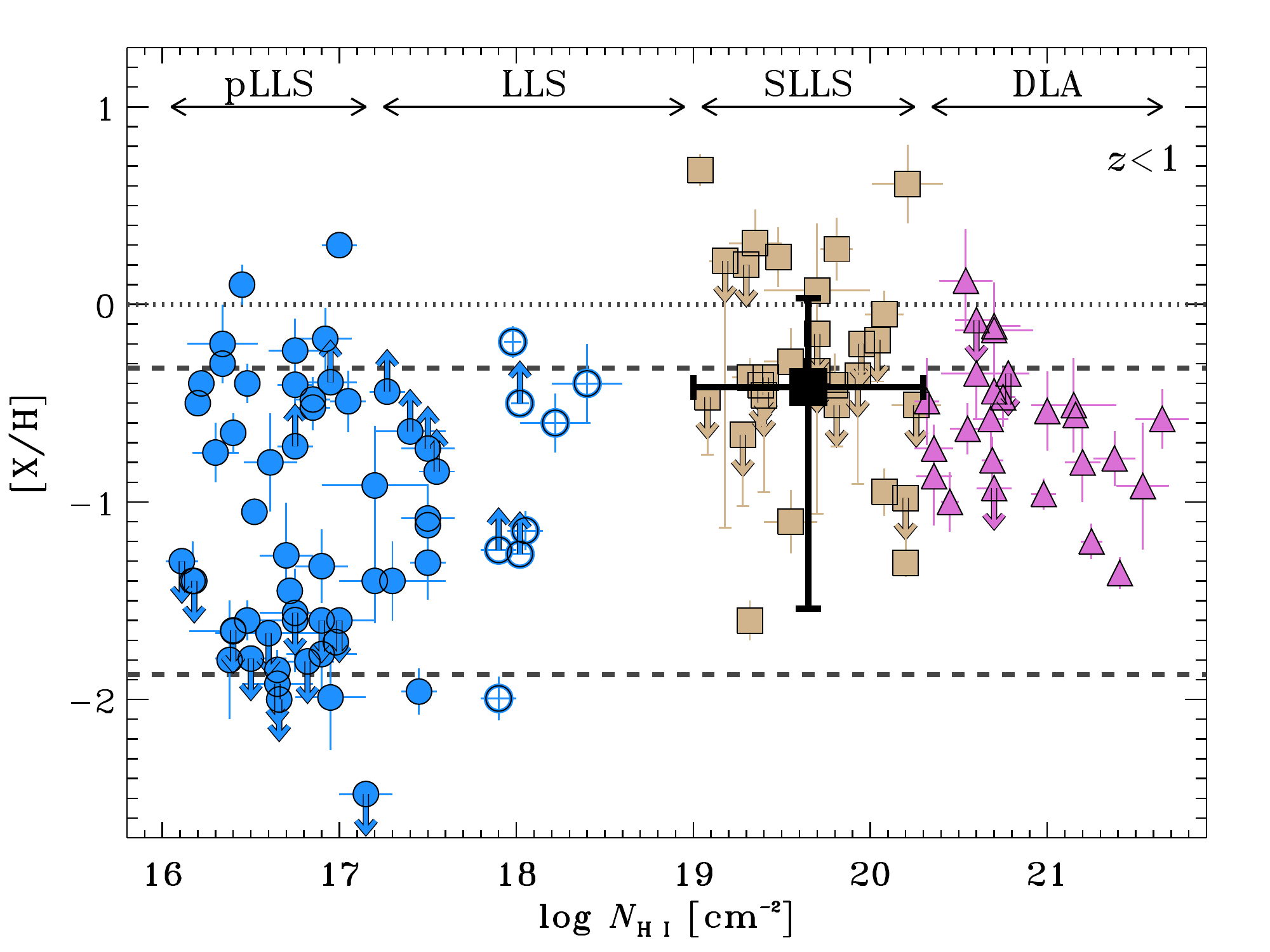}
    \caption{Metallicities of pLLSs, LLSs, SLLSs, and DLAs at $z\lesssim1$ as a function of \nhi. The dotted line represents solar metallicity; the dashed lines represent the means of the \Lthirteen+W16 pLLS MDF, \LWPLLSLowmetPercent\ and \LWPLLSHighmetPercent\ solar metallicities. The bimodality and very low metallicities ($<$$-1.4$) in the MDF of the pLLSs (and somewhat of the LLSs) is clearly seen, while the unimodal MDFs of the SLLSs and DLAs starkly contrast. The LLSs excluded from the \Lthirteen+W16 sample (see \S\ref{s-adoptedsample}) are shown as open circles. Several SLLSs have upper limits on their metallicities from their [Zn/H] ratios. We can constrain these to the high-metallicity regime by using their [Fe/H] metallicity as a lower limit (depicted in the figure as a negative error bar with a hat), noting that these lower limits are all at about $[{\rm X/H}]\sim-1$ or higher. The large, black data point represents the mean of $z<1$ SLLSs from \citet{Fumagalli2016}, where the horizontal bar indicates the range of the \nhi\ bin and the vertical error bar represents the 25th/75th percentile of the composite posterior PDF.
    \label{f-metvsh1}}
\end{figure}

In Figure~\ref{f-metvsh1} we show the metallicity distribution of the pLLSs and LLSs from the \Lthirteen+W16 sample (filled blue circles) as a function of \nhi. The unbinned data show a clear deficit of pLLSs near $[{\rm X/H}]\sim-1$ and two clusters centered around $-$0.3 and $-$1.7 dex, fully consistent with our conclusions in \S\ref{s-results-L13_W16}. The open blue circles represent the data excluded from our statistical sample (see \S\ref{s-adoptedsample}) for which we are able to derive the metallicity (or limit on the metallicity). In this figure we also show the metallicities of the SLLSs and DLAs at $z\lesssim1$, which are from the compilation made by \citetalias{Lehner2013} (see their Table 1 and references therein). The black square in the figure represents the mean metallicity of $z<1$ SLLSs from the new analysis of \citet{Fumagalli2016}, where the horizontal bar indicates the range of the \nhi\ bin and the vertical error bar represents the 25th/75th percentile of the composite posterior PDF. There is no clear indication that the MDF of \citet{Fumagalli2016} is inconsistent with a unimodal distribution.

The distribution of the data points for the SLLSs and DLAs in Figure~\ref{f-metvsh1} differ significantly from the pLLSs and LLSs, as already noted by \citetalias{Lehner2013}. The metallicities of the SLLSs and DLAs are consistent with unimodal distributions, and their distributions largely overlap. There is a clear absence of DLAs and SLLSs with $[{\rm X/H}]<-1.4$, where a large number of pLLSs are observed. We thus define ``very metal-poor'' absorbers as those with metallicities 2$\sigma$ below the mean of the DLAs compiled by \citetalias{Lehner2013}; i.e., those at $[{\rm X/H}]<-1.4$, where we expect to find just 2.5\% of DLAs. Examining Figure~\ref{f-metvsh1}, we also note that the transition between the bimodal and unimodal MDFs likely occurs between $17.2\lesssim\log\mathnhi<19.0$, since we have shown the distributions toward lower and higher \nhi\ to be bimodal and unimodal, respectively.

To quantify the similarities and differences between different types of absorbers, we now estimate and compare the means and ranges of the metallicity distributions of pLLSs, LLSs, SLLSs, and DLAs at $z\lesssim1$. The mean of the DLA sample, treating the upper limits with the Kaplan--Meier statistic, is $\left\langle[{\rm X/H}]\right\rangle = -0.66\pm0.07$. The mean metallicity of the SLLSs assembled by \citetalias{Lehner2013} is difficult to assess owing to the larger number of upper and lower limits, so we consider instead the results of \citet{Fumagalli2016}, $\left\langle[{\rm X/H}]\right\rangle = -0.42^{+0.45}_{-1.12}$, where the errors represent the 25th/75th percentile of the composite posterior PDF. The mean metallicity of the LLSs using the Kaplan--Meier statistic is $\left\langle[{\rm X/H}]\right\rangle = \LWLLSAllmet\pm\LWLLSAllmetError$. The mean is biased low because all four of the highest-metallicity LLSs are lower limits; additionally, the Kaplan--Meier statistic necessarily treats the highest point (a lower limit) as a detection. There is an overall agreement between the means of the DLAs, SLLSs \citep[contrasting the results of][]{Som2015}, and the upper end of the LLSs (the mean of the LLSs may be biased low). The mean metallicity of the high-metallicity branch of the pLLSs using the Kaplan--Meier statistic is $\left\langle[{\rm X/H}]\right\rangle = \LWPLLSHighmet\pm\LWPLLSHighmetError$. Thus, the high-metallicity branch of the pLLSs has a statistically higher mean than that of the DLAs and LLSs (and a higher mean than that of the SLLSs, though this is not statistically-significant), which is explained by the lack of pLLSs with $-1.2<[{\rm X/H}]<-0.8$ (the dip of the MDF of the pLLSs) where many DLAs and LLSs are observed (see Figure~\ref{f-metvsh1}).

To understand how the distribution of low-metallicity and high-metallicity gas might change with environment, we compare in Table~\ref{t-fmet} the fractions of pLLSs, LLSs, SLLSs, and DLAs that are below or above various metallicity thresholds.\footnote{While the upper limits above (or lower limits below) these thresholds are necessarily excluded (since we cannot determine on which side of the threshold they exist), including them would change the fractions only sightly in all but one case; due to the large number of LLSs with lower limits just below $[{\rm X/H}]=-0.5$, the fraction of LLSs above this threshold could be as high as $f([{\rm X/H}]>-0.5)=(36^{+19}_{-17})\%$.} There is a smaller fraction of very metal-poor ($[{\rm X/H}]\lesssim-1.4$) LLSs than of very metal-poor pLLSs (1--27\% vs.\ 35--51\%, respectively, at the 68\% confidence level). However, the fractions of LLSs and pLLSs with $[{\rm X/H}]<-1$ (or $>$$-$1) are quite similar, as are the fractions with $[{\rm X/H}]>-0.5$ (given the above caveat) or $-$1.0. Thus, very metal-poor gas with $[{\rm X/H}]\lesssim-1.4$ is seen more frequently in pLLSs than in LLSs.

In Table~\ref{t-fmet} we also list these fractions for the SLLSs and DLAs. The lack of low-metallicity DLAs at $z\lesssim1$ is striking, with very few having $[{\rm X/H}]<-1$. This is in clear contrast with the pLLSs and LLSs, where many very metal-poor absorbers are found. The results are less clear for the SLLSs, though it is apparent there is also an overall lack of SLLSs with $[{\rm X/H}]<-1.4$ (see Table~\ref{t-fmet} and Figure~\ref{f-metvsh1}; see also the recent work at $z<1.25$ by \citealt{Quiret2016}). The overall picture from this analysis at $z\lesssim1$ is that very metal-poor gas ($[{\rm X/H}]\lesssim-1.4$) is mostly confined to absorbers with $\log\mathnhi\lesssim17.2$ and metal-poor gas ($[{\rm X/H}]\lesssim-1$) to absorbers with $\log\mathnhi\lesssim18.5$.

\begin{deluxetable*}{lcccccc}
\tabcolsep=3pt
\tablecolumns{7}
\tablewidth{0pc}
\tablecaption{Fraction of Metal-Enriched Systems across \nhi\label{t-fmet}}
\tabletypesize{\footnotesize}
\tablehead{\colhead{Sample} & \colhead{$n$} & \colhead{log \nhi\ [cm$^{-2}$]} & \colhead{$f([{\rm X/H}] < -1.4)$} & \colhead{$f([{\rm X/H}] < -1.0)$} & \colhead{$f([{\rm X/H}] > -1.0)$} & \colhead{$f([{\rm X/H}] > -0.5)$}}
\startdata
    pLLSs                        &  44  &  $[16.4, 17.2)$   &  $\left(43\pm8\right)\%$         &  $\left(57\pm8\right)\%$         &  $\left(43\pm8\right)\%$         &  $\left(30\pm8\right)\%$          \\
    LLSs                         &  11  &  $[17.2, 18.5)$   &  $\left(9^{+18}_{-8}\right)\%$   &  $\left(55\pm19\right)\%$        &  $\left(45\pm19\right)\%$        &  $\left(9^{+18}_{-8}\right)\%$    \\
    SLLSs\tablenotemark{a}       &  29  &  $[19.0, 20.3)$   &  $\left(3^{+7}_{-3}\right)\%$    &  $\left(10^{+9}_{-6}\right)\%$   &  $\left(90^{+5}_{-9}\right)\%$   &  $\left(73\pm10\right)\%$         \\
    SLLSs\tablenotemark{b}       &  29  &  $[19.0, 20.3)$   &  $\left(3^{+7}_{-3}\right)\%$    &  $\left(27\pm9\right)\%$         &  $\left(73\pm9\right)\%$         &  $\left(53\pm10\right)\%$         \\
    DLAs                         &  26  &  $[20.3, 22.0]$   &  $\left(3^{+8}_{-3}\right)\%$    &  $\left(8^{+9}_{-5}\right)\%$    &  $\left(88^{+6}_{-10}\right)\%$  &  $\left(35\pm11\right)\%$
\enddata
\tablecomments{
The number of absorbers in each \nhi\ bin is given by $n$. The fractions, $f$, are given at the 68\% confidence level using the Wilson score interval, and exclude systems with lower (upper) limits that lie below (above) the metallicity boundary of each column (e.g., the first column excludes lower (upper) limits that lie below (above) $[{\rm X/H}]=-1.4$).
}
\tablenotetext{a}{Uses upper limits derived by [Zn/H].}
\tablenotetext{b}{Uses lower limits derived by [Fe/H].}
\end{deluxetable*}


\section{Discussion}\label{s-discussion}

We have studied the MDF of an \hi-selected sample of absorbers with $16.4\lesssim\log\mathnhi\le17.7$ because these are the most likely to be associated with flows through the CGM at $z\lesssim1$ \citep[e.g.,][]{Hafen2016,vandeVoort2012a,Fumagalli2011a}. \citetalias{Lehner2013} have shown the gas associated with the \hi\ in these systems has temperatures of a few$\times10^4$ K, and 9/10 of their pLLSs and LLSs with a measured impact parameter lie within $\rho\sim125$ kpc of galaxies (the host galaxy of 1 pLLS has perhaps not yet been correctly identified, and the remaining 18 have not been identified; see \citetalias{Lehner2013}). Further, the overdensity probed by a pLLS/LLS at $z=0.7$ ($\delta\rho_{m}\sim900$; see \S\ref{s-intro}) is in line with the overdensities of galaxy halos seen in simulations and analytically \citep{Schaye2001b}, with $\delta\rho_{m}\sim{\rm few}\times10^2$--$10^3$.

This work has doubled the sample of \hi-selected $0.1\lesssim z\lesssim1.1$ pLLSs+LLSs with measured metallicities, in particular probing a more representative \nhi\ distribution above $\log \mathnhi \sim 16.8$ than the previous work of \citetalias{Lehner2013}. The MDF of the combined \Lthirteen+W16 pLLS+LLS sample is still statistically consistent with a bimodal distribution. We find a unimodal distribution is rejected at the \pGMMLW\ confidence level, and there is a \pDipLW\ probability that the distribution is bimodal, consistent with the results of \citetalias{Lehner2013}. Further, \numLWLowmet\ of the \numLW\ absorbers, \fLowmetLW, are associated with the low-metallicity branch ($[{\rm X/H}]<-1$). Absorbers in the low-metallicity branch have on average a metallicity $\sim$20$\times$ lower than those in the high-metallicity branch.


\subsection{The Bimodality of CGM Metallicities at $\lowercase{z}\lesssim1$}

When we look separately at the pLLSs from the combined \Lthirteen+W16 sample, there is a \pDipLWPLLS\ probability that the pLLS MDF is bimodal, according to the dip test. We find that \fLowmetLWPLLS\ of the pLLSs are associated with the low-metallicity branch ($[{\rm X/H}]<-1$). A similar fraction of LLSs have metallicities $[{\rm X/H}]<-1$. However, only 1/11 LLSs has a metallicity $[{\rm X/H}]<-1.4$ (2$\sigma$ below the mean metallicity of the DLAs; see \S\ref{s-results-met_vs_h1}), while $\sim$43\% of pLLSs have such low metallicities. Hence, while the sample of LLSs is quite small, there is a strong suggestion with our new sample that the LLS MDF is different from the pLLS MDF. At even higher \hi\ column densities, the SLLS and DLA populations have unimodal distributions with relatively high mean metallicities, similar to that of the high-metallicity branch of the pLLSs. However, the high-\nhi\ systems lack the low-metallicity gas with $[{\rm X/H}]<-1$ that is common in the lower \hi\ column density material. Thus, Figure~\ref{f-metvsh1} and Table~\ref{t-fmet} demonstrate that the metallicity distribution of absorbers across the column density regime $16.4\le\log\mathnhi\lesssim22.0$ varies significantly. It implies there is a fundamental change in the physical origins with \hi\ column density.

The bimodal distribution must transition to a unimodal distribution in the regime where we have relatively few absorbers, $17.2\lesssim\log\mathnhi\lesssim19.0$. Since the MDF of the LLSs does not appear to match that of the pLLSs, our data suggest that the transition is beginning in the LLS regime. This transition is likely associated with changes in the structures probed on average by CGM absorbers as one looks to higher and higher column densities. We expect that the lower column density systems in Figure~\ref{f-metvsh1} are probing more diffuse material farther from the central galaxy, whereas the SLLSs and DLAs are probing material at smaller impact parameters (e.g., \citetalias{Lehner2013}; \citealt{Meiring2011}; \citealt{Battisti2012}; \citealt{Tumlinson2013}).

With our new survey, we have also discovered the lowest-metallicity gas yet observed at $z<1$, with no detection of \mgii. This pLLS is found toward J1500+4836 at $z\sim0.9$ and has a metallicity $[{\rm X/H}]<-2.48$ or $Z<0.3\%Z_\sun$. This is a factor $>$5 lower than the mean metallicities of the low-metallicity branch of the pLLSs and a factor $>$100 lower than the mean metallicity of the DLAs at $z<1$. With future high-resolution, high-S/N ground-based and {\it HST} observations, we will search for any evidence of metals (e.g., \mgii, \ciii), but this limit already hints that there are regions in the universe that have remained uncontaminated for over 6 billion years.


\subsection{The High-Metallicity Branch of the Bimodal MDF}

The MDF of the pLLSs suggests the absorbers probe two distinct gas populations. The high-metallicity population has a mean metallicity of $\sim$50\% solar metallicity, similar to the present-day metallicity of the Large Magellanic Cloud \citep{Russell1992}. This significantly metal-enriched gas has properties consistent with those expected for cold matter entrained in feedback-driven outflows \citep[similar to the outflows traced in galaxy spectroscopy, e.g., by][and others]{Rubin2014, Chisholm2016}, for matter being tidally-stripped from satellite galaxies \citep{Lehner2009}, and with simulated material tracing the remnants of earlier outflows that are being ``recycled'' through the CGM \citep[e.g.,][]{Oppenheimer2010, Ford2013}.

This broad range of potential origins for the high-metallicity absorbers reflects both the uncertainties in the properties of the galaxies associated with each absorber (esp.\ their metallicities) and the wide range of metallicities spanned by this peak, from 0.1 to 2 $Z_{\sun}$. There are individual cases where specific mechanisms have been associated with the absorbers. For example, \citet{Tripp2011} have argued an LLS is tracing cold gas in an outflow from a $z\approx0.92$ galaxy on the basis of a very large velocity spread ($\Delta v \approx 1200$ km s$^{-1}$), a super-solar metallicity, and the presence of large amounts of warm-hot gas traced by \ovi\ and \neviii. \citet{Lehner2009} similarly identified a metal-rich LLS ($[{\rm O/H}] = -0.6$) that they argued traced tidal detritus about a galaxy at $z\approx0.2$. In simulations at $z\sim2$--4 \citep[e.g.,][]{Fumagalli2011a,Kimm2011,Shen2013}, outflows and recycling gas show a range of \hi\ column densities, from those of pLLSs to DLAs. The expectation is that these outflows should have metallicities comparable to their host galaxies or higher, although this depends on the amount of coronal matter that is cooled as it interacts with the outflowing material. For example, the outflowing gas in the simulations of \citet{Shen2013} has a mean metallicity of $[{\rm X/H}]=-0.25$, comparable to our results, albeit at $z\sim3$. The very broad metallicity range exhibited by the pLLSs in the metal-rich branch of our MDF could be a consequence both of contributions from a range of these phenomena, as well as a potentially broad range of metallicities in the host galaxies driving outflows or losing gas to stripping.


\subsection{The Low-Metallicity Branch of the Bimodal MDF}

The other half of the pLLSs has a mean metallicity of a few percent solar, extending to metallicities well below 1\% solar. This material has experienced very little enrichment, although in every case yet tested (10 absorbers) they reside within the virial radius of a relatively luminous galaxy \citepalias[see][]{Lehner2013}. Like the more metal-rich population, the \hi\ in these systems is cool, of order $\sim$$10^4$ K, and certainly much less than the virial temperature \citep[expected to be of order $T_{\rm vir}\sim10^6$ K for galaxies like those in the \Lthirteen\ sample;][]{Dekel2006}. This matter is very unlikely to be part of a galaxy in its current form. The \hi\ column densities are too low to be part of the interstellar medium of a dwarf galaxy \citep{Ott2012}, and the metallicities are so low that they would generally not be associated with typical dwarf galaxies on the standard mass-metallicity relationship \citep{Lian2016,Jimmy2015}. While there are some extremely low-metallicity dwarfs \citep{Skillman2013}, these systems tend to be exceedingly rare \citep{Izotov2012} and would produce very few absorbers unless their cross-sections represented many times their virial radii.

As discussed by \citetalias{Lehner2013}, the low-metallicity branch has column densities, temperatures, and enrichment levels consistent with cold accretion. Recent simulations have emphasized that the \hi\ column density range expected for cold accretion streams overlaps the pLLS/LLS regime studied here \citep[e.g.,][]{Hafen2016,Fumagalli2011a,vandeVoort2012a,Shen2013,Stewart2011a}. The metallicity of such streams is expected to be low, but not zero. Dwarf galaxies are expected to enrich the gas that ultimately accretes onto central galaxies beyond the enrichment expected from Population III, with metallicities of a few percent solar predicted. For example, \citet{Fumagalli2011a} find that cold accretion streams at $z=1.3$--4 in the form of LLSs have metallicities of $[{\rm X/H}]\approx-1.8\pm0.5$, while \citet{vandeVoort2012a} found LLSs at $z=3$ tracing cold accretion at metallicities $[{\rm X/H}]\lesssim-1.5$. \citet{Shen2013} find that the metallicities of inflowing pLLSs and LLSs at $z\sim3$ are centered around $[{\rm X/H}]=-1.3$, higher than the low-metallicity absorbers in our sample, though the infall seen in these simulations may include a substantial fraction of high-metallicity gas recycling back onto the disk. Recently, \citet{Hafen2016} used zoom-in Feedback In Realistic Environments (FIRE) simulations at $z<1$ to study the metallicity distribution of the pLLSs and LLSs in a redshift interval similar to our survey. A comparison between their MDF and the observed MDF shows striking differences: (1) in their simulation the MDF is not bimodal and has a metallicity plateau between about $-1.3$ and $-0.5$, (2) low metallicity pLLSs and LLSs are prevalent in the observations but not in their simulations, and a large majority of the simulated pLLSs and LLSs have metallicities between $-1.5$ and $-0.3$ dex.

While it is beyond the scope of this paper to understand the origin(s) of the differences between observations and simulations, it seems that the recycling and mixing of the gas are too efficient in the FIRE simulations to allow for the presence of widespread low-metallicity pLLSs. Inflowing and outflowing gas are seen in these simulations, notably in the high-metallicity gas, which is consistent with our interpretation that high-metallicity gas probed by pLLSs and LLSs must trace outflowing gas and recycling (infalling) gas. In their FIRE simulations, there is about equal probability of metal-rich outflowing and inflowing gas. For the low-metallicity pLLSs/LLSs ($[{\rm X/H}]\lesssim-0.9$), the gas seen in these simulations is more likely to probe metal-poor inflows, but there is also an outflow component.

However, cold accretion --- accretion from the intergalactic medium (IGM) that is never heated to the virial temperature of a galaxy --- is not the only interpretation of the low-metallicity branch absorbers. If accreted material is shock-heated while encountering a galaxy's halo, it may eventually cool from the hot gas and fall toward the center of the galaxy \citep[e.g.,][]{Voit2015}. Such material could be detected in the form of cool \hi\ with column densities similar to pLLSs/LLSs. However, the metallicities we detect are so low that the cooling times of the gas at these metallicities may be exceedingly long. While the cooling times for gas at $T=10^6$ K and densities characteristic of LLSs \citepalias{Lehner2013} are only $t_{\rm cool}\sim1$--3 Gyr \citep[depending on the assumptions and with cooling efficiencies from][]{Oppenheimer2013, Gnat2007}, these have densities well in excess of that expected for the hot gas from which they may have cooled. In fact, we expect this gas to cool isobarically given its very long cooling times relative to reasonable crossing times \citep{Gnat2007}; thus the cooling time from its original state is likely to be a factor of $100\times$ larger than this. Therefore, cooling of such low-metallicity gas from a hot state (e.g., $T=T_{\rm vir}$) is likely to be very difficult. It seems more likely that it has stayed cold from the outset. Whatever the source of this population, it is clear that dense, low-metallicity gas is prevalent in the halos of low-redshift galaxies. The global mass density of gas with low metallicity is at least as much as that of gas tracing outflow phenomena (given the relative number of low- and high-metallicity systems in our MDF).


\subsection{Paucity of \lowercase{p}LLS\lowercase{s}+LLS\lowercase{s} at $[{\rm X/H}]=-1$}

One of the striking elements of the MDF shown in Figures~\ref{f-zdistmnl} and \ref{f-zdistmnlnhistack} is the paucity of pLLSs at $[{\rm X/H}]=-1$, which is also seen in the \Lthirteen\ sample alone. As discussed in \citetalias{Lehner2013} and \citet{Shen2013}, the interpretation of the gap in this log-scale plot has to be interpreted with some care. For example, the mixture of equal amounts of zero- and solar-metallicity gas would wind up well within the high-metallicity branch of the MDF. Additionally, low-metallicity gas may be effectively hidden by overlapping high-metallicity gas unless the \hi\ columns are much stronger in the low-metallicity matter. However, the lack of systems in the $[{\rm X/H}]=-1$ regime raises the question of whether or not there is strong mixing between the two populations. At the very least, the low-metallicity gas likely traces a population of unmixed gas.

The lack of obvious mixing between the two gas populations could be caused by a temporal separation of infalling and outflowing gas around individual galaxies. If accretion fuels star formation that drives large-scale outflows, accreting gas would be present in the galaxy's halo prior to a star formation event that drives strong outflows. If the accretion then either simply ends prior to outflow or is choked off by shock-heating shortly after interacting with outflows, it would no longer be observable as a cool, low-metallicity pLLS or LLS. This clean separation between the metal-rich and -poor absorbers could also be related to a true physical separation for the low- and high-metallicity gas. For example, the metal-rich population could contain a significant contribution from outflows along the host galaxies' minor axes (as expected for outflows and winds), while the metal-poor population could trace matter preferentially infalling along the major axes \citep[along the disk plane; see][]{Stewart2011a}. We note, however, that if tidally-stripped material (or perhaps even recycling material) contributes significantly, there should be little reason to expect an azimuthal dependence in the distribution of that gas.

This potential azimuthal dependence brings to mind recent studies that have identified an azimuthal dependence (with respect to the host galaxy) of absorber properties. For example, the \mgii-selected observations of \citet{Kacprzak2011a, Bordoloi2011, Bordoloi2014, Bouche2012} suggest an azimuthal dependence of CGM absorption properties, with a preference for \mgii\ absorption along the major and minor axes of galaxies. Indeed, \citet{Kacprzak2012a} find that 40\% of their observed \mgii\ absorption is along the galaxies' major axes, and they interpret these as infalling systems. Similarly, \citet{Bouche2013} have studied major-axis absorption from a high-redshift galaxy where the metal-line absorption is offset from the systemic velocity in the same sense as the rotation of the galaxy. They argue this gas is being accreted along the disk plane (although their absorber is more metal-rich than one might naively expect for infall from the IGM).


\subsection{Low-Resolution Method}

In this work we have presented a new approach for estimating the metallicities of pLLSs and LLSs at $z\lesssim1$ that relies on the adoption of an empirical distribution of ionization parameters to estimate metallicities using only \mgii\ and \hi\ observations. Its primary advantage is the greatly-reduced observational cost, requiring a factor of $\sim$10 less space-based telescope time than the method typically employed (i.e., high-resolution UV spectroscopy of many ions). It requires only low-resolution UV spectra to obtain the \hi\ column density and low- or medium-resolution optical spectra to estimate the \mgii\ column density. As demonstrated in Figure~\ref{f-nlvscbw}, our ``low-resolution'' method is able to accurately reproduce the metallicities derived using the standard method of ionization modeling using high-resolution observations. This method works both because the distribution in $\log U$ is relatively narrow and because the ICFs to transform [\mgii/\hi] to [Mg/H] do not vary largely over the range of ionization parameters observed in LLSs. While this statement and our derived ICFs depend somewhat on the choice for the shape of the ultraviolet background (UVB) radiation field, choosing among the several currently-popular radiation fields does not alter our results significantly \citepalias[see the Appendix and][]{Lehner2013}. Furthermore, \citet{Fumagalli2016} investigated this problem in some detail and found that their derived LLS metallicities do not depend sensitively on their choice of UVB, so uncertainties in our method should have a similarly-small effect.

The low-resolution technique is more limited when there are poor estimates of either \hi\ or \mgii\ column density (esp.\ due to non-detections or saturated lines). First, this technique would be more challenging for $z\gtrsim1$ pLLSs and LLSs. At higher redshift, the metallicity distribution of these systems tends toward lower metallicities \citep{Lehner2016}. As such, it requires higher S/N \mgii\ spectra to place stringent limits on very low-metallicity absorbers with $[{\rm X/H}]<-2$ that are common at higher redshifts. Second, we have a limited range over which we can make use of saturated \mgii\ systems for determining lower limits due to the saturation of the break at the Lyman limit. Our method is more susceptible to metal saturation issues than the standard technique because we use only one metal ion. \mgii\ is strong, leading to saturation at lower metallicity, especially in the LLS regime. For example, there are no lower limits in the \Lthirteen\ sample, while there are 19 lower limits in this work (6 in the statistical W16 sample); all are due to \mgii\ saturation. It may also skew the statistics of the metallicity peaks, though the use of survival analysis should negate this effect in part. Finally, we note also that determining \hi\ for $18.0\lesssim\log\mathnhi\lesssim19.0$ limits the \hi\ column density range over which we can study metallicities. Overall, we excluded \numWUpperLower\ systems due to simultaneous lower limits on \nhi\ and \nmgii. Finally, using this method we must assume there is no variation of the metallicity in blended absorbers. However, as discussed above, this effect typically leads to differences of $<$0.2--0.3 dex on the metallicity.

All this being said, our low-resolution metallicity determination potentially has several important uses. In particular, because one only needs low-resolution UV spectroscopy, the number of QSOs against which it is possible to study the metallicities of LLSs is greatly increased. This should allow us to study the LLSs around rarer classes of galaxies, ones for which the number of UV-bright QSO--galaxy pairs is otherwise very low. Our approach is limited to studying absorbers in the column density range $16.4\lesssim\log\mathnhi\lesssim17.7$ or so, but this range in \hi\ column densities has proven to be particularly interesting given the observed bimodal MDF and the potential association with inflows and outflows through the CGM.


\section{Summary}\label{s-summary}
Using our snapshot {\it HST} COS G140L survey of 61 QSOs, we have identified \numW\ new pLLSs and LLSs at $z\lesssim1$ for which we could reliably estimate the \hi\ column density from the break at the Lyman limit. All of these systems are \hi-selected using the break at the Lyman limit to identify them. We complemented the COS observations with ground-based spectroscopic observations of \mgii\ to estimate the metallicities of the pLLSs and LLSs. We show that with a prior knowledge of the ionization parameter ($U$) distribution (from \citetalias{Lehner2013}), we can accurately estimate or place a constraining limit on the metallicities of the pLLSs and LLSs at $z\lesssim1$. This new approach (the ``low-resolution'' method) is powerful for estimating their metallicities, requiring a factor of $\sim$10 less space-based observing time. Combined with the \Lthirteen\ sample, we now have \numLW\ pLLSs+LLSs with $16.1<\log\mathnhi\lesssim17.7$ (\numLWPLLS\ pLLSs and \numLWLLS\ LLSs). Our main results are as follows:

\begin{enumerate}
\item With a sample twice as large as that of \citetalias{Lehner2013}, we show that the metallicity distribution of the optically-thin pLLSs at $0.1\lesssim z\lesssim1.1$ is bimodal, with a \pDipLWPLLS\ confidence and a clear dip between the two branches at $[{\rm X/H}]\simeq-1$. The two peaks are centered at $\left\langle[{\rm X/H}]\right\rangle = \LWPLLSLowmet \pm \LWPLLSLowmetError$ (\LWPLLSLowmetPercent\ solar metallicity) and $\left\langle[{\rm X/H}]\right\rangle = \LWPLLSHighmet \pm \LWPLLSHighmetError$ (\LWPLLSHighmetPercent\ solar metallicity). There are \fLowmetLWPLLS\ of the pLLSs in the low-metallicity peak, suggesting that cool, dense, low-metallicity CGM gas is not uncommon within the virial radius of $z\lesssim1$ galaxies.

\item The metallicity distribution of the optically-thick LLSs is more difficult to assess owing to the small sample size (11 LLSs) and prevalence of lower limits (4/11). However, there is only 1 LLS that is very metal-poor (i.e., with $[{\rm X/H}]<-1.4$), while for the pLLSs there are 19/44. Additionally, there are 4/11 LLSs between $-1.2<[{\rm X/H}]<-0.8$, i.e., in the metallicity gap of the pLLSs. Together, these suggest there may be a difference between the metallicity distributions of the pLLSs and the LLSs.

\item The bimodal MDF of the pLLSs contrasts remarkably with the MDF of the DLAs ($\log\mathnhi\ge20.3$) and SLLSs (a.k.a.\ sub-DLAs; $19\lesssim\log\mathnhi<20.3$), which have unimodal distributions that overlap the high-metallicity branch of the pLLSs. There tend to be fewer very metal-poor (with $[{\rm X/H}]<-1.4$) LLSs than pLLSs, but the LLSs still exhibit a different metallicity distribution from the DLAs, with a large fraction ($\sim$55\%) of LLSs with $[{\rm X/H}]<-1$ compared to 8\% for the DLAs. Hence the MDF of gas in and around galaxies depends sensitively on the \hi\ column density. The transition between the bimodal and unimodal MDF is perhaps beginning as low as $\log\mathnhi=17.2$ and must be in place by $\log\mathnhi\sim19.0$.

\item There is a substantial fraction (35--51\% at the 68\% confidence interval) of very metal-poor pLLSs with $[{\rm X/H}]\lesssim-1.4$, 2$\sigma$ below the mean of the DLAs at $z<1$. The pLLSs are therefore unique probes of very metal-poor gas at $z\lesssim1$. Since typically SLLSs and DLAs probe gas closer to (or in) galaxies than do pLLSs, the change of $[{\rm X/H}]$ with \nhi\ likely suggests a changing mixture of the physical origins of the gas with distance from galaxies' centers.

\item We conclude that the high-metallicity branch of the pLLS MDF likely traces galactic winds, recycled gas, and tidally-stripped gas; i.e., it traces gas that had previously been processed for a while in a galaxy. The low-metallicity pLLSs are probably the best candidates of metal-poor accretion seen in cosmological simulations: they have all the properties expected for metal-poor infalling matter, including the temperature, ionization structure, kinematic properties, and metallicity. Regardless of the exact origin of the low-metallicity pLLSs/LLSs with $[{\rm X/H}]<-1$, their large fraction ($\sim$56\%) implies there is a significant mass of cool, dense, low-metallicity gas at $z\lesssim1$ that may be available as fuel for continuing star formation in galaxies over cosmic time and to control the abundance trends seen in galaxies.
\end{enumerate}


\section*{Acknowledgments}

We thank the referee for providing thoughtful and useful feedback, which helped improve this work. Support for this research was provided by NASA through grants HST-GO-11741, HST-GO-11598, and HST-AR-12854 from the Space Telescope Science Institute, which is operated by the Association of Universities for Research in Astronomy, Incorporated, under NASA contract NAS5-26555. This material is also based upon work supported by the National Science Foundation under Grant AST-1212012. This research has made use of NASA's Astrophysics Data System and the SIMBAD database, operated at CDS, Strasbourg, France. This paper made use of the modsIDL spectral data reduction reduction pipeline developed in part with funds provided by NSF Grant AST-1108693. This paper used data obtained with the MODS spectrographs built with funding from NSF grant AST-9987045 and the NSF Telescope System Instrumentation Program (TSIP), with additional funds from the Ohio Board of Regents and the Ohio State University Office of Research. This paper includes data gathered with the 6.5 meter Magellan Telescopes located at Las Campinas Observatory, Chile. Some of the data presented herein were obtained at the W.M.\ Keck Observatory, which is operated as a scientific partnership among the California Institute of Technology, the University of California and the National Aeronautics and Space Administration. The Observatory was made possible by the generous financial support of the W.M.\ Keck Foundation. The authors wish to recognize and acknowledge the very significant cultural role and reverence that the summit of Mauna Kea has always had within the indigenous Hawaiian community. We are most fortunate to have the opportunity to conduct observations from this mountain.

\facilities{LBT (MODS), Magellan:Clay (MagE), Keck:I HIRES, HST (COS G140L)}



\appendix
\makeatletter
\renewcommand{\thefigure}{A\@arabic\c@figure}
\renewcommand{\thetable}{A\@arabic\c@table}
\makeatother

In determining the metallicities, a crucial step is understanding the uncertainty in the ICF. The ICF relies on a spectrum of ionizing radiation from a UV Background (UVB), including quasar and/or galaxy emission capable of ionizing intervening gas. Changing the shape of the UVB spectrum in the Cloudy photoionization models could have an impact on the derived metallicities, as first discussed in \citetalias{Lehner2013}. Recently, \citet{Fumagalli2016} showed that at $z=2$--3.5 the shape of the UVB spectrum had a small effect on the derived metallicities of the LLSs. Here we explore more quantitatively the effect of changing the UVB spectrum on the metallicities of the pLLSs and LLSs at $z\lesssim1$ derived in \citetalias{Lehner2013}. We use two revisions of the widely-used UVB first introduced in \citet{Haardt1996} to determine the change in the ICF and hence in the metallicity. The important distinction between the 2005 revision of \citet[hereafter \citetalias{Haardt1996}]{Haardt1996}, used in \citetalias{Lehner2013}, and \citet[hereafter \citetalias{Haardt2012}]{Haardt2012} is the greatly-reduced escape fraction of radiation from galaxies to match high-redshift data, leading to a harder UVB spectrum (see Figure 13 in \citealt{Werk2014}). A harder spectrum, where the flux falls off more quickly toward higher energies, systematically increases the calculated metallicities.

To determine whether the shape of the UVB has an impact on our results, we analyze 13 absorbers (9 with $[{\rm X/H}]<-1.0$ and 4 with $[{\rm X/H}]\ge-1.0$) from the \Lthirteen\ sample using the \citetalias{Haardt1996} and \citetalias{Haardt2012} UVBs. These 13 pLLSs+LLSs were selected because they have detailed information regarding the column densities and absorption profiles spanning the range of metallicities observed in the MDF of the pLLSs and LLSs (for more information and for the velocity profiles of these absorbers, see \citetalias{Lehner2013}). We constrain the $\log U$ and $[{\rm X/H}]$ of each absorber using the detailed method described in \S\ref{s-typicalioncorrect} and \citetalias{Lehner2013}, utilizing the available low-ionization state ions (i.e., singly- and doubly-ionized species). We primarily use low-ions, since they typically match the velocity structure of the \hi\ absorption that traces the pLLS or LLS (see \citetalias{Lehner2013}). Higher-ionization state absorption often traces gas in an ionization phase different from that of the pLLS/LLS. For example, \ovi\ associated with pLLSs/LLSs typically traces much more highly ionized gas than the cool gas probed by \hi\ and the low ions \citep{Fox2013}. In our analysis, when \ovi\ is detected, the models underproduced the \ovi\ column densities relative to the observations by several orders of magnitude in every absorber, supporting a multiphase medium. Finally, since [C/$\alpha$] is not necessarily solar in low-metallicity gas owing to different nucleosynthetic histories (see \citetalias{Lehner2013}), we allow this ratio to vary in our models.

In Table~\ref{t-hm05vshm12} we list for each absorber the redshift; the metallicity and ions used to constrain the model using the \citetalias{Haardt1996} UVB spectrum, as provided by \citetalias{Lehner2013}; and the metallicity and ions used to constrain the model using the \citetalias{Haardt2012} UVB spectrum, done in this work. We also report the differences in the derived metallicities between these two UVBs as $\Delta[{\rm X/H}]_{\rm HM} \equiv [{\rm X/H}]_{\rm HM12} - [{\rm X/H}]_{\rm HM05}$. The comparison of the results demonstrates that the harder spectrum of \citetalias{Haardt2012} increases our derived ICF by $\Delta{\rm ICF} \sim 0.0$ to $0.6$ dex, which is reflected in the metallicities. On average, the metallicities changed by by $\left\langle\Delta[{\rm X/H}]_{\rm HM}\right\rangle=+0.3$ dex for the low-metallicity ($[{\rm X/H}]<-1$) and high-metallicity ($[{\rm X/H}]\ge-1$) pLLSs+LLSs. The ionization parameter is different by $\Delta\log U \sim -0.2$ to $0.0$ dex. The amplitude of this systematic error is well within those reported in \citetalias{Lehner2013}. Since the systematic errors are the same for both high and low metallicity, they do not affect the shape of the MDF.

Note that for two absorbers, it was difficult to find an adequate solution using the harder spectrum of \citetalias{Haardt2012} because we were unable to reconcile the ion column densities (and the ratios of column densities between ions of the same element) satisfactorily (see Table~\ref{t-hm05vshm12}). For that reason and because the pLLSs+LLSs are more likely to trace the CGM than the IGM, i.e., gas affected by ionizing radiation from the galaxies, we feel that the \citetalias{Haardt1996} is more suited for our analysis since the \citetalias{Haardt1996} UVB spectrum includes a higher galaxy escape fraction of flux than the \citetalias{Haardt2012} spectrum.


\clearpage
\rotatetablecbw
\begin{deluxetable*}{lcc|cc|cc|cc}
\tabcolsep=3pt
\tablecolumns{9}
\tablewidth{0pc}
\tablecaption{Comparison of the metallicities derived using the \citetalias{Haardt1996} and \citetalias{Haardt2012} UVB spectra\label{t-hm05vshm12}}
\tabletypesize{\scriptsize}
\tablehead{\colhead{Sightline} & \colhead{$z_{\rm abs}$} & \colhead{$\mathnhi$} & \multicolumn{2}{c}{\citetalias{Haardt1996}} & \multicolumn{2}{c}{\citetalias{Haardt2012}} & \multicolumn{2}{c}{\citetalias{Haardt1996} vs.\ \citetalias{Haardt2012}}}
\startdata
    &  &  & $[{\rm X/H}]_{\Lthirteen}$ & Ions Matched \citepalias{Lehner2013} & [{\rm X/H}] & Ions Matched & $\Delta[{\rm X/H}]_{\rm HM}$ & Notes \\
\cline{4-9}
    PG1522+101   &  0.7289  &  16.66  &  $-$2.00                   &  \mgiit, \oiiit, \Siiit\ (\ciit, \ciiit, \oiit, \sivt, \svt)  &  $-$2.00                   &  \oiiit, \Siiit\ (\ciit, \ciiit, \mgiit, \oiit, \sivt, \svt)   &  0.00 & \tablenotemark{a}  \\ 
    J1435+3604   &  0.3730  &  16.65  &  $-$1.85                   &  \ciiit, \mgiit, \siiiit\ (\oiit, \oiiit, \siiit)              &  $-$1.55                   &  \ciiit, \mgiit, \siiiit\ (\oiit, \oiiit, \siiit)              & +0.30 & \tablenotemark{a}  \\
    PG1407+265   &  0.6828  &  16.38  &  $-$1.80                   &  \mgiit, \oiiit, \Siiit, \sivt\ (\ciit, \ciiit, \oiit)        &  $-$1.10\tablenotemark{b}  &  \oivt, \Siiit, \sivt, \svt, \svit\ (\ciit, \ciiit, \oiit, \mgiit)  & +0.70 & \tablenotemark{c}  \\ 
    PG1630+377   &  0.2740  &  16.98  &  $-$1.71                   &  \mgiit, \siiit, \siiiit, \siivt\ (\ciit, \ciiit)             &  $-$1.24                   &  \mgiit, \siiiit, \siivt\ (\ciit, \ciiit)                      & +0.47 & \tablenotemark{a,d}  \\ 
    PG1216+069   &  0.2823  &  16.40  &  $-$1.65                   &  \siiit, \siiiit\ (\ciit, \ciiit, \niit)                      &  $-$1.35                   &  \siiit, \siiiit\ (\ciit, \ciiit, \niit)                       & +0.30 & \tablenotemark{a}  \\  
    J1619+3342   &  0.2694  &  16.48  &  $-$1.60\tablenotemark{b}  &  \mgiit, \siiiit\ (\ciit, \ciiit, \siiit, \siivt)             &  $-$1.00\tablenotemark{b}  &  \mgiit, \siiiit\ (\ciit, \ciiit, \siiit, \siivt)              & +0.60 & --  \\
    PHL1377      &  0.7390  &  16.72  &  $-$1.45                   &  \mgiit, \Siiit\ (\aliit, \ciit, \ciiit)                      &  $-$1.15\tablenotemark{b}  &  \mgiit, \Siiit\ (\aliit, \ciit, \ciiit)                       & +0.30 & \tablenotemark{e}  \\ 
    J1435+3604   &  0.3878  &  16.18  &  $<$$-$1.40                &  \mgiit, \siiiit\ (\oiit)                                     &  $<$$-$1.40                &  \mgiit, \siiiit\ (\ciit, \oiit, \siiit)                       & $<$+0.00 & \tablenotemark{a,f}  \\ 
    SBS1122+594  &  0.5574  &  16.24  &  $-$1.05\tablenotemark{b}  &  \oiit, \oiiit\ (\ciit, \ciiit, \oivt, \sivt)                 &  $-$1.05\tablenotemark{b}  &  \oiit, \oiiit, \oivt\ (\ciit, \ciiit, \sivt)                  &  0.00 & --  \\
    PKS0637-752  &  0.4685  &  16.48  &  $-$0.50                   &  \mgiit, \oiit\ (\ciit, \sivt)                                &    +0.20                   &  \mgiit, \oiit\ (\ciit, \sivt)                                 & +0.70 & \tablenotemark{a,g}  \\ 
    PG1522+101   &  0.5185  &  16.22  &  $-$0.40\tablenotemark{b}  &  \mgiit, \oiit, \oiiit\ (\ciit, \ciiit, \oivt, \ovit)         &  $-$0.38\tablenotemark{b}  &  \mgiit, \oiit, \oiiit\ (\ciit, \ciiit, \oivt, \ovit)          & +0.03 & --  \\
    HE0439-5254  &  0.6153  &  16.28  &  $-$0.30\tablenotemark{h}  &  \oiit, \Siiit\ (\ciit, \ciiit, \oiiit)                       &  $-$0.15                   &  \oiit, \Siiit\ (\ciit, \ciiit, \oiiit)                        & +0.15 & \tablenotemark{i}  \\ 
    PG1338+416   &  0.6865  &  16.45  &    +0.10\tablenotemark{h}  &  \mgiit, \oiit\ (\ciit)                                       &    +0.50\tablenotemark{h}  &  \mgiit, \oiit\ (\ciit)                                        & +0.40 & \tablenotemark{j} 
\enddata
\tablecomments{
The metallicity difference is $\Delta[{\rm X/H}]_{\rm HM}\equiv[{\rm X/H}]_{\rm HM12}-[{\rm X/H}]_{\rm HM05}$. The ions listed outside of parentheses are detections or limits that were the most constraining\\in determining $\log U$ in the models, while the ions listed inside parentheses were detections or limits that were consistent with observations, but less constraining.
}
\tablenotetext{a}{A solar $[{\rm C}/\alpha]$ ratio was derived using both \citetalias{Haardt1996} and \citetalias{Haardt2012}.}
\tablenotetext{b}{A subsolar $[{\rm C}/\alpha]$ ratio was derived using the respective UVB.}
\tablenotetext{c}{The models using the \citetalias{Haardt1996} UVB underproduced \oivt, \svt, and \svit.}
\tablenotetext{d}{The models using the \citetalias{Haardt2012} UVB are unable to reproduce \siiit.}
\tablenotetext{e}{Neither UVB model is able to reproduce \civt\ for this absorber.}
\tablenotetext{f}{The column densities of \ciiit\ and \oiiit\ are uncertain because of a small velocity shift relative to \hit\ in this\\absorber (see \citetalias{Lehner2013}), so we only derive an upper limit on the metallicity.}
\tablenotetext{g}{This absorber is multiphase. The models are unable to reproduce \ciiit, \oiiit, \oivt, \ovit, \svt.}
\tablenotetext{h}{A supersolar $[{\rm C}/\alpha]$ ratio was derived using the respective UVB.}
\tablenotetext{i}{This absorber is multiphase. The models are unable to reproduce \oivt, \sivt, \svt, \svit.}
\tablenotetext{j}{This absorber is multiphase. The models are only able to reproduce the low-ions; they underproduce \Siiit, \ciiit.}
\end{deluxetable*}

\endrotatetablecbw
\clearpage


\end{document}